\documentclass[oneside,english,reqno]{amsbook}
\usepackage{lmodern}

\usepackage[T1]{fontenc}
\usepackage[latin9]{inputenc}
\usepackage{geometry}
\usepackage{color}
\usepackage{babel}
\usepackage{textcomp}
\usepackage{amsthm}
\usepackage{amssymb}
\usepackage{graphicx}
\usepackage{setspace}
\usepackage{esint}
\usepackage[unicode=true,pdfusetitle,
 bookmarks=true,bookmarksnumbered=false,bookmarksopen=false,
 breaklinks=false,pdfborder={0 0 1},backref=false,colorlinks=false]
 {hyperref}

\makeatletter

\providecommand{\tabularnewline}{\\}

\numberwithin{section}{chapter}
\numberwithin{equation}{section}
\numberwithin{figure}{section}

\makeatother

\begin{document}
\begin{center}
\textbf{\Huge{Theory of adiabatic quantum control in the presence
of cavity-photon shot noise}}
\par\end{center}{\Huge \par}

\bigskip{}

\begin{center}
\textbf{\LARGE{Christopher Chamberland}}\\
\textbf{\LARGE{Department of Physics}}\\
\textbf{\LARGE{McGill University, Montreal}}
\par\end{center}{\LARGE \par}

\medskip{}

\begin{center}
\textbf{\LARGE{March 2014}}
\par\end{center}{\LARGE \par}

\bigskip{}

\bigskip{}
\bigskip{}
\bigskip{}
\bigskip{}
\bigskip{}
\bigskip{}
\bigskip{}
\bigskip{}
\bigskip{}
\bigskip{}
\bigskip{}
\bigskip{}
\bigskip{}
\bigskip{}
\bigskip{}
\bigskip{}
\bigskip{}
\bigskip{}
\bigskip{}
\bigskip{}

\begin{center}
\textbf{\Large{A thesis submitted to McGill University in partial
fulfillment of the requirements of the degree of M.Sc.}}
\par\end{center}{\Large \par}

\bigskip{}
\bigskip{}
\bigskip{}

\begin{center}
© \textbf{\large{Christopher Chamberland 2014}}
\par\end{center}{\large \par}

\newpage{}

\textit{\large{À mes parents, à Marie-Michelle et à mon oncle Claude.}}\newpage{}

\begin{center}
\tableofcontents{}
\par\end{center}

\newpage{}

\listoffigures

\begin{center}
\newpage{}
\par\end{center}

\chapter*{Abstract}

\begin{doublespace}
Many areas of physics rely upon adiabatic state transfer protocols,
allowing a quantum state to be moved between different physical systems
for storage and retrieval or state manipulation. However, these state-transfer
protocols suffer from dephasing and dissipation. In this thesis we
go beyond the standard open-systems treatment of quantum dissipation
allowing us to consider non-Markovian environments. We use adiabatic
perturbation theory in order to give analytic descriptions for various
quantum state-transfer protocols. The leading-order corrections will
give rise to additional terms adding to the geometric phase preventing
us from achieving a perfect fidelity. We obtain analytical descriptions
for the effects of the geometric phase in non-Markovian regimes. The
Markovian regime is usually treated by solving a standard Bloch-Redfield
master equation, while in the non-Markovian regime, we perform a secular
approximation allowing us to obtain a solution to the density matrix
without solving master equations. This solution contains all the relevant
phase information for our state-transfer protocol. After developing
the general theoretical tools, we apply our methods to adiabatic state
transfer between a four-level atom in a driven cavity. We explicitly
consider dephasing effects due to unavoidable photon shot noise and
give a protocol for performing a phase gate. These results will be
useful to ongoing experiments in circuit quantum electrodynamics (QED)
systems.
\end{doublespace}

\newpage{}

\chapter*{Résumé}

\begin{doublespace}
Plusieurs domaine de la physique ce fit sur des protocoles adiabatique
de transfert d'état, permettant un état quantique de se déplacer entre
diffèrent système physique afin de les emmagasiner, les récupéré ou
de les manipuler. Par contre, ces protocoles de transferts d'état
souffrent du phénomène de déphasage et de dissipation. Dans ce mémoire,
nous allons au-delà des traitements standards de système ouverts décrivant
la dissipation quantique nous permettant de considérer des environnements
non-Markoviens. Nous utilisons la théorie de perturbation adiabatique
afin de donner une description analytique à plusieurs protocoles de
transfert quantique. Les corrections d\textquoteright{}ordres principaux
donnent lieu à une phase géométrique de système ouvert, ce qui nous
empêche d'atteindre une fidélité parfaite. Nous obtenons une description
analytique pour les effets de phase géométrique dans des régimes non-Markovien.
Les régimes Markovien sont habituellement traités en trouvant une
solution à une équation de maitresse Bloch-Redfield, tandis que pour
les régimes non-Markovien, nous allons performer une approximation
séculaire nous permettant d'obtenir une solution exacte de la matrice
de densité qui contient toute l'information de phase pertinente pour
nôtre protocole de transfert d'état. Après avoir développé les outils
théoriques généraux, nous allons appliquer nos méthodes à un transfert
d'état adiabatique entre les états d'un atom à quatre niveau dans
une cavité entraînée. Nous considérons explicitement les effets de
déphasage du au bruit photonique inévitable. Ces résultats seront
utiles pour des expériences en cours dans des systèmes de circuits
QED.
\end{doublespace}

\newpage{}

\chapter*{Acknowledgments}

\begin{doublespace}
I would like to sincerely thank both of my supervisors, Aash Clerk
and Bill Coish for their continued support and guidance throughout
my masters degree. Your never ending patience and very generous office
hours have proven to be extremely valuable in making progress during
my masters degree. I am also very grateful to have had the chance
to work with two extremely talented and visionary physicists. This
has without a doubt been the most inspiring part of my masters degree.
Both of you have been very helpful in making me become a more independent
researcher which are skills that I will carry with me for the rest
of my life.

I would like to give many thanks to Tami Pereg-Barnea for sharing
much of her research ideas and opening doors into a new world of physics
that I would have never thought I would be interested in. 

I would like to thank Benjamin D'Anjou, Félix Beaudoin, Marc-Antoine
Lemonde, Saeed Khan, Judy Wang, Abhishek Kumar and Pawel Mazurek for
your guidance and many scientific discussions that have not only been
essential but very enlightening throughout my masters degree. I would
also like to thank Aaron Farrell, Pericles Philippopoulos and Ben
Levitan for your many collaborations and scientific ideas. 

I would like to thank Wei Chen, Stefano Chesi, Anja Metelmann and
Nicolas Didier for their very sound advice and help in guiding me
throughout my masters degree.

I owe a very large thank you to Marius Oltean. Your constant presence
and support has been extremely valuable. Your help in teaching me
LaTex and numeric's has been essential on many occasions and I could
not have done it without you. I would also like to thank you for our
many deep scientific dicussions about the very foundation of physics
and especially quantum mechanics. There are very few people who worry
about these types of questions but they are of uttermost importance.
I am very grateful for having had a friend like you and it has certainly
opened my eyes to some of the most fascinating features that physics
has to offer. 

J'aimerais remercier mon oncle Claude pour son hébergement, son support
constant et ses conseils extrêmement pratique. Merci aussi pour m'avoir
guidé à travers la ville de Montréal et pour les délicieux repas.

Finally, I owe my infinite gratitude to my parents and sister for
their continued support throughout my time at McGill. Thank you for
taking the time to put up with me during hard times and for your continued
moral support. Also, thank you for always looking out for me and making
sure sure that I always focus on the most important things.
\end{doublespace}

\newpage{}

\chapter{Introduction}

\rule[0.5ex]{1\columnwidth}{1pt}

\begin{doublespace}
Quantum state transfer arises when one would like to transfer a quantum
state from one system to another with the highest possible fidelity.
There are many state-transfer protocols that are of practical significance.
For example, if one is interested in building a secure quantum system
for private communication or for long-distance quantum communication,
it is necessary to transfer information between quantum-memory atoms
and photons \cite{key-1,key-2,key-3,key-4,key-5,key-6}. In other
situations, it is crucial to be able to preserve a quantum state for
as long as possible in order to trade states back and fourth between
an ensemble of atoms and photons \cite{key-7,key-8,key-9,key-10,key-11,key-12,key-13}.
Quantum state transfer can also lead to striking results. For instance,
when applied to quantum-logic clocks, they become so precise that
they are able to keep time to within 1 second every 3.7 billion years
\cite{key-14}. In reality, when one performs a state-transfer protocol,
the quantum systems under consideration are never completely isolated
from an exterior environment. Consequently, regardless of which state-transfer
protocol is being employed, one will always be faced with the challenge
of fighting decoherence. 

\begin{figure}
\begin{centering}
\includegraphics[scale=0.5]{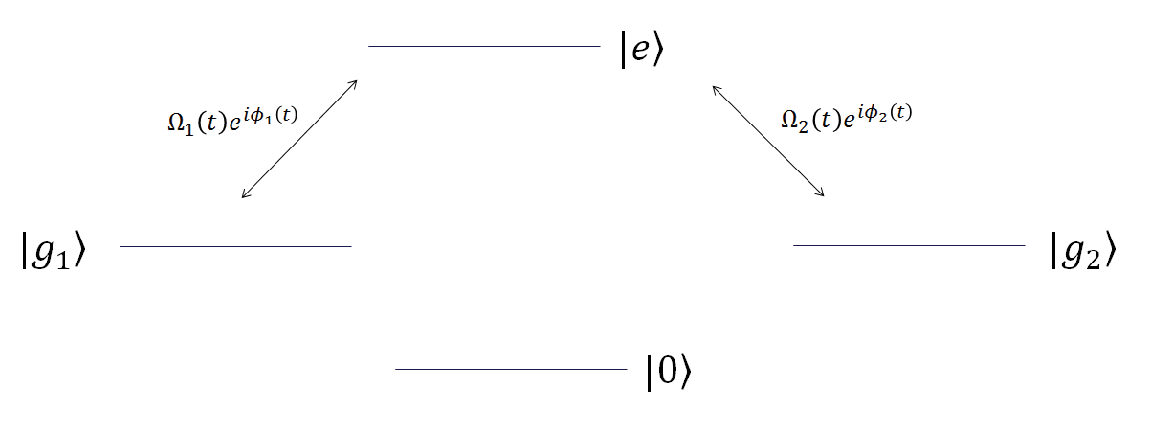}
\par\end{centering}

\caption{\label{fig:State-transfer-diagram}\ State-transfer diagram}
\textit{Schematic representation of the state-transfer protocol described
by the Hamiltonian of Eq. (\ref{eq:2.2.9}). The two ground states
$\left.|g_{1}\right\rangle $ and $\left.|g_{2}\right\rangle $ are
coupled to the excited state $\left.|e\right\rangle $ via two time-dependent
classical laser fields. The spectator state $\left.|0\right\rangle $
does not couple to any other state (which is why we call it a spectator
state). }

\end{figure}

During the early days of quantum mechanics, the adiabatic theorem
was a perturbative tool that was developed to deal with slowly-varying
time-dependent Hamiltonians \cite{key-15,key-16}. The Hamiltonian
$H\left(t\right)$ must vary slowly compared to internal time scales
of the system quantified by a dimensionless parameter $A$. For example,
the parameter $A$ for a spin-half coupled to a time-dependent magnetic
field would be $Bt_{p}$ where $t_{p}$ is the total time it takes
the magnetic field to complete a closed loop trajectory in 3-D space
\cite{key-29}. To leading order in adiabatic perturbation theory,
the instantaneous eigenstates of the Hamiltonian acquire a geometric
phase (along with its usual dynamical phase). The term ``geometric
phase'' comes from the fact that this phase depends on the path traversed
by the system in Hilbert space but not on the time to traverse the
path \cite{key-17}. Recently, it has been shown that in the framework
of quantum information and quantum computation, the geometric phase
is robust in the presence of some sources of decoherence \cite{key-18,key-19,key-20,key-21}.
Thus, it may be useful to take advantage of this robustness to perform
a phase gate in the presence of a dissipative quantum system. On the
other hand, corrections to the geometric phase in the presence of
a dissipative environment will be the leading source preventing one
from doing a state-transfer protocol with perfect fidelity. Given
these considerations, it is of paramount importance to study geometric
phases for open quantum systems. Many authors \cite{key-22,key-23,key-24,key-25,key-26,key-27,key-28,key-29}
have considered geometric phases for systems coupled to classical
or quantum environments. The authors of \cite{key-29} consider a
spin-1/2 system which is both subject to a slowly-varying magnetic
field and weakly coupled to a dissipative environment (which the authors
choose to be quantum). The authors obtained a modification to the
closed-system geometric phase due to the coupling to the environment
and found that this phase is also geometrical. However, the authors'
results are limited by the fact that they consider only weak dissipation
and an environment giving rise to Markovian evolution. In this thesis,
we develop a general theory that gives rise to non-Markovian evolution
by performing a secular approximation following the criteria of \cite{key-30}.
As was shown in \cite{key-30}, performing a secular approximation
on time-dependent systems can often be problematic and must be done
with care. Consequently, we will go into some detail in order to ensure
that the secular approximation is done correctly. Given that our main
motivation is to perform a phase gate under a quantum state-transfer
protocol while minimizing dephasing effects, we will develop our theory
under the framework of a Stimulated Raman Adiabatic Passage (STIRAP)
applied to a tripod system (Fig. \ref{fig:State-transfer-diagram}).
A STIRAP system is an adiabatic process that is efficient at transferring
population in a three-level system \cite{key-31,key-32,key-33,key-34,key-35,key-36,key-37,key-38}.
When applied to a tripod system, one now deals with four levels (instead
of three) allowing the possibility to transfer population amongst
the two ground state levels (the levels spacing $\left.|g_{1}\right\rangle $
and$\left.|g_{2}\right\rangle $ in Fig. \ref{fig:State-transfer-diagram}).
Our task will be twofold. First we will perform a state-transfer protocol
in the presence of a dissipative environment. Secondly, we will perform
a phase-gate state-transfer protocol by transferring quantum information
between the four-level system the presence of a dissipative environment.
The fourth level (the state $\left.|0\right\rangle $ in Fig. \ref{fig:State-transfer-diagram})
in our tripod system will act as a spectator state meaning that no
driving terms will couple it to an excited state (unlike the two other
levels of the tripod system which are coupled to an excited state
via time-dependent classical fields). In this case, the spectator
state will not acquire a geometric phase. However, the other instantaneous
eigenstates of our system will acquire an environment-induced (as
well as a closed-system) geometric phase. It will thus be crucial
to understand how non-Markovian environments modify the closed-system
geometric phase.

In chapter 2 we give an overview of adiabatic perturbation theory
and of the dynamics of the geometric phase. We will then apply adiabatic
perturbation theory to describe the closed-system four-level adiabatic
state-transfer protocol described above. We will also show how all
the phase information is encoded in the off-diagonal components of
the density matrix and how our state-transfer protocol can be described
in the language of density matrices. We will conclude chapter 2 by
giving an overview of cavity-photon shot noise. In chapter 3, we develop
a theory allowing us to calculate dephasing effects in our state-transfer
protocol for the case where the system is coupled to a quantum dissipative
bath. We will do this by performing a secular approximation in a superadiabatic
basis allowing us to obtain the relevant off-diagonal component of
the density matrix. We will conclude chapter 3 by applying our theory
to a system of independent bosonic modes coupled to our four-level
system. In chapter 4, we will apply our methods to adiabatic state
transfer of a four-level system in a driven cavity. We will explicitly
consider dephasing effects due to unavoidable photon shot noise. Once
the dephasing effects have been accounted for, we will show how it
is possible to perform a phase gate by choosing specific functional
forms for the phase difference between the two laser fields driving
the cavity. These results will be useful to ongoing experiments in
circuit QED systems.
\end{doublespace}

\newpage{}

\chapter{Adiabatic evolution and noise}

\rule[0.5ex]{1\columnwidth}{1pt}

\section{Adiabatic approximation and the Berry's phase\label{sec:Adiabatic-approximation-and}}

\begin{doublespace}
When dealing with physical systems where the Hamiltonian takes on
an explicit time dependence, it is often very difficult to solve the
Schroedinger equation exactly. One must often approach the problem
using perturbative techniques to have some hope of getting analytical
results. The adiabatic approximation is very useful for cases where
the Hamiltonian changes slowly in time. However, at this stage, one
must correctly define what we mean by changing ``slowly in time''.
To address this issue on analytical grounds, we can start by looking
at the energy eigenvalues of the Hamiltonian. For a fixed value of
time, it is always possible to obtain the energy spectrum of the Hamiltonian
which depends on a set of parameters. Consequently, when we say that
the Hamiltonian changes slowly in time, we mean that the set of parameters
change on a time scale $T$ that is much larger than $2\pi/E_{ab}$
where $E_{ab}$ are the energy differences between two levels. To
describe this mechanism mathematically, we follow closely the treatment
given in \cite{key-39}. The first step is to find the instantaneous
eigenstates of the Hamiltonian. Since the eigenstates are defined
for each moment in time, we can represent them as 
\begin{equation}
\hat{H}\left(t\right)\left.|\phi_{n}\left(t\right)\right\rangle =E\left(t\right)\left.|\phi_{n}\left(t\right)\right\rangle .\label{eq:2.1.1}
\end{equation}
Here, $\left.|\phi_{n}\left(t\right)\right\rangle $ correspond to
the instantaneous eigenstates of the Hamiltonian and $E\left(t\right)$
are the corresponding instantaneous eigenenergies. We can write the
usual Schroedinger equation as (setting $\hbar=1$) 
\begin{equation}
i\frac{\partial}{\partial t}\left.|\psi\left(t\right)\right\rangle =\hat{H}\left(t\right)\left.|\psi\left(t\right)\right\rangle .\label{eq:2.1.2}
\end{equation}
The trick is to expand the wave function satisfying the Schroedinger
equation as 
\begin{equation}
\left.|\psi\left(t\right)\right\rangle =\sum_{n}c_{n}\left(t\right)e^{i\theta_{n}\left(t\right)}\left.|\phi_{n}\left(t\right)\right\rangle ,\label{eq:2.1.3}
\end{equation}
where we define
\begin{equation}
\theta_{n}\left(t\right)\equiv-\int_{0}^{t}E_{n}\left(t'\right)dt'.\label{eq:2.1.4}
\end{equation}
Note that in (\ref{eq:2.1.3}) the coefficients $c_{n}\left(t\right)$
are taken to be real. The phase $\theta_{n}\left(t\right)$ is referred
to as the dynamical phase. Next we substitute the expansion (\ref{eq:2.1.3})
into (\ref{eq:2.1.2}) and use the property (\ref{eq:2.1.1}). After
a little bit of algebra and using the orthonormality property of the
instantaneous eigenstates, we find the following differential equation
for the expansion coefficients 
\begin{equation}
\dot{c}_{m}\left(t\right)=-\sum_{n}c_{n}\left(t\right)e^{i\left[\theta_{n}\left(t\right)-\theta_{m}\left(t\right)\right]}\left\langle \phi_{m}\left(t\right)|\right.\left[\partial_{t}\left.|\phi_{n}\left(t\right)\right\rangle \right].\label{eq:2.1.5}
\end{equation}
Notice the inner product $\left\langle \phi_{m}\left(t\right)|\right.\left[\partial_{t}\left.|\phi_{n}\left(t\right)\right\rangle \right]$
appearing in equation (\ref{eq:2.1.5}). Since the Hamiltonian is
time dependent, this quantity will not vanish in general. Thus, we
must find a way to calculate it. This can be done by going back to
the eigenvalue equation (\ref{eq:2.1.1}) and taking a time derivative
on both sides of it. If we restrict ourselves to the case where $m\neq n$,
we find that 
\begin{equation}
\left\langle \phi_{m}\left(t\right)|\right.\dot{\hat{H}}\left(t\right)\left.|\phi_{n}\left(t\right)\right\rangle =\left[E_{n}\left(t\right)-E_{m}\left(t\right)\right]\left\langle \phi_{m}\left(t\right)|\right.\left[\partial_{t}\left.|\phi_{n}\left(t\right)\right\rangle \right].\label{eq:2.1.6}
\end{equation}
We can replace (\ref{eq:2.1.6}) into (\ref{eq:2.1.5}) for the term
where $m\neq n$ and this term can thus be written as 
\begin{equation}
\dot{c}_{m}\left(t\right)=-c_{m}\left(t\right)\left\langle \phi_{m}\left(t\right)|\right.\left[\partial_{t}\left.|\phi_{m}\left(t\right)\right\rangle \right]-\sum_{n\neq m}c_{n}\left(t\right)e^{i\left[\theta_{n}\left(t\right)-\theta_{m}\left(t\right)\right]}\frac{\left\langle \phi_{m}\left(t\right)|\right.\dot{\hat{H}}\left(t\right)\left.|\phi_{n}\left(t\right)\right\rangle }{E_{n}-E_{m}}.\label{eq:2.1.7}
\end{equation}
So far everything is exact. Due to the second term in (\ref{eq:2.1.7}),
as the system evolves in time, the states $\left.|\phi_{n}\left(t\right)\right\rangle $
(with $n\neq m$) will mix with $\left.|\phi_{m}\left(t\right)\right\rangle $.
The adiabatic approximation consists of neglecting the second term
in (\ref{eq:2.1.7}). In other words, we must require that 
\begin{equation}
\left|\frac{\left\langle \phi_{m}\left(t\right)|\right.\dot{\hat{H}}\left(t\right)\left.|\phi_{n}\left(t\right)\right\rangle }{E_{n}-E_{m}}\right|\ll\left|\left\langle \phi_{m}\left(t\right)|\right.\left[\partial_{t}\left.|\phi_{m}\left(t\right)\right\rangle \right]\right|.\label{eq:2.1.8}
\end{equation}
If we identify the term on the left of (\ref{eq:2.1.8}) as $1/\tau$,
where $\tau$ describes the characteristic time scale for changes
in the Hamiltonian and the term on the right as $E_{m}$ (natural
frequency of the state-phase factor), then we require that $\tau\gg1/E_{m}$
\cite{key-39}. In other words, the characteristic time scale for
changes in the Hamiltonian must be much larger than the inverse natural
frequency of the state-phase factor. However, we must be careful in
using (\ref{eq:2.1.8}) to establish the validity of the adiabatic
approximation. To see this, we consider the following argument. Consider
an equation of motion of the form 
\begin{equation}
\dot{c}_{m}(t)=\delta c_{m}^{0}(t)+\delta c_{m}^{\nu}(t),\label{eq:2.1.9}
\end{equation}
with the condition 
\begin{equation}
|\delta c_{m}^{0}(t)|\gg|\delta c_{m}^{\nu}(t)|,\label{eq:2.1.10}
\end{equation}
 (here the coefficients $\delta c_{m}^{0}(t)$ and $\delta c_{m}^{\nu}(t)$
are time-dependent functions analogous to the first and second term
on the right hand side of (\ref{eq:2.1.7})). We have to be very careful
when we say that we can neglect the second term under the condition
(\ref{eq:2.1.10}). If we integrate (\ref{eq:2.1.9}) we get
\begin{equation}
c_{m}(t)=c_{m}(0)+\int_{0}^{t}\delta c_{m}^{0}(t')dt'+\int_{0}^{t}\delta c_{m}^{\nu}(t')dt'.\label{eq:2.1.11}
\end{equation}
After a time scale $\tilde{t}$ the second term in (\ref{eq:2.1.11})
could be of order one and thus could no longer be neglected. The important
information to gather out of this argument is that we can always use
(\ref{eq:2.1.8}) as a criterion for the adiabatic approximation as
long as we are within a time scale which guarantees that the term
\begin{equation}
\int_{0}^{t}\sum_{n}c_{n}(t')e^{i\left[\theta_{n}(t')-\theta_{m}(t')\right]}\frac{\left\langle \phi_{m}\left(t'\right)|\right.\dot{\hat{H}}\left(t\right)\left.|\phi_{n}\left(t'\right)\right\rangle }{\left[E_{n}(t')-E_{m}(t')\right]}dt'\lesssim1,\label{eq:2.1.12}
\end{equation}
 is below of order one. Above this time scale we can no longer use
adiabatic perturbation theory with confidence. Assuming we are within
a time scale such that (\ref{eq:2.1.8}) is valid, (\ref{eq:2.1.7})
will simplify to 
\begin{equation}
c_{n}\left(t\right)\cong e^{i\gamma_{n}\left(t\right)}c_{n}\left(0\right),\label{eq:2.1.13}
\end{equation}
where 
\begin{equation}
\gamma_{n}\left(t\right)\equiv i\int_{0}^{t}\left\langle \phi_{n}\left(t'\right)|\right.\left[\partial_{t}\left.|\phi_{n}\left(t'\right)\right\rangle \right]dt'.\label{eq:2.1.14}
\end{equation}
The function $\gamma_{n}\left(t\right)$ is called Berry's phase and
it satisfies some nice properties. As a first note, $\gamma_{n}\left(t\right)$
is real which can be seen from the fact that differentiating the inner
product of $\left.|\phi_{n}\left(t\right)\right\rangle $ with itself
gives zero so that 
\begin{equation}
\left\langle \phi_{n}\left(t\right)|\right.\left[\partial_{t}\left.|\phi_{n}\left(t\right)\right\rangle \right]=-\left(\left\langle \phi_{n}\left(t'\right)|\right.\left[\partial_{t}\left.|\phi_{n}\left(t'\right)\right\rangle \right]\right)^{*}.\label{eq:2.1.15}
\end{equation}
The integrand of (\ref{eq:2.1.14}) is then purely imaginary. Coming
back to the state (\ref{eq:2.1.3}) and using (\ref{eq:2.1.13}),
we find that 
\begin{equation}
\left.|\psi_{n}\left(t\right)\right\rangle =e^{i\gamma_{n}\left(t\right)+i\theta_{n}\left(t\right)}\left.|\phi_{n}\left(t\right)\right\rangle .\label{eq:2.1.16}
\end{equation}
We can thus conclude that in the adiabatic approximation, the general
solution to the Schroedinger equation is given by a linear combination
of instantaneous eigenstates modulated by a dynamical plus geometric
phase. As will be seen below, Berry's phase (also known as the geometric
phase) plays a crucial role for systems that are cyclic in time. Note
that the derivation of the geometric phase has been for closed quantum
systems. In chapter 3 we will consider quantum systems interacting
with an environment. The interaction with the environment will induce
corrections to the geometric phase and part of the goal of chapter
3 will be to develop methods that will allow us to calculate and interpret
these corrections. 
\end{doublespace}

\section{Quantum state transfer protocol}

\begin{doublespace}
In this section we will give a quantitative description of the physical
system describing how information will be transferred between two
qubit states. Throughout this thesis we will often perform unitary
transformations on the Hamiltonian allowing us to go into a rotating
frame. For instance, this will be essential when writing our Hamiltonian
in a superadiabatic basis allowing us to perform a secular approximation
\cite{key-30}. To describe the transformation of the Hamiltonian
when going into a rotating frame, we start with the Schroedinger equation
\begin{equation}
i\frac{\partial}{\partial t}\left.|\psi\left(t\right)\right\rangle =\hat{H}\left(t\right)\left.|\psi\left(t\right)\right\rangle .\label{eq:2.2.1}
\end{equation}
Next we apply a unitary transformation to the Schroedinger picture
eigenstates 
\begin{equation}
\left.|\tilde{\psi}\left(t\right)\right\rangle =\hat{U}\left(t\right)\left.|\psi\left(t\right)\right\rangle .\label{eq:2.2.2}
\end{equation}
At this stage we would like to write down a Schroedinger equation
for the transformed eigenstates $\left.|\tilde{\psi}\left(t\right)\right\rangle $.
We can use (\ref{eq:2.2.1}) by writing $\left.|\psi\left(t\right)\right\rangle =\hat{U}\left(t\right)^{-1}\left.|\tilde{\psi}\left(t\right)\right\rangle $
(note also the time dependence in the unitary operator $\hat{U}\left(t\right)$).
We then have 
\begin{equation}
i\frac{\partial}{\partial t}\left(\hat{U}\left(t\right)^{-1}\left.|\tilde{\psi}\left(t\right)\right\rangle \right)=\hat{H}\left(t\right)\left(\hat{U}\left(t\right)^{-1}\left.|\tilde{\psi}\left(t\right)\right\rangle \right).\label{eq:2.2.3}
\end{equation}
Using the fact that $\hat{U}\left(t\right)$ is time dependent (\ref{eq:2.2.3})
becomes 
\begin{align}
\left(\frac{\partial\hat{U}\left(t\right)^{-1}}{\partial t}\left.|\tilde{\psi}\left(t\right)\right\rangle +\hat{U}\left(t\right)^{-1}\frac{\partial}{\partial t}\left.|\tilde{\psi}\left(t\right)\right\rangle \right) & =\hat{H}\left(t\right)\hat{U}\left(t\right)^{-1}\left.|\tilde{\psi}\left(t\right)\right\rangle \nonumber \\
i\frac{\partial}{\partial t}\left.|\tilde{\psi}\left(t\right)\right\rangle  & =\hat{U}\left(t\right)\hat{H}\left(t\right)\hat{U}\left(t\right)^{-1}\left.|\tilde{\psi}\left(t\right)\right\rangle -i\hat{U}\left(t\right)\frac{\partial\hat{U}\left(t\right)^{-1}}{\partial t}\left.|\tilde{\psi}\left(t\right)\right\rangle \nonumber \\
i\frac{\partial}{\partial t}\left.|\tilde{\psi}\left(t\right)\right\rangle  & =\left(\hat{U}\left(t\right)\hat{H}\left(t\right)\hat{U}\left(t\right)^{-1}-i\hat{U}\left(t\right)\frac{\partial\hat{U}\left(t\right)^{-1}}{\partial t}\right)\left.|\tilde{\psi}\left(t\right)\right\rangle .\label{eq:2.2.4}
\end{align}
We have written the Schroedinger equation for the states $\left.|\tilde{\psi}\left(t\right)\right\rangle $
in the rotating frame with the new Hamiltonian
\begin{equation}
\tilde{\hat{H}}\left(t\right)=\hat{U}\left(t\right)\hat{H}\left(t\right)\hat{U}\left(t\right)^{-1}-i\hat{U}\left(t\right)\frac{\partial\hat{U}\left(t\right)^{-1}}{\partial t}.\label{eq:2.2.5}
\end{equation}
Using (\ref{eq:2.2.5}), it is now possible to define what we mean
by writing the Hamiltonian in a superadiabatic basis. The key is to
find the instantaneous eigenstates of the Hamiltonian in the original
frame. Then we choose the unitary operator to be 
\begin{equation}
\hat{U}\left(t\right)=\sum_{i}\left.|n_{i}\right\rangle \left\langle n_{i}\left(t\right)|\right..\label{eq:2.2.6}
\end{equation}
Here $\left.|n_{i}\left(t\right)\right\rangle $ correspond to the
instantaneous eigenstates of the Hamiltonian at time $t$ and $\left.|n_{i}\right\rangle $
(often referred to as reference states) are simply the instantaneous
eigenstates evaluated at time $t=0$. By applying the unitary transformation
(\ref{eq:2.2.6}) in (\ref{eq:2.2.5}), the transformed Hamiltonian
will be given in the superadiabatic basis \cite{key-30}. 

With these theoretical tools it is now possible to describe the state-transfer
protocol that will be used throughout this thesis. The idea will be
to consider a four-level system consisting of two ground states (which
we will call $\left.|g_{1}\right\rangle $ and $\left.|g_{2}\right\rangle $),
an excited state $\left.|e\right\rangle $, and a spectator state
$\left.|0\right\rangle $. Initially, at time $t=0$, we will consider
a qubit state written as 
\begin{equation}
\left.|\psi\left(0\right)\right\rangle =\alpha\left.|0\right\rangle +\beta\left.|g_{1}\right\rangle .\label{eq:2.2.7}
\end{equation}
The above state will represent our initial qubit state. The state-transfer
protocol will consist of transferring this state into an intermediate
qubit state, and then transferring the intermediate state back to
the original state. There are a variety of reasons for doing this.
For example, suppose the initial qubit state has a very short coherence
time. We could transfer this quantum state to an intermediate state
with a much longer coherence time to ensure that the state be preserved
for longer periods of time. Another application would be to perform
a phase gate. In this scenario, the goal is to transform the state
$\left.|\psi\left(0\right)\right\rangle $ as 
\begin{equation}
\left.|\psi\left(t_{f}\right)\right\rangle =\alpha\left.|0\right\rangle +\beta e^{i\phi}\left.|g_{1}\right\rangle .\label{eq:2.2.8}
\end{equation}
Notice that in this case the state $\left.|g_{1}\right\rangle $ picks
up a phase. After writing down the Hamiltonian describing the couplings
between the ground and excited states, we will show that this phase
indeed corresponds to the geometric phase acquired by performing a
cyclic evolution. In our case, the cyclic evolution will correspond
to transferring our qubit state in (\ref{eq:2.2.7}) to an intermediate
state and then back onto itself. 

We now write down the Hamiltonian describing the couplings between
the states mentioned above
\begin{align}
\hat{H}\left(t\right) & =E_{0}\left.|0\right\rangle \left\langle 0|\right.+\epsilon_{g_{1}}\left.|g_{1}\right\rangle \left\langle g_{1}|\right.+\epsilon_{e}\left.|e\right\rangle \left\langle e|\right.+\epsilon_{g_{2}}\left.|g_{2}\right\rangle \left\langle g_{2}|\right.\nonumber \\
 & +i\frac{\Omega_{1}\left(t\right)}{2}\left(\left.|e\right\rangle \left\langle g_{1}|\right.e^{-i\omega_{1}t}-h.c.\right)+i\frac{\Omega_{2}\left(t\right)}{2}\left(e^{-i\phi\left(t\right)}\left.|e\right\rangle \left\langle g_{2}|\right.e^{-i\omega_{2}t}-h.c.\right).\label{eq:2.2.9}
\end{align}
The first line in (\ref{eq:2.2.9}) corresponds to the energy levels
of the bare (undriven) states. The time-dependent functions $\Omega_{1}\left(t\right)$
and $\Omega_{2}\left(t\right)$ correspond to the classical laser
field amplitudes which couple the ground states $\left.|g_{1}\right\rangle $
and $\left.|g_{2}\right\rangle $ to the excited state $\left.|e\right\rangle $.
We take $\left.|0\right\rangle $ to be uncoupled and call this the
spectator state. The time-dependent phase $\phi\left(t\right)$ represents
the phase difference between our two drives. As will be shown below,
it is very important that this phase be time-dependent to perform
a phase gate. If this were not the case, the geometric phase would
vanish. 
\end{doublespace}

We can greatly simplify the Hamiltonian of Eq. (\ref{eq:2.2.9}) if
we choose to go into an interaction picture. By doing so, this will
allow us to choose resonant frequencies for our lasers that cancel
the frequencies $\omega_{1}$ and $\omega_{2}$ present in the coupling
terms of the Hamiltonian. To see this, we define 
\begin{equation}
\hat{H}_{0}=E_{0}\left.|0\right\rangle \left\langle 0|\right.+\epsilon_{g_{1}}\left.|g_{1}\right\rangle \left\langle g_{1}|\right.+\epsilon_{e}\left.|e\right\rangle \left\langle e|\right.+\epsilon_{g_{2}}\left.|g_{2}\right\rangle \left\langle g_{2}|\right.,\label{eq:2.2.10}
\end{equation}
and
\begin{equation}
\hat{H}_{1}\left(t\right)=i\frac{\Omega_{1}\left(t\right)}{2}\left(\left.|e\right\rangle \left\langle g_{1}|\right.e^{-i\omega_{1}t}-h.c.\right)+i\frac{\Omega_{2}\left(t\right)}{2}\left(e^{-i\phi\left(t\right)}\left.|e\right\rangle \left\langle g_{2}|\right.e^{-i\omega_{2}t}-h.c.\right).\label{eq:2.2.11}
\end{equation}
The interaction-picture Hamiltonian will then be given by
\begin{equation}
\hat{V}_{I}\left(t\right)=e^{i\hat{H}_{0}t}\hat{H}_{1}\left(t\right)e^{-i\hat{H}_{0}t}.\label{eq:2.2.12}
\end{equation}
After performing this transformation, we obtain
\begin{equation}
\hat{V}_{I}\left(t\right)=i\frac{\Omega_{1}\left(t\right)}{2}\left(\left.|e\right\rangle \left\langle g_{1}|\right.e^{-i\left(\omega_{1}+\omega_{g_{1}}-\omega_{e}\right)t}-h.c.\right)+i\frac{\Omega_{2}\left(t\right)}{2}\left(e^{-i\phi\left(t\right)}\left.|e\right\rangle \left\langle g_{2}|\right.e^{-i\left(\omega_{2}+\omega_{g_{2}}-\omega_{e}\right)t}-h.c.\right).\label{eq:2.2.13}
\end{equation}
It can immediately be seen from (\ref{eq:2.2.13}) that by choosing
$\omega_{1}=\omega_{e}-\omega_{g_{1}}$ and $\omega_{2}=\omega_{e}-\omega_{g_{2}}$
(which is the required condition for resonance), we get a simple form
for our interaction-picture Hamiltonian given by
\begin{equation}
\hat{V}_{I}\left(t\right)=i\frac{\Omega_{1}\left(t\right)}{2}\left(\left.|e\right\rangle \left\langle g_{1}|\right.-h.c.\right)+i\frac{\Omega_{2}\left(t\right)}{2}\left(e^{-i\phi\left(t\right)}\left.|e\right\rangle \left\langle g_{2}|\right.-h.c.\right).\label{eq:2.2.14}
\end{equation}

\begin{doublespace}
As we showed in Eq. (\ref{eq:2.1.3}), the solution to the Schroedinger
equation under an adiabatic approximation is given by a linear combination
of the instantaneous eigenstates of the Hamiltonian with the appropriate
phase factors (including both the dynamical and geometric phases).
The next step is to find the eigenstates of the Hamiltonian (\ref{eq:2.2.14}).
Out of the four eigenstates of this Hamiltonian, there will only be
one (apart from the spectator state $\left.|0\right\rangle $) which
does not couple to the excited state. We will call this state the
``dark state'' represented by $\left.|d\left(t\right)\right\rangle $.
Diagonalizing (\ref{eq:2.2.14}) in the $\left\{ \left.|0\right\rangle ,\left.|g_{1}\right\rangle ,\left.|g_{2}\right\rangle ,\left.|e\right\rangle \right\} $
basis, the normalized eigenstates are given by 
\begin{equation}
\left.|0\right\rangle ,\label{eq:2.2.15}
\end{equation}
\begin{equation}
\left.|d(t)\right\rangle =-\frac{\Omega_{2}(t)}{\sqrt{\Omega_{1}^{2}(t)+\Omega_{2}^{2}(t)}}\left.|g_{1}\right\rangle +\frac{\Omega_{1}(t)e^{i\phi\left(t\right)}}{\sqrt{\Omega_{1}^{2}(t)+\Omega_{2}^{2}(t)}}\left.|g_{2}\right\rangle ,\label{eq:2.2.16}
\end{equation}
\begin{equation}
\left.|+(t)\right\rangle =\frac{1}{\sqrt{2}}\left(\frac{\Omega_{1}(t)}{\sqrt{\Omega_{1}^{2}(t)+\Omega_{2}^{2}(t)}}\left.|g_{1}\right\rangle +i\left.|e\right\rangle +\frac{\Omega_{2}(t)e^{i\phi\left(t\right)}}{\sqrt{\Omega_{1}^{2}(t)+\Omega_{2}^{2}(t)}}\left.|g_{2}\right\rangle \right),\label{eq:2.2.17}
\end{equation}
\begin{equation}
\left.|-(t)\right\rangle =\frac{1}{\sqrt{2}}\left(\frac{\Omega_{1}(t)}{\sqrt{\Omega_{1}^{2}(t)+\Omega_{2}^{2}(t)}}\left.|g_{1}\right\rangle -i\left.|e\right\rangle +\frac{\Omega_{2}(t)e^{i\phi\left(t\right)}}{\sqrt{\Omega_{1}^{2}(t)+\Omega_{2}^{2}(t)}}\left.|g_{2}\right\rangle \right).\label{eq:2.2.18}
\end{equation}
The above eigenstates can be greatly simplified and written in a more
intuitive way by performing a change of variables. We define the angle
$\theta\left(t\right)$ and the parameter $G\left(t\right)$ by 
\begin{equation}
\tan\theta\left(t\right)\equiv-\frac{\Omega_{1}\left(t\right)}{\Omega_{2}\left(t\right)},\label{eq:2.2.19}
\end{equation}
\begin{equation}
G(t)\equiv\frac{1}{2}\sqrt{\Omega_{1}^{2}\left(t\right)+\Omega_{2}^{2}\left(t\right)}.\label{eq:2.2.20}
\end{equation}
With these definitions it is possible to rewrite the eigenstates as
\begin{equation}
\left.|d(t)\right\rangle =\cos\theta(t)\left.|g_{1}\right\rangle +e^{i\phi\left(t\right)}\sin\theta(t)\left.|g_{2}\right\rangle ,\label{eq:2.2.21}
\end{equation}
\begin{equation}
\left.|+(t)\right\rangle =\frac{1}{\sqrt{2}}\left(\sin\theta(t)\left.|g_{1}\right\rangle +i\left.|e\right\rangle -e^{i\phi\left(t\right)}\cos\theta(t)\left.|g_{2}\right\rangle \right),\label{eq:2.2.22}
\end{equation}
\begin{equation}
\left.|-(t)\right\rangle =\frac{1}{\sqrt{2}}\left(\sin\theta(t)\left.|g_{1}\right\rangle -i\left.|e\right\rangle -e^{i\phi\left(t\right)}\cos\theta(t)\left.|g_{2}\right\rangle \right).\label{eq:2.2.23}
\end{equation}
The corresponding energies are 
\begin{equation}
E_{d}=0,\label{eq:2.2.24}
\end{equation}
 and 
\begin{equation}
E_{\pm}=\pm G(t).\label{eq:2.2.25}
\end{equation}
Notice that the states $\left.|\pm(t)\right\rangle $ contain the
excited state. For this reason these states will be called bright
states \cite{key-31}. In the interaction picture, the dark state
has no dynamical phase since its energy eigenvalue vanishes. We now
have all the ingredients to describe the closed-system quantum state-transfer
protocol. Since $\left.|d(t)\right\rangle $ has no amplitude to be
in one of the excited states, then from (\ref{eq:2.1.16}) the transfer
of information between any two qubit states can be generally described
as 
\begin{equation}
\left.|\psi\left(t\right)\right\rangle =\alpha\left.|0\right\rangle +\beta e^{i\gamma_{d}\left(t\right)}\left.|d\left(t\right)\right\rangle .\label{eq:2.2.26}
\end{equation}
Here, $\gamma_{d}\left(t\right)$ is the geometric phase associated
with the dark state. Using (\ref{eq:2.1.14}) and (\ref{eq:2.2.21}),
the geometric phase is given by 
\begin{equation}
\gamma_{d}\left(t\right)=\int_{0}^{t}\dot{\phi}\left(t'\right)\sin^{2}\theta\left(t'\right)dt'.\label{eq:2.2.27}
\end{equation}
It can be seen that, if the laser phases were independent of time
($\dot{\phi}=0$), then the geometric phase would vanish. Now, at
the initial time, $t_{i}=0$, we retrieve the qubit state of Eq. (\ref{eq:2.2.7}).
At an intermediate time ($t_{i}<t_{int}<t_{f}$), $\theta\left(t_{int}\right)=\left(2n+1\right)\frac{\pi}{2}$
with $n$ an integer, (\ref{eq:2.2.26}) becomes
\begin{equation}
\left.|\psi\left(t_{int}\right)\right\rangle =\alpha\left.|0\right\rangle +\beta e^{i\gamma_{d}\left(t_{int}\right)}\left.|g_{2}\right\rangle .\label{eq:2.2.28}
\end{equation}
The above expression shows that when the initial qubit state ($\alpha\left.|0\right\rangle +\beta\left.|g_{1}\right\rangle $)
is transferred to the second qubit state (\ref{eq:2.2.28}), the state
$\left.|g_{2}\right\rangle $ picks up a geometric phase (which is
why at the beginning of this section we mentioned that this state
transfer protocol would be well-suited for performing a phase gate).
The integer $n\in\mathbb{N}$ describes the number of times the state
is transferred (later it will be shown that performing the state-transfer
protocol over many cycles will result in an improved fidelity). At
the final time ($t_{f}$), the angle $\theta\left(t_{f}\right)=n\pi$
so that 
\begin{equation}
\left.|\psi\left(t_{f}\right)\right\rangle =\alpha\left.|0\right\rangle +\left(-1\right)^{n}\beta e^{i\gamma_{d}\left(t_{f}\right)}\left.|g_{1}\right\rangle .\label{eq:2.2.29}
\end{equation}
Note that the relative sign difference between the spectator state
($\left.|0\right\rangle $) and the first ground state ($\left.|g_{1}\right\rangle $)
is irrelevant since it can be absorbed into the phase $\gamma_{d}\left(t_{f}\right)$.
In the next section we will see how to formulate the state-transfer
protocol described above using a density-matrix approach. The main
advantage to using density matrices will be clear when we consider
open quantum systems. Since the off-diagonal components of the density
matrix contain all the phase information (geometric and dynamical),
it will be possible to obtain corrections to the geometric phase arising
from the coupling to a quantum environment. 
\end{doublespace}

\section{Density matrix approach}

\begin{doublespace}
In this section we will show the general methods for obtaining the
density matrix for closed-system dynamics. We will see that the geometric
phase calculated in (\ref{eq:2.2.27}) will arise in computing the
relevant off-diagonal element of the density matrix (also known as
coherence). 

For a pure state, the general form of the density matrix is given
by 
\begin{equation}
\hat{\rho}\left(t\right)=\left.|\psi\left(t\right)\right\rangle \left\langle \psi\left(t\right)|\right..\label{eq:2.3.1}
\end{equation}
Using (\ref{eq:2.2.26}), the density matrix for our system can be
written as
\begin{equation}
\hat{\rho}\left(t\right)=\left|\alpha\right|^{2}\left.|0\right\rangle \left\langle 0|\right.+\alpha\beta^{*}e^{-i\gamma_{d}\left(t\right)}\left.|0\right\rangle \left\langle d\left(t\right)|\right.+\alpha^{*}\beta e^{i\gamma_{d}\left(t\right)}\left.|d\left(t\right)\right\rangle \left\langle 0|\right.+\left|\beta\right|^{2}\left.|d\left(t\right)\right\rangle \left\langle d\left(t\right)|\right..\label{eq:2.3.2}
\end{equation}
Now, suppose we want to diagonalize our initial Hamiltonian. This
diagonalization can be done by going into a rotating frame with the
unitary operator 
\begin{equation}
\hat{U}\left(t\right)=\left.|0\right\rangle \left\langle 0|\right.+\left.|d\right\rangle \left\langle d(t)|\right.+\left.|+\right\rangle \left\langle +(t)|\right.+\left.|-\right\rangle \left\langle -(t)|\right..\label{eq:2.3.3}
\end{equation}
Applying the operation $\tilde{\hat{H}}\left(t\right)=\hat{U}\left(t\right)\hat{H}\left(t\right)\hat{U}^{\dagger}\left(t\right)-i\hat{U}\left(t\right)\dot{\hat{U}}^{\dagger}\left(t\right)$,
the new Hamiltonian takes the form 
\begin{align}
\tilde{\hat{H}}(t) & =G(t)\left\{ \left.|+\right\rangle \left\langle +|\right.-\left.|-\right\rangle \left\langle -|\right.\right\} +\dot{\phi}\sin^{2}\theta\left(t\right)\left.|d\right\rangle \left\langle d|\right.\nonumber \\
 & +i\frac{\dot{\theta}}{\sqrt{2}}\left(\left.|+\right\rangle \left\langle d|\right.-\left.|d\right\rangle \left\langle +|\right.+\left.|-\right\rangle \left\langle d|\right.-\left.|d\right\rangle \left\langle -|\right.\right)-\frac{\dot{\phi}}{\sqrt{2}}\left(\left.|+\right\rangle \left\langle d|\right.+\left.|d\right\rangle \left\langle +|\right.+\left.|-\right\rangle \left\langle d|\right.+\left.|d\right\rangle \left\langle -|\right.\right)\nonumber \\
 & +\frac{\dot{\phi}\cos^{2}\theta\left(t\right)}{2}\left(\left.|+\right\rangle \left\langle +|\right.+\left.|-\right\rangle \left\langle +|\right.+\left.|+\right\rangle \left\langle -|\right.+\left.|-\right\rangle \left\langle -|\right.\right).\label{eq:2.3.4}
\end{align}
The above expression corresponds to the Hamiltonian written in the
first superadiabatic eigenbasis. We say ``first'' superadiabatic
eigenbasis because we could repeat this process to $j^{th}$order.
If, for example, we were to find the eigenstates of the Hamiltonian
in Eq. (\ref{eq:2.3.4}), we could write down a unitary operator of
the same form as in (\ref{eq:2.3.3}) but written in terms of the
instantaneous eigenstates of (\ref{eq:2.3.4}) instead of (\ref{eq:2.2.14}).
We would perform the unitary transformation of (\ref{eq:2.2.5}) on
(\ref{eq:2.3.4}) to find a new Hamiltonian written in the second-order
adiabatic basis, and so on.

To obtain the density matrix in the rotating frame, one must apply
the transformation $\tilde{\hat{\rho}}\left(t\right)=\hat{U}\left(t\right)\hat{\rho}\left(t\right)\hat{U}^{\dagger}\left(t\right)$.
Using (\ref{eq:2.3.2}) and (\ref{eq:2.3.3}) we find that
\begin{equation}
\tilde{\hat{\rho}}\left(t\right)=\left|\alpha\right|^{2}\left.|0\right\rangle \left\langle 0|\right.+\alpha\beta^{*}e^{-i\gamma_{d}\left(t\right)}\left.|0\right\rangle \left\langle d|\right.+\alpha^{*}\beta e^{i\gamma_{d}\left(t\right)}\left.|d\right\rangle \left\langle 0|\right.+\left|\beta\right|^{2}\left.|d\right\rangle \left\langle d|\right..\label{eq:2.3.5}
\end{equation}
Since only the dark ($\left.|d\right\rangle $) and spectator ($\left.|0\right\rangle $)
states are present in our state transfer protocol, we will only be
interested in the ``$od$'' component of the density matrix. In
this case it will be useful to define 
\begin{equation}
\tilde{\hat{\Pi}}_{0d}\equiv\left.|0\right\rangle \left\langle d|\right..\label{eq:2.3.6}
\end{equation}
Throughout the entire thesis, whenever we take an average of an arbitrary
operator $\hat{A}$, it will always be with respect to $\tilde{\hat{\rho}}\left(t\right)$
in the following way
\begin{equation}
\left\langle \hat{A}\left(t\right)\right\rangle =\mathrm{Tr}\left\{ \tilde{\hat{\rho}}\left(t\right)\hat{A}\right\} \label{eq:2.3.7-1}
\end{equation}
Using (\ref{eq:2.3.5}), the expectation value of the relevant component
of the density matrix will be given by 
\begin{align}
\left\langle \tilde{\hat{\Pi}}_{0d}\left(t\right)\right\rangle  & =\mathrm{Tr}\left\{ \tilde{\hat{\rho}}\left(t\right)\tilde{\hat{\Pi}}_{0d}\right\} =\mathrm{Tr}\left\{ \hat{\rho}\left(t\right)\left.|0\right\rangle \left\langle d\left(t\right)|\right.\right\} \nonumber \\
 & =\alpha\beta^{*}e^{-i\gamma_{d}\left(t\right)}.\label{eq:2.3.7}
\end{align}
Note that $\tilde{\hat{\Pi}}_{0d}\left(t\right)$ is in the Heisenberg
picture with respect to the Hamiltonian $\tilde{\hat{H}}(t)$. 
\end{doublespace}

At the time $t=0$, (\ref{eq:2.3.7}) reduces to
\begin{equation}
\left\langle \tilde{\hat{\Pi}}_{0d}\left(0\right)\right\rangle =\alpha\beta^{*}.\label{eq:2.3.8}
\end{equation}
To find the ``$0d$'' component of the density matrix in the lab
frame we can obtain this term by ``undoing'' the unitary transformation
we performed to get $\tilde{\hat{\rho}}\left(t\right)$ or we can
read it off of (\ref{eq:2.3.2}) so that 
\begin{equation}
\hat{\Pi}_{0d}\left(t\right)=\left.|0\right\rangle \left\langle d\left(t\right)|\right..\label{eq:2.3.9}
\end{equation}
As we mentioned earlier, all the relevant phase information is contained
in the ``$0d$'' component of the density matrix. As expected, only
the geometric phase is present since in the interaction picture the
dark-state eigenenergy vanishes. If this were not the case, then one
would expect that (\ref{eq:2.3.7}) also have a dynamical phase dependence.

When we consider open quantum systems, we can use the exact same procedure
as used in this section to extract all the relevant phase information.

\section{\label{sec:Review-of-photon}Review of cavity-photon shot noise}

\begin{doublespace}
After developing the theoretical tools allowing us to compute corrections
to the geometric phase for open quantum systems, we will apply our
methods to adiabatic state transfer between qubits in a driven cavity.
We will explicitly consider dephasing effects due to unavoidable photon
shot noise. Consequently, it is necessary to review the physics behind
cavity-photon shot noise. Much of what follows will be based on \cite{key-40}. 

We consider a cavity driven by an external field. For a cavity to
be driven, it is necessary to open one if its ports. Consequently,
this will enable the cavity to leak energy in the surrounding bath.
For high-Q cavities it is possible to make a distinction between the
internal cavity modes and the external bath modes. We can write the
Hamiltonian as 
\begin{equation}
\hat{H}=\hat{H}_{sys}+\hat{H}_{bath}+\hat{H}_{int}.\label{eq:2.4.1}
\end{equation}
The bath Hamiltonian will be described by a collection of harmonic
modes 
\begin{equation}
\hat{H}_{bath}=\sum_{q}\hbar\omega_{q}\hat{b}_{q}^{\dagger}\hat{b}_{q}.\label{eq:2.4.2}
\end{equation}
The bath modes obey the commutation relations 
\begin{equation}
\left[\hat{b}_{q},\hat{b}_{q'}^{\dagger}\right]=\delta_{q,q'}.\label{eq:2.4.3}
\end{equation}
Within a rotating-wave approximation, the coupling Hamiltonian is
described as 
\begin{equation}
\hat{H}_{int}=-i\hbar\sum_{q}\left[f_{q}\hat{a}^{\dagger}\hat{b}_{q}-f_{q}^{*}\hat{b}_{q}^{\dagger}\hat{a}\right].\label{eq:2.4.4}
\end{equation}
We neglected terms such as $\hat{a}\hat{b}_{q}$ and $\hat{a}^{\dagger}\hat{b}_{q}^{\dagger}$
since in a rotating-wave approximation they oscillate at high frequency
and so have little effects on the dynamics. The cavity will be specified
by a single degree of freedom obeying the bosonic commutation relation
\begin{equation}
\left[\hat{a},\hat{a}^{\dagger}\right]=1.\label{eq:2.4.5}
\end{equation}
The Heisenberg equation of motion for the bosonic bath modes is given
by 
\begin{equation}
\dot{\hat{b}}_{q}=-i\omega_{q}\hat{b}_{q}+f_{q}^{*}\hat{a}.\label{eq:2.4.6}
\end{equation}
The second term on the right-hand side of (\ref{eq:2.4.6}) corresponds
to a forcing term due to the motion of the cavity degree-of-freedom.
Considering $t_{0}<t$ to be a time in the past before a wave packet
launched at the cavity has reached it, we can solve (\ref{eq:2.4.6})
exactly to obtain 
\begin{equation}
\hat{b}_{q}\left(t\right)=e^{-i\omega_{q}\left(t-t_{0}\right)}\hat{b}_{q}\left(t_{0}\right)+\int_{t_{0}}^{t}d\tau e^{-i\omega_{q}\left(t-\tau\right)}f_{q}^{*}\hat{a}\left(\tau\right).\label{eq:2.4.7}
\end{equation}
The second term on the right corresponds to a wave radiated by the
cavity into the bath. If we had considered $t_{1}>t$ to be a time
in the future after the input field had interacted with the cavity,
then the solution to (\ref{eq:2.4.6}) would instead take the form
\begin{equation}
\hat{b}_{q}\left(t\right)=e^{-i\omega_{q}\left(t-t_{1}\right)}\hat{b}_{q}\left(t_{1}\right)-\int_{t}^{t_{1}}d\tau e^{-i\omega_{q}\left(t-\tau\right)}f_{q}^{*}\hat{a}\left(\tau\right).\label{eq:2.4.8}
\end{equation}
Notice that there is a sign difference in the second term compared
to (\ref{eq:2.4.7}) arising from the fact that $t_{1}>t$. Now we
can write down the equation-of-motion for the cavity degree of freedom
\begin{equation}
\dot{\hat{a}}=\frac{i}{\hbar}\left[\hat{H}_{sys},\hat{a}\right]-\sum_{q}f_{q}\hat{b}_{q}.\label{eq:2.4.9}
\end{equation}
It is important to note that so far we have not specified the dynamics
of the cavity, so that the first term in (\ref{eq:2.4.9}) is left
completely general. We can use (\ref{eq:2.4.7}) in the second term
of (\ref{eq:2.4.9}) to get 
\begin{align}
\sum_{q}f_{q}\hat{b}_{q} & =\sum_{q}f_{q}e^{-i\omega_{q}\left(t-t_{0}\right)}\hat{b}_{q}\left(t_{0}\right)\nonumber \\
 & +\sum_{q}\left|f_{q}\right|^{2}\int_{t_{0}}^{t}d\tau e^{-i\left(\omega_{q}-\omega_{c}\right)\left(t-\tau\right)}\left[e^{i\omega_{c}\left(t-\tau\right)}\hat{a}\left(\tau\right)\right].\label{eq:2.4.10}
\end{align}
We can simplify the last result by noting that if we considered the
cavity to be a simple harmonic mode of frequency $\omega_{c}$, then
we could represent the decay rate from the $n=1$ single-photon excited
state to the $n=0$ ground state by a Fermi golden rule expression
\begin{equation}
\kappa\left(\omega_{c}\right)=2\pi\sum_{q}\left|f_{q}\right|^{2}\delta\left(\omega_{c}-\omega_{q}\right).\label{eq:2.4.11}
\end{equation}
Fourier transforming, we find 
\begin{equation}
\int_{-\infty}^{\infty}\frac{d\nu}{2\pi}\kappa\left(\omega_{c}+\nu\right)e^{-i\nu\left(t-\tau\right)}=\sum_{q}\left|f_{q}\right|^{2}e^{-i\left(\omega_{q}-\omega_{c}\right)\left(t-\tau\right)}.\label{eq:2.4.12}
\end{equation}
In a Markov approximation, we set $\kappa\left(\nu\right)=\kappa$
to be a constant over the cavity frequencies. Using the fact that
$\int_{-\infty}^{\infty}\frac{d\nu}{2\pi}e^{-i\nu\left(t-\tau\right)}=\delta\left(t-\tau\right)$,
(\ref{eq:2.4.12}) can be simplified to 
\begin{equation}
\sum_{q}\left|f_{q}\right|^{2}e^{-i\left(\omega_{q}-\omega_{c}\right)\left(t-\tau\right)}=\kappa\delta\left(t-\tau\right).\label{eq:2.4.13}
\end{equation}
Within the range of validity of the Markov approximation, we can also
set $t_{0}=-\infty$. Using $\int_{-\infty}^{\tau}\delta\left(t'-\tau\right)dt'=\frac{1}{2}$,
the equation-of-motion for the cavity degree of freedom simplifies
to 
\begin{equation}
\dot{\hat{a}}=\frac{i}{\hbar}\left[\hat{H}_{sys},\hat{a}\right]-\frac{\kappa}{2}\hat{a}-\sum_{q}f_{q}e^{-i\omega_{q}\left(t-t_{0}\right)}\hat{b}_{q}\left(t_{0}\right).\label{eq:2.4.14}
\end{equation}
The term representing the wave radiated by the cavity is now a simple
linear damping term under a Markov approximation. The factor of 2
indicates that the amplitude decays at half the rate of the intensity.
By performing a Markov approximation, we can approximate $f\equiv\sqrt{\left|f_{q}\right|^{2}}$
to be a constant and also set the density of states $\rho=\sum_{q}\delta\left(\omega_{c}-\omega_{q}\right)$
to be a constant \cite{key-40}. Then from (\ref{eq:2.4.11}) we get
the very simple result 
\begin{equation}
\kappa=2\pi f^{2}\rho.\label{eq:2.4.15}
\end{equation}
At this stage it will prove convenient to define the ``input mode''
by 
\begin{equation}
\hat{b}_{in}\left(t\right)\equiv\frac{1}{\sqrt{2\pi\rho}}\sum_{q}e^{-i\omega_{q}\left(t-t_{0}\right)}\hat{b}_{q}\left(t_{0}\right).\label{eq:2.4.16}
\end{equation}
Then with this definition, we can rewrite the equation-of-motion for
the cavity mode to be 
\begin{equation}
\dot{\hat{a}}=\frac{i}{\hbar}\left[\hat{H}_{sys},\hat{a}\right]-\frac{\kappa}{2}\hat{a}-\sqrt{\kappa}\hat{b}_{in}\left(t\right).\label{eq:2.4.17}
\end{equation}
The input mode will evolve freely until it comes into contact with
the cavity at which point it will begin driving the cavity. It is
sound to interpret $\hat{b}_{in}\left(t\right)$ as an input mode
since it evolves under the free bath Hamiltonian and acts as a driving
term in the equation-of-motion for the cavity mode. Following along
the same reasoning as described above, we can also define an output
mode from Eq. (\ref{eq:2.4.8}) by
\begin{equation}
\hat{b}_{out}\left(t\right)\equiv\frac{1}{\sqrt{2\pi\rho}}\sum_{q}e^{-i\omega_{q}\left(t-t_{1}\right)}\hat{b}_{q}\left(t_{1}\right).\label{eq:2.4.18}
\end{equation}
We can interpret this as an output mode since it is simply the free
evolution of the bath modes in the distant future after interacting
with the cavity. Note that it is also possible to write an equation-of-motion
for the cavity mode in terms of the output field. This is given by
\begin{equation}
\dot{\hat{a}}=\frac{i}{\hbar}\left[\hat{H}_{sys},\hat{a}\right]+\frac{\kappa}{2}\hat{a}-\sqrt{\kappa}\hat{b}_{out}\left(t\right).\label{eq:2.4.19}
\end{equation}
If we subtract (\ref{eq:2.4.19}) from (\ref{eq:2.4.17}), the we
can write the output field in terms of the input field as 
\begin{equation}
\hat{b}_{out}\left(t\right)=\hat{b}_{in}\left(t\right)+\sqrt{\kappa}\hat{a}\left(t\right).\label{eq:2.4.20}
\end{equation}
This is consistent with the view that the output field should be a
sum of a reflected incoming field plus the field radiated by the cavity. 

So far we have considered a completely general Hamiltonian for the
cavity dynamics (apart from being restricted to a single mode). We
can now consider the specific case where the cavity is comprised of
a single harmonic oscillator with frequency $\omega_{c}$. In this
case the system Hamiltonian would be given by 
\begin{equation}
\hat{H}_{sys}=\hbar\omega_{c}\hat{a}^{\dagger}\hat{a}.\label{eq:2.4.21}
\end{equation}
The commutator in (\ref{eq:2.4.17}) is now straightforward to compute
and so the cavity equation-of-motion reduces to 
\begin{equation}
\dot{\hat{a}}=-i\omega_{c}\hat{a}-\frac{\kappa}{2}\hat{a}-\sqrt{\kappa}\hat{b}_{in}\left(t\right).\label{eq:2.4.22}
\end{equation}
A convenient trick for solving this equation is to first perform a
Fourier transform to write a solution for $\hat{a}\left[\omega\right]=\int_{-\infty}^{\infty}dt\hat{a}\left(t\right)e^{i\omega t}$.
Proceeding in this manner we find 
\begin{equation}
\hat{a}\left[\omega\right]=-\sqrt{\kappa}\chi_{c}\left[\omega-\omega_{c}\right]\hat{b}_{in}\left[\omega\right].\label{eq:2.4.23}
\end{equation}
Here we defined the susceptibility of the cavity by 
\begin{equation}
\chi_{c}\left[\omega-\omega_{c}\right]\equiv\frac{1}{-i\left(\omega-\omega_{c}\right)+\kappa/2}.\label{eq:2.4.24}
\end{equation}
We can also solve for the output field in terms of the input field
which will be given by 
\begin{equation}
\hat{b}_{out}\left[\omega\right]=\frac{\omega-\omega_{c}-i\kappa/2}{\omega-\omega_{c}+i\kappa/2}\hat{b}_{in}\left[\omega\right].\label{eq:2.4.25}
\end{equation}
If we drive the cavity on resonance so that $\omega=\omega_{c}$,
the output field assumes the simple form 
\begin{equation}
\hat{b}_{out}\left[\omega\right]=\frac{\sqrt{\kappa}}{2}\hat{a}\left[\omega\right].\label{eq:2.4.26}
\end{equation}
The equation-of-motion for the cavity mode can also be solved in the
time domain. The solution is given by 
\begin{equation}
\hat{a}\left(t\right)=e^{-\left(i\omega_{c}+\kappa/2\right)\left(t-t_{0}\right)}\hat{a}\left(t_{0}\right)-\sqrt{\kappa}\int_{t_{0}}^{t}d\tau e^{-\left(i\omega_{c}+\kappa/2\right)\left(t-\tau\right)}\hat{b}_{in}\left(\tau\right).\label{eq:2.4.27}
\end{equation}
We can specialize to the case where the input field is a coherent
drive at a frequency $\omega_{L}=\omega_{c}+\Delta$ with a classical
and quantum amplitude given by 
\begin{equation}
\hat{b}_{in}\left(t\right)=e^{-i\omega_{L}t}\left[\bar{b}_{in}+\hat{\xi}\left(t\right)\right].\label{eq:2.4.28}
\end{equation}
Here $\bar{b}_{in}$ is the classical amplitude and $\hat{\xi}\left(t\right)$
is the quantum amplitude. We can also take $t_{0}\rightarrow\infty$
in (\ref{eq:2.4.27}) which corresponds to having the initial transient
in the cavity damped out. In this case the solution to (\ref{eq:2.4.27})
will be given by 
\begin{equation}
\hat{a}\left(t\right)=e^{-i\omega_{L}t}\left[\bar{a}+\hat{d}\left(t\right)\right],\label{eq:2.4.29}
\end{equation}
with the classical contribution $\bar{a}$ given by 
\begin{equation}
\bar{a}=-\frac{\sqrt{\kappa}}{-i\Delta+\kappa/2}\bar{b}_{in}.\label{eq:2.4.30}
\end{equation}
where $\Delta$ is the detuning frequency. In the frame rotating at
the drive frequency, the quantum contribution will be given by 
\begin{equation}
\hat{d}\left(t\right)=-\sqrt{\kappa}\int_{-\infty}^{t}d\tau e^{\left(i\Delta-\kappa/2\right)\left(t-\tau\right)}\hat{\xi}\left(\tau\right).\label{eq:2.4.31}
\end{equation}
Using (\ref{eq:2.4.16}) and (\ref{eq:2.4.28}), we can obtain the
commutation relations for the fields $\hat{\xi}\left(t\right)$
\begin{align}
\left[\hat{b}_{in}\left(t\right),\hat{b}_{in}^{\dagger}\left(t'\right)\right] & =\left[\hat{\xi}\left(t\right),\hat{\xi}^{\dagger}\left(t'\right)\right]\nonumber \\
 & =\frac{1}{2\pi\rho}\sum_{q}e^{-i\left(\omega_{q}-\omega_{L}\right)\left(t-t'\right)}\nonumber \\
 & =\delta\left(t-t'\right).\label{eq:2.4.32}
\end{align}
We can also verify that indeed the correct commutation relations for
the time-dependent cavity modes are satisfied by using (\ref{eq:2.4.29}),
(\ref{eq:2.4.31}) and (\ref{eq:2.4.32})
\begin{align}
\left[\hat{a}\left(t\right),\hat{a}^{\dagger}\left(t\right)\right] & =\left[\hat{d}\left(t\right),\hat{d}^{\dagger}\left(t\right)\right]\nonumber \\
 & =\kappa\int_{-\infty}^{t}d\tau\int_{-\infty}^{t}d\tau'e^{-\left(-i\Delta+\kappa/2\right)\left(t-\tau\right)}e^{-\left(i\Delta+\kappa/2\right)\left(t-\tau'\right)}\delta\left(\tau-\tau'\right)\nonumber \\
 & =1.\label{eq:2.4.33}
\end{align}
Since the port of the cavity is open, vacuum noise will enter the
cavity creating zero-point fluctuations \cite{key-40}. 
\end{doublespace}

It is now possible to give an explanation for the quantum noise in
the number of photons inside the cavity. Based on the picture that
we have developed, this noise will be due to the vacuum noise that
enters through the cavity port that was brought in by the classical
field. There will be interference between the vacuum noise and the
classical drive that will lead to fluctuations in the number of photons
inside the cavity. 

\begin{doublespace}
As a last remark, we should also think about temperature. The field
$\hat{\xi}$ will contain thermal radiation when in thermal equilibrium.
Recall that when making the Markov approximation, we assumed that
the bath was being probed over a very broad range of frequencies centered
on $\omega_{c}$. To a good approximation \cite{key-40}, we have
\begin{equation}
\left\langle \hat{\xi}^{\dagger}\left(t\right)\hat{\xi}\left(t'\right)\right\rangle =N_{th}\delta\left(t-t'\right),\label{eq:2.4.34}
\end{equation}
\begin{equation}
\left\langle \hat{\xi}\left(t\right)\hat{\xi}^{\dagger}\left(t'\right)\right\rangle =\left(N_{th}+1\right)\delta\left(t-t'\right),\label{eq:2.4.35}
\end{equation}
where $N_{th}=n_{B}\left(\hbar\omega_{c}\right)$ is the bosonic thermal
equilibrium occupation number of the mode at the frequency of interest.
The relations (\ref{eq:2.4.34}) and (\ref{eq:2.4.35}) will be used
extensively in chapter 4 when we apply our state transfer methods
to cavity-photon shot noise. As a final note, using (\ref{eq:2.4.31}),
(\ref{eq:2.4.34}) and (\ref{eq:2.4.35}), we find that the correlation
function of the $\hat{d}$ operator is given by 
\begin{equation}
\left\langle \hat{d}^{\dagger}\left(t\right)\hat{d}\left(t'\right)\right\rangle =N_{th}e^{i\Delta\left(t-t'\right)}e^{-\frac{\kappa}{2}\left|t-t'\right|}.\label{eq:2.4.36}
\end{equation}

\end{doublespace}

\newpage{}

\chapter{Open-system evolution under the secular approximation}

\rule[0.5ex]{1\columnwidth}{1pt}

\section{\label{sec:Density-matrix-phase}Density matrix phase}

\begin{doublespace}
In this section we will develop the general theoretical tools that
are required to calculate dephasing effects for the case where our
four-level system is coupled to a quantum dissipative bath. The starting
point will be to consider the Hamiltonian that we obtained in (\ref{eq:2.2.14})
and add the noisy terms $\delta\hat{\omega}_{1}\left.|g_{1}\right\rangle \left\langle g_{1}|\right.+\delta\hat{\omega}_{2}\left.|g_{2}\right\rangle \left\langle g_{2}|\right.$.
The functions $\delta\hat{\omega}_{1}$ and $\delta\hat{\omega}_{2}$
correspond to bath degrees-of-freedom which couple to the ground states
of our four-level system. The interaction-picture Hamiltonian will
now be given by 
\begin{align}
\hat{H}\left(t\right) & =i\frac{\Omega_{1}\left(t\right)}{2}\left(\left.|e\right\rangle \left\langle g_{1}|\right.-\left.|g_{1}\right\rangle \left\langle e|\right.\right)+i\frac{\Omega_{2}\left(t\right)}{2}\left(e^{-i\phi\left(t\right)}\left.|e\right\rangle \left\langle g_{2}|\right.-e^{i\phi\left(t\right)}\left.|g_{2}\right\rangle \left\langle e|\right.\right)\nonumber \\
 & +\delta\hat{\omega}_{1}\left.|g_{1}\right\rangle \left\langle g_{1}|\right.+\delta\hat{\omega}_{2}\left.|g_{2}\right\rangle \left\langle g_{2}|\right.+\hat{H}_{env}\left(\delta\hat{\omega}_{1},\delta\hat{\omega}_{2}\right).\label{eq:3.1.1}
\end{align}
The term $\hat{H}_{env}\left(\delta\hat{\omega}_{1},\delta\hat{\omega}_{2}\right)$
in (\ref{eq:3.1.1}) dictates the dynamics of the bath degrees of
freedom. For now we will consider this to be completely general since
its specific operator dependence will not be relevant. 

The first step is to go into a rotating frame by choosing a unitary
transformation $\hat{U}$ which diagonalizes $\hat{H}_{0}\left(t\right)$
where
\begin{equation}
\hat{H}_{0}\left(t\right)=i\frac{\Omega_{1}\left(t\right)}{2}\left(\left.|e\right\rangle \left\langle g_{1}|\right.-\left.|g_{1}\right\rangle \left\langle e|\right.\right)+i\frac{\Omega_{2}\left(t\right)}{2}\left(e^{-i\phi\left(t\right)}\left.|e\right\rangle \left\langle g_{2}|\right.-e^{i\phi\left(t\right)}\left.|g_{2}\right\rangle \left\langle e|\right.\right).\label{eq:3.1.2}
\end{equation}
 In order to do so we choose the unitary transformation of (\ref{eq:2.3.3}).
The states $\left.|d\right\rangle $, $\left.|+\right\rangle $ and
$\left.|-\right\rangle $ are reference states chosen to be the dark
and bright states evaluated at time $t=0$. Notice that the situation
is different than what we were dealing with in chapter 2. For a purely
closed system, we chose a unitary transformation that diagonalized
the entire Hamiltonian. For the open-system case, we don't consider
a unitary transformation given in terms of the eigenstates of the
entire Hamiltonian. Instead, we only choose the eigenstates of $\hat{H}_{0}\left(t\right)$.
The transformed Hamiltonian will take the form $\tilde{\hat{H}}\left(t\right)=\hat{U}\hat{H}\left(t\right)\hat{U}^{\dagger}-i\hat{U}\dot{\hat{U}}^{\dagger}$.
The result of this transformation is given by
\begin{align}
\tilde{\hat{H}}(t) & =G(t)\left\{ \left.|+\right\rangle \left\langle +|\right.-\left.|-\right\rangle \left\langle -|\right.\right\} +\left(\dot{\phi}\sin^{2}\theta\left(t\right)+\delta\hat{\omega}_{1}\cos^{2}\theta\left(t\right)+\delta\hat{\omega}_{2}\sin^{2}\theta\left(t\right)\right)\left.|d\right\rangle \left\langle d|\right.\nonumber \\
 & +\left(\delta\hat{\omega}_{1}-\delta\hat{\omega}_{2}\right)\frac{\cos\theta(t)\sin\theta(t)}{\sqrt{2}}\left(\left.|+\right\rangle \left\langle d|\right.+\left.|-\right\rangle \left\langle d|\right.+\left.|d\right\rangle \left\langle +|\right.+\left.|d\right\rangle \left\langle -|\right.\right)\nonumber \\
 & +\frac{\left(\delta\hat{\omega}_{1}\sin^{2}\theta\left(t\right)+\delta\hat{\omega}_{2}\cos^{2}\theta\left(t\right)\right)}{2}\left(\left.|+\right\rangle \left\langle +|\right.+\left.|-\right\rangle \left\langle +|\right.+\left.|+\right\rangle \left\langle -|\right.+\left.|-\right\rangle \left\langle -|\right.\right)\nonumber \\
 & +i\frac{\dot{\theta}}{\sqrt{2}}\left(\left.|+\right\rangle \left\langle d|\right.-\left.|d\right\rangle \left\langle +|\right.+\left.|-\right\rangle \left\langle d|\right.-\left.|d\right\rangle \left\langle -|\right.\right)-\frac{\dot{\phi}}{\sqrt{2}}\left(\left.|+\right\rangle \left\langle d|\right.+\left.|d\right\rangle \left\langle +|\right.+\left.|-\right\rangle \left\langle d|\right.+\left.|d\right\rangle \left\langle -|\right.\right)\nonumber \\
 & +\frac{\dot{\phi}\cos^{2}\theta\left(t\right)}{2}\left(\left.|+\right\rangle \left\langle +|\right.+\left.|-\right\rangle \left\langle +|\right.+\left.|+\right\rangle \left\langle -|\right.+\left.|-\right\rangle \left\langle -|\right.\right)+\hat{H}_{env}\left(\delta\hat{\omega}_{1},\delta\hat{\omega}_{2}\right).\label{eq:3.1.3}
\end{align}
At first it might seem hopeless to make any progress with this Hamiltonian
without resorting to approximation methods such as Bloch-Redfield
theory or other similar means which would limit our results to Markovian
environments. However, the Hamiltonian can be greatly simplified by
performing a secular approximation. All the off-diagonal terms of
the Hamiltonian will enter in perturbation theory with suppression
factors $\sim\frac{1}{G}$ compared to the diagonal terms. If we consider
the limit where $\frac{\dot{\theta}}{G}\ll1$, $\frac{\dot{\phi}}{G}\ll1$
and $\frac{\left|\delta\hat{\omega}_{i}\right|}{G}\ll1$ (which can
be achieved by considering very large laser amplitudes relative to
the other parameters of the system), then we can drop all the off-diagonal
terms of the Hamiltonian. This defines the secular approximation.
Hence we get 
\begin{equation}
\tilde{\hat{H}}_{sec}(t)=\hat{V}_{sec}\left(t\right)+\hat{H}_{env}\left(\delta\hat{\omega}_{1},\delta\hat{\omega}_{2}\right),
\end{equation}
where
\begin{align}
\hat{V}_{sec}\left(t\right) & \equiv G(t)\left\{ \left.|+\right\rangle \left\langle +|\right.-\left.|-\right\rangle \left\langle -|\right.\right\} +\left(\dot{\phi}\sin^{2}\theta\left(t\right)+\delta\hat{\omega}_{1}\cos^{2}\theta\left(t\right)+\delta\hat{\omega}_{2}\sin^{2}\theta\left(t\right)\right)\left.|d\right\rangle \left\langle d|\right.\nonumber \\
 & +\frac{\left(\dot{\phi}\cos^{2}\theta\left(t\right)+\delta\hat{\omega}_{1}\sin^{2}\theta\left(t\right)+\delta\hat{\omega}_{2}\cos^{2}\theta\left(t\right)\right)}{2}\left(\left.|+\right\rangle \left\langle +|\right.+\left.|-\right\rangle \left\langle -|\right.\right)\label{eq:3.1.4-1}
\end{align}
This Hamiltonian is clearly much simpler than the previous one. As
will be shown below, the great technical advantage in performing the
secular approximation is that we won't need to solve a master equation
using Bloch-Redfield theory to obtain the phase information of the
density matrix. Furthermore, Bloch-Redfield theory would only be valid
for systems weakly coupled to an environment with a short correlation
time (i.e. a Markovian environment) relative to the decay time of
the off-diagonal component of the density matrix \cite{key-41}. In
our approach, the phase information will be obtained from the equation-of-motion
for the off-diagonal component of the density matrix. Since the Hamiltonian
written in the superadiabatic basis is purely diagonal after performing
a secular approximation, its commutation relations with the component
of interest of the density matrix will turn out to be fairly simple. 

We start by calculating the average of $\left\langle \tilde{\hat{\Pi}}_{0d}\left(t\right)\right\rangle $.
After obtaining the thermal average of $\left\langle \tilde{\hat{\Pi}}_{0d}\left(t\right)\right\rangle $
in the rotating frame (see (\ref{eq:3.1.5}) written below), it will
be a simple matter of undoing the unitary transformations following
(\ref{eq:2.3.9}) to obtain the desired component in the lab frame.
For a system-bath coupling, the previous average can be written as
\begin{equation}
\left\langle \tilde{\hat{\Pi}}_{0d}\left(t\right)\right\rangle =\mathrm{Tr}\left\{ \tilde{\hat{\Pi}}_{0d}\left(t\right)\tilde{\hat{\rho}}\left(0\right)\right\} ,\label{eq:3.1.5}
\end{equation}
where $\tilde{\hat{\rho}}\left(0\right)=\tilde{\hat{\rho}}_{S}\left(0\right)\otimes\hat{\rho}_{env}$
with $\hat{\rho}_{env}=\frac{1}{Z_{env}}e^{-\beta\hat{H}_{env}}$
and $Z_{env}=\mathrm{Tr}\left\{ e^{-\beta\hat{H}_{env}}\right\} $
(note that we can shift the time dependence between $\tilde{\hat{\Pi}}_{0d}$
and $\tilde{\hat{\rho}}$ by using the cyclic permutation properties
of the trace). This assumes that at the initial time, the system and
bath are uncorrelated and that the environment is in thermal equilibrium.
Later in chapter 4, we will consider cases where the environment is
not in thermal equilibrium and so $\hat{\rho}_{env}$ will have to
be modified. The term $\tilde{\hat{\rho}}_{S}\left(0\right)$ corresponds
to the density matrix of our system of interest. 

To make further progress, we write down the equation-of-motion for
$\tilde{\hat{\Pi}}_{0d}\left(t\right)$ in the interaction picture
with respect to $\hat{H}_{env}\left(\delta\hat{\omega}_{1},\delta\hat{\omega}_{2}\right)$.
The trace in (\ref{eq:3.1.5}) can be written as 
\begin{equation}
\left\langle \tilde{\hat{\Pi}}_{0d}\left(t\right)\right\rangle =\mathrm{Tr}\left\{ \tilde{\hat{\Pi}}_{I'0d}\left(t\right)\tilde{\hat{\rho}}\left(0\right)\right\} ,\label{eq:3.1.6}
\end{equation}
 where the subscript $I'$ denotes an operator that is in the interaction
picture with respect to $\hat{H}_{env}\left(\delta\hat{\omega}_{1},\delta\hat{\omega}_{2}\right)$.
The equation-of-motion is then given by
\begin{equation}
\frac{d}{dt}\tilde{\hat{\Pi}}_{I'0d}\left(t\right)=i\left[\hat{V}_{secI'}\left(t\right),\tilde{\hat{\Pi}}_{I'0d}\left(t\right)\right].\label{eq:3.1.7}
\end{equation}
Notice that in (\ref{eq:3.1.7}), $\tilde{\hat{\Pi}}_{I'0d}\left(t\right)$
has explicit time dependence. One possible trick for getting rid of
this time dependence is to integrate the equation of motion using
an iterative procedure. Integrating the equation-of-motion we get
\begin{equation}
\tilde{\hat{\Pi}}_{I'0d}\left(t\right)=\tilde{\hat{\Pi}}_{I'0d}\left(0\right)+i\int_{0}^{t}dt'\left[\hat{V}_{secI'}\left(t'\right),\tilde{\hat{\Pi}}_{I'0d}\left(t'\right)\right].\label{eq:3.1.8}
\end{equation}
We can iterate this result by inserting the first term as a zeroth-order
solution and repeating this process for all higher orders. Doing this
we find 
\begin{equation}
\tilde{\hat{\Pi}}_{I'0d}\left(t\right)=\tilde{\hat{\Pi}}_{I'0d}\left(0\right)+i\int_{0}^{t}dt_{1}\left[\hat{V}_{secI'}\left(t_{1}\right),\tilde{\hat{\Pi}}_{I'0d}\left(0\right)\right]+\left(i\right)^{2}\int_{0}^{t}dt_{1}\int_{0}^{t_{1}}dt_{2}\left[\hat{V}_{secI'}\left(t_{1}\right),\left[\hat{V}_{secI'}\left(t_{2}\right),\tilde{\hat{\Pi}}_{I'0d}\left(0\right)\right]\right]+...\label{eq:3.1.9}
\end{equation}
As can be seen from (\ref{eq:3.1.9}), $\tilde{\hat{\Pi}}_{I'0d}$
no longer depends on time and it is a straightforward matter to perform
the commutation relation using (\ref{eq:2.3.6}). First, however,
we can rewrite (\ref{eq:3.1.9}) in a much simpler form as 
\begin{equation}
\tilde{\hat{\Pi}}_{I'0d}\left(t\right)=T_{t}e^{i\int_{0}^{t}dt'L_{1}\left(t'\right)}\tilde{\hat{\Pi}}_{I'0d}\left(0\right),\label{eq:3.1.10}
\end{equation}
where $T_{t}$ is the time-ordering operator. By time-ordering operator,
we mean that all operators evaluated at later times appear to the
left of those evaluated at earlier times. So as an example, for a
product of two operators we could write
\begin{equation}
T_{t}\hat{A}\left(t_{1}\right)\hat{B}\left(t_{2}\right)=\hat{A}\left(t_{1}\right)\hat{B}\left(t_{2}\right)\Theta\left(t_{1}-t_{2}\right)\pm\hat{B}\left(t_{2}\right)\hat{A}\left(t_{1}\right)\Theta\left(t_{2}-t_{1}\right).\label{eq:3.1.11}
\end{equation}
The superoperator $L_{1}\left(t\right)$ acts in the following way
\begin{equation}
L_{1}\left(t\right)\tilde{\hat{\Pi}}_{I'0d}\left(0\right)=\left[\hat{V}_{secI'}\left(t\right),\tilde{\hat{\Pi}}_{I'0d}\left(0\right)\right].\label{eq:3.1.12}
\end{equation}
Using the fact that $\left[\left.|d\right\rangle \left\langle d|\right.,\left.|0\right\rangle \left\langle d|\right.\right]=-\left.|0\right\rangle \left\langle d|\right.$,
the commutator is evaluated to be
\begin{equation}
\left[\hat{V}_{secI'}\left(t\right),\tilde{\hat{\Pi}}_{I'0d}\left(0\right)\right]=-\left\{ \dot{\phi}\sin^{2}\theta\left(t\right)+\delta\hat{\omega}_{1}\left(t\right)\cos^{2}\theta\left(t\right)+\delta\hat{\omega}_{2}\left(t\right)\sin^{2}\theta\left(t\right)\right\} \left.|0\right\rangle \left\langle d|\right.,\label{eq:3.1.13}
\end{equation}
where $\delta\hat{\omega}_{1}\left(t\right)$ and $\delta\hat{\omega}_{2}\left(t\right)$
are in the interaction picture with respect to $\hat{H}_{env}\left(\delta\hat{\omega}_{1},\delta\hat{\omega}_{2}\right)$.
Consequently, we find that 
\begin{equation}
\tilde{\hat{\Pi}}_{I'0d}\left(t\right)=T_{t}e^{-i\int_{0}^{t}dt'\left(\dot{\phi}\sin^{2}\theta\left(t'\right)+\delta\hat{\omega}_{1}\left(t'\right)\cos^{2}\theta\left(t'\right)+\delta\hat{\omega}_{2}\left(t'\right)\sin^{2}\theta\left(t'\right)\right)}\tilde{\hat{\Pi}}_{0d},\label{eq:3.1.14}
\end{equation}
where $\tilde{\hat{\Pi}}_{0d}$ was defined in (\ref{eq:2.3.6}).
Inserting this result into the trace of Eq. (\ref{eq:3.1.6}), we
find that 
\begin{equation}
\left\langle \tilde{\hat{\Pi}}_{0d}\left(t\right)\right\rangle =\mathrm{Tr}\left\{ T_{t}e^{-i\int_{0}^{t}dt'\left(\dot{\phi}\sin^{2}\theta\left(t'\right)+\delta\hat{\omega}_{1}\left(t'\right)\cos^{2}\theta\left(t'\right)+\delta\hat{\omega}_{2}\left(t'\right)\sin^{2}\theta\left(t'\right)\right)}\tilde{\hat{\Pi}}_{0d}\tilde{\hat{\rho}}_{S}\left(0\right)\otimes\hat{\rho}_{env}\right\} .\label{eq:3.1.15}
\end{equation}
Using the identity $\mathrm{Tr}\left\{ \hat{A}\otimes\hat{B}\right\} =\mathrm{Tr}\left\{ \hat{A}\right\} \mathrm{Tr}\left\{ \hat{B}\right\} $
equation (\ref{eq:3.1.15}) reduces to
\begin{equation}
\left\langle \tilde{\hat{\Pi}}_{0d}\left(t\right)\right\rangle =\left\langle \tilde{\hat{\Pi}}_{0d}\left(0\right)\right\rangle \left\langle T_{t}e^{-i\int_{0}^{t}dt'\left(\dot{\phi}\sin^{2}\theta\left(t'\right)+\delta\hat{\omega}_{1}\left(t'\right)\cos^{2}\theta\left(t'\right)+\delta\hat{\omega}_{2}\left(t'\right)\sin^{2}\theta\left(t'\right)\right)}\right\rangle .\label{eq:3.1.16}
\end{equation}
Since the average in the second term of equation (\ref{eq:3.1.16})
is with respect to the environmental bath modes, we can simplify our
expression by writing
\begin{equation}
\left\langle \tilde{\hat{\Pi}}_{0d}\left(t\right)\right\rangle =e^{-i\gamma_{d}\left(t\right)}\left\langle \tilde{\hat{\Pi}}_{0d}\left(0\right)\right\rangle \left\langle T_{t}e^{-i\int_{0}^{t}dt'\left(\delta\hat{\omega}_{1}\left(t'\right)\cos^{2}\theta\left(t'\right)+\delta\hat{\omega}_{2}\left(t'\right)\sin^{2}\theta\left(t'\right)\right)}\right\rangle .\label{eq:3.1.17}
\end{equation}
Notice that the correction to the geometric phase arising from the
system-bath coupling is now transparent. The term inside the thermal
average corresponds to corrections to the geometric phase which will
give rise to dephasing (which is why these types of contributions
are called geometric dephasing \cite{key-29}). We also mention that
the average $\left\langle \tilde{\hat{\Pi}}_{0d}\left(0\right)\right\rangle $
is given by (\ref{eq:2.3.8}).
\end{doublespace}

\section{\label{sec:Antisymmetric-noise}Antisymmetric noise $\delta\hat{\omega}_{1}=-\delta\hat{\omega}_{2}$}

In this section we consider the case of purely antisymmetric noise
so that $\delta\hat{\omega}_{1}=-\delta\hat{\omega}_{2}$. In this
case (\ref{eq:3.1.17}) will simplify to

\begin{equation}
\left\langle \tilde{\hat{\Pi}}_{0d}\left(t\right)\right\rangle =\left\langle \tilde{\hat{\Pi}}_{0d}\left(0\right)\right\rangle e^{-i\gamma_{d}\left(t\right)}\left\langle T_{t}e^{-i\int_{0}^{t}dt'\cos\left(2\theta\left(t'\right)\right)\delta\hat{\omega}_{1}\left(t'\right)}\right\rangle .\label{eq:3.1.18}
\end{equation}
This is a crucial step for later calculations to come. By picking
$\delta\hat{\omega}_{1}=-\delta\hat{\omega}_{2}$, we are considering
a situation where there are strong correlations between the noise
hitting the two levels $\left.|g_{1}\right\rangle $ and $\left.|g_{2}\right\rangle $.
This opens up a very interesting possibility, the idea of using this
correlation to suppress decoherence while one is still doing the desired
adiabatic evolution protocol. We refer the reader to chapter \ref{chap:Atom-coupled-to}
for more details. Note that in chapter \ref{chap:Atom-coupled-to},
it will be possible to choose reasonable parameters so that the cavity
shot noise will be in the regime satisfying $\delta\hat{\omega}_{1}=-\delta\hat{\omega}_{2}$. 

It is also important to understand the origin of the function $\cos\left(2\theta\left(t\right)\right)$
appearing in (\ref{eq:3.1.18}). If $\theta\left(t\right)=0$ or $\theta\left(t\right)=\frac{\pi}{2}$,
it is clear from (\ref{eq:2.2.21}) and (\ref{eq:2.2.26}) that the
state is either all $\left.|g_{1}\right\rangle $ or all $\left.|g_{2}\right\rangle $.
In this case, we would only see the $\delta\hat{\omega}_{1}$ or $\delta\hat{\omega}_{2}$
noise and hence the correlations aren't important. Therefore, we would
expect the noise to be maximal for $\theta\left(t\right)=0$ or $\theta\left(t\right)=\frac{\pi}{2}$.
On the other hand, for $\theta\left(t\right)=\frac{\pi}{4}$, the
state (\ref{eq:2.2.26}) is in an equal superposition of $\left.|g_{1}\right\rangle $
and $\left.|g_{2}\right\rangle $. In this case the average of the
noise Hamiltonian is $\delta\hat{\omega}_{1}+\delta\hat{\omega}_{2}=0$.
Thus the state (\ref{eq:2.2.26}) sees both the $\delta\hat{\omega}_{1}$
and $\delta\hat{\omega}_{2}$ noises equally; as they are perfectly
anti-correlated, the net contribution is always zero. All these arguments
are quantitatively understood from the function $\cos\left(2\theta\left(t\right)\right)$. 

The quantity that we will mostly be interested in when performing
our state-transfer protocols is the fidelity. This quantity, which
takes on values between 0 and 1, will describe how ``close'' our
final state is to the original one. So for a perfect state transfer
with no accumulated phase difference between the spectator and ground
state, the fidelity would be unity. Formally, the fidelity is defined
as 
\begin{equation}
F\equiv Tr\left\{ \tilde{\hat{\rho}}\left(t_{f}\right)\left.|\psi_{s}\left(t_{i}\right)\right\rangle \left\langle \psi_{s}\left(t_{i}\right)|\right.\right\} .\label{eq:3.1.19}
\end{equation}
The state $\left.|\psi_{s}\left(t_{i}\right)\right\rangle $ is given
by (\ref{eq:2.2.7}) and we used the fact that at the initial time,
the density matrix is described by a pure state and can thus be written
as 
\begin{equation}
\tilde{\hat{\rho}}_{s}\left(t_{i}\right)=\left.|\psi_{s}\left(t_{i}\right)\right\rangle \left\langle \psi_{s}\left(t_{i}\right)|\right.\label{eq:3.1.20}
\end{equation}
 The fidelity can be written in terms of the relevant component of
the density matrix using the following procedure. In general, the
average $\left\langle \tilde{\hat{\Pi}}_{\alpha\beta}\left(t_{i}\right)\right\rangle $
is given by 
\begin{equation}
\left\langle \tilde{\hat{\Pi}}_{\alpha\beta}\left(t\right)\right\rangle \equiv Tr\left\{ \left.|\alpha\right\rangle \left\langle \beta|\right.\tilde{\hat{\rho}}\left(t\right)\right\} .\label{eq:3.1.21}
\end{equation}
Using the relation (\ref{eq:2.3.5}) and (\ref{eq:3.1.21}), the fidelity
reduces to 
\begin{equation}
F=|\alpha|^{2}\left\langle \tilde{\hat{\Pi}}_{00}\left(t_{f}\right)\right\rangle +2Re\left[\alpha^{*}\beta\left\langle \tilde{\hat{\Pi}}_{0d}(t_{f})\right\rangle \right]+|\beta|^{2}\left\langle \tilde{\hat{\Pi}}_{dd}\left(t_{f}\right)\right\rangle ,\label{eq:3.1.22}
\end{equation}
where we used (\ref{eq:2.3.5}) to obtain $\left\langle \tilde{\hat{\Pi}}_{00}(t_{i})\right\rangle =|\alpha|^{2}$
and $\left\langle \tilde{\hat{\Pi}}_{dd}(t_{i})\right\rangle =|\beta|^{2}$.
Since $\left[\left.|0\right\rangle \left\langle 0|\right.,\hat{V}_{secI'}\left(t\right)\right]=\left[\left.|d\right\rangle \left\langle d|\right.,\hat{V}_{secI'}\left(t\right)\right]=0$,
then we can follow the same logic that was used to go from (\ref{eq:3.1.5})
to (\ref{eq:3.1.16}). In this case we find that 
\begin{equation}
\left\langle \tilde{\hat{\Pi}}_{00}\left(t_{f}\right)\right\rangle =|\alpha|^{2},\label{eq:3.1.23}
\end{equation}
and
\begin{equation}
\left\langle \tilde{\hat{\Pi}}_{dd}\left(t_{f}\right)\right\rangle =|\beta|^{2}.\label{eq:3.1.24}
\end{equation}
 The fidelity then reduces to 
\begin{equation}
F=|\alpha|^{4}+|\beta|^{4}+2Re\left[\alpha^{*}\beta\left\langle \tilde{\hat{\Pi}}_{0d}(t_{f})\right\rangle \right].\label{eq:3.1.25}
\end{equation}
We can use (\ref{eq:3.1.18}) to write the fidelity as 
\begin{equation}
F=\left|\alpha\right|^{4}+\left|\beta\right|^{4}+2\left|\alpha\right|^{2}\left|\beta\right|^{2}\mathrm{Re}\left[e^{-i\gamma_{d}\left(t_{f}\right)}\left\langle T_{t}e^{-i\int_{0}^{t_{f}}dt'\cos\left(2\theta\left(t'\right)\right)\delta\hat{\omega}_{1}\left(t'\right)}\right\rangle \right].\label{eq:3.1.26}
\end{equation}
This is one of the central results of our work and will be used throughout
this thesis. If the system were not coupled to a dissipative quantum
environment, then the fidelity would simply be given by 
\begin{equation}
F=\left|\alpha\right|^{4}+\left|\beta\right|^{4}+2\left|\alpha\right|^{2}\left|\beta\right|^{2}\cos\gamma_{d}\left(t_{f}\right)\label{eq:3.1.27}
\end{equation}
When the system is coupled to a quantum dissipative environment, the
fidelity will be given by (\ref{eq:3.1.26}) which has an extra contribution
which will depend on the spectral density of the bath. The reason
is that we must perform an average over the bath degrees of freedom
and so will be faced with calculating correlation functions of the
form $\left\langle \delta\hat{\omega}_{1}\left(t_{1}\right)\delta\hat{\omega}_{1}\left(t_{2}\right)\right\rangle $.
Consequently, the dynamics of the bath will determine how the geometric
phase will be modified due to the quantum system-bath coupling. Recall
that these results are valid in the large-$G$ limit. If this were
not the case, we would need to include all the off-diagonal components
of the superadiabatic Hamiltonian which would then give extra contributions
to the geometric phase which are not accounted for in (\ref{eq:3.1.26}). 

\begin{doublespace}
We conclude this section with a neat result that stems from taking
a particular spectral density for the bath degrees-of-freedom. To
simplify the notation, we define 
\begin{equation}
\hat{X}\left(t\right)\equiv\int_{0}^{t}dt'\cos\left(2\theta\left(t'\right)\right)\delta\hat{\omega}_{1}\left(t'\right).\label{eq:3.1.28}
\end{equation}
When performing a statistical ensemble average over the realizations
of the bath degrees-of-freedom, we assume that the central-limit theorem
applies so that 
\begin{equation}
\left\langle e^{-i\hat{X}\left(t\right)}\right\rangle \approx e^{-\frac{1}{2}\left\langle \hat{X}^{2}\left(t\right)\right\rangle }.\label{eq:3.1.29}
\end{equation}
In getting the above result, we effectively treat $\hat{X}\left(t\right)$
as Gaussian, which implies that $t$ in (\ref{eq:3.1.29}) is in general
much longer than the correlation length of the noise coming from $\delta\hat{\omega}_{1}\left(t\right)$
(or that $\delta\hat{\omega}_{1}\left(t\right)$ is itself Gaussian).
In general, we can write 
\begin{equation}
\left\langle \delta\hat{\omega}_{1}\left(t\right)\delta\hat{\omega}_{1}\right\rangle =\mathrm{Re}\left[\int_{-\infty}^{\infty}\frac{d\omega}{2\pi}e^{-i\omega t}J\left(\omega\right)\right],\label{eq:3.1.30}
\end{equation}
where we take the real part since this is the only relevant contribution
that will appear in the fidelity. We will consider the case where
the spectral density $J\left(\omega\right)$ is given by a Lorentzian
peaked at a non-zero frequency $\nu_{0}$. Consequently, it can be
written as 
\begin{equation}
J\left(\omega\right)=\frac{\sigma^{2}\Gamma}{\left(\omega-\nu_{0}\right)^{2}+\Gamma^{2}}.\label{eq:3.1.31}
\end{equation}
Here, $\Gamma$ specifies the width of the Lorentzian and $\sigma$
specifies its amplitude. Performing the integral in (\ref{eq:3.1.30})
using (\ref{eq:3.1.31}), we find that 
\begin{equation}
\left\langle \delta\hat{\omega}_{1}\left(t\right)\delta\hat{\omega}_{1}\right\rangle =\sigma^{2}\cos\left(\nu_{0}t\right)e^{-\Gamma\left|t\right|}.\label{eq:3.1.32}
\end{equation}

The goal is to compute the fidelity for our state-transfer protocol
for the case where the spectral density of the bath degrees-of-freedom
is given by the Lorentzian described above. From (\ref{eq:3.1.28})
and (\ref{eq:3.1.29}), the relevant quantity to calculate is 
\begin{equation}
\left\langle X^{2}\left(t\right)\right\rangle =\int_{0}^{t}dt_{1}\int_{0}^{t}dt_{2}\cos\left(2\theta\left(t_{1}\right)\right)\cos\left(2\theta\left(t_{2}\right)\right)\left\langle \delta\hat{\omega}_{1}\left(t_{1}\right)\delta\hat{\omega}_{1}\left(t_{2}\right)\right\rangle .\label{eq:3.1.33}
\end{equation}
Using (\ref{eq:3.1.32}), this reduces to 
\begin{equation}
\left\langle X^{2}\left(t\right)\right\rangle =\sigma^{2}\int_{0}^{t}dt_{1}\int_{0}^{t}dt_{2}\cos\left(2\theta\left(t_{1}\right)\right)\cos\left(2\theta\left(t_{2}\right)\right)\cos\left(\nu_{0}\left(t_{1}-t_{2}\right)\right)e^{-\Gamma\left|t_{1}-t_{2}\right|}.\label{eq:3.1.34}
\end{equation}
So we see from (\ref{eq:3.1.34}) that $\Gamma$ acts as a damping
rate. We also need to specify the path for our state transfer protocol
in $\left\{ \Omega_{1},\Omega_{2}\right\} $ space that runs from
$t_{i}=0$ to $t_{f}$. One of the simplest possible paths (and one
that will be used throughout this thesis) is a circular path. To achieve
this, we choose 
\begin{equation}
\Omega_{1}\left(t\right)=-A\sin\left(\frac{n\pi t}{t_{f}}\right),\label{eq:3.1.35}
\end{equation}
\begin{equation}
\Omega_{2}\left(t\right)=A\cos\left(\frac{n\pi t}{t_{f}}\right),\label{eq:3.1.36}
\end{equation}
\begin{figure}[h]
\begin{centering}
\includegraphics[scale=0.5]{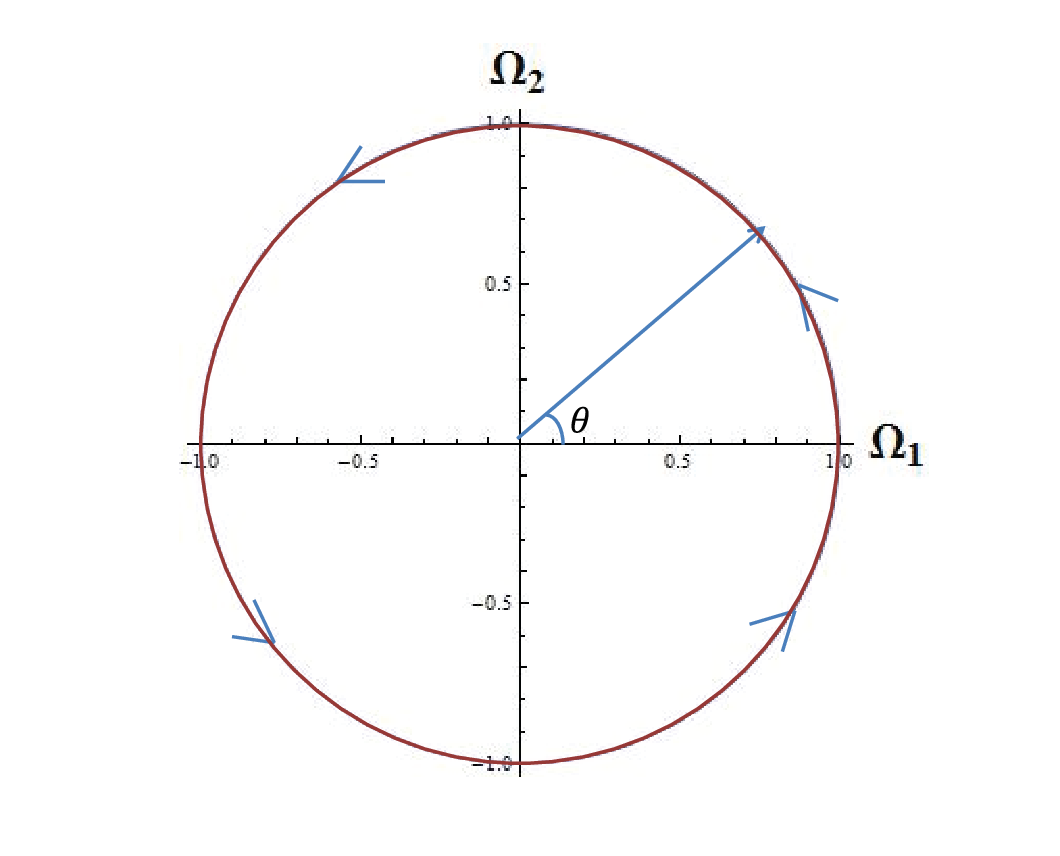}
\par\end{centering}

\caption{\ Circular path}

\begin{doublespace}
\centering{}\textit{Plot of the path for our quantum state transfer
protocol in $\left\{ \Omega_{1},\Omega_{2}\right\} $ space. The functions
chosen for our laser amplitudes are given in Eq. (\ref{eq:3.1.35})
and (\ref{eq:3.1.36}). }\end{doublespace}
\end{figure}
 where $n$ is a parameter that determines how many loops we do during
the evolution. Then, from (\ref{eq:2.2.19}), we get a simple linear
relation for the angle $\theta\left(t\right)$ given by 
\begin{equation}
\theta\left(t\right)=\frac{n\pi t}{t_{f}}.\label{eq:3.1.37}
\end{equation}
This choice also keeps the gap $G\left(t\right)$ constant. In the
ideal case where the damping rate $\Gamma$ is set to zero, then the
integral in (\ref{eq:3.1.34}) is straightforward to calculate and
will lead to some interesting properties for special values of the
frequency $\nu_{0}$. We remind the reader that $\nu_{0}$ is the
frequency of the peak in the noise spectral density, see equation
(\ref{eq:3.1.31}). Evaluating the integral for a single cycle in
parameter space ($n=1$) and for a total time $t_{f}$ leads to 
\begin{equation}
\left\langle X^{2}\left(t_{f}\right)\right\rangle \left|_{\Gamma=0,n=1}\right.=\frac{2\sigma^{2}t_{f}^{4}\nu_{0}^{2}\left(1-\cos\left(\nu_{0}t_{f}\right)\right)}{\left(\nu_{0}^{2}t_{f}^{2}-4\pi^{2}\right)^{2}}.\label{eq:3.1.38}
\end{equation}
At this stage, one can immediately see that for the values 
\begin{equation}
\nu_{0}=\frac{2m\pi}{t_{f}},\,\left\{ m\neq2\right\} \label{eq:3.1.39}
\end{equation}
where $m$ is a positive integer not equal to two, the function $\left\langle X^{2}\left(t_{f}\right)\right\rangle \left|_{\Gamma=0,n=1}\right.$
is identically zero. Thus, choosing these specific values for the
frequency $\nu_{0}$ (or equivalently, choosing $t_{f}$ for fixed
$\nu_{0}$) would cancel the effect of having our system coupled to
a quantum dissipative bath. It will be instructive to draw plots of
$\left\langle \tilde{\hat{\Pi}}_{0d}\left(t\right)\right\rangle $,
also called coherence, for values of time that start at zero and end
at $t_{f}$. To obtain these, we integrate equation (\ref{eq:3.1.34})
from zero to $t$. From (\ref{eq:3.1.18}) we would have
\begin{equation}
\left\langle \tilde{\hat{\Pi}}_{0d}\left(t\right)\right\rangle =\left\langle \tilde{\hat{\Pi}}_{0d}\left(0\right)\right\rangle e^{-i\gamma_{d}\left(t\right)}e^{-\frac{1}{2}\left\langle X^{2}\left(t\right)\right\rangle \left|_{\Gamma=0,n=1}\right.}\label{eq:3.1.40}
\end{equation}

\end{doublespace}

\begin{center}
\begin{figure}[h]
\begin{centering}
\includegraphics[scale=0.9]{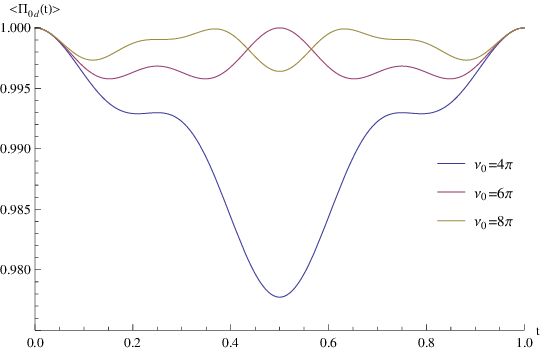}
\par\end{centering}

\caption{\ Density matrix revivals}
\label{Revivals}

\begin{doublespace}
\centering{}\textit{Plot of $\left\langle \tilde{\hat{\Pi}}_{0d}\left(t\right)\right\rangle $
as a function of time. We set $t_{f}=1$, $\sigma=1$ and $\Gamma=0$.
Since we chose values of $\nu_{0}$ which correspond to those of (\ref{eq:3.1.39}),
we observe revivals at the end of the state transfer time. We also
set the geometric phase $\gamma_{d}\left(t\right)=0$. }\end{doublespace}
\end{figure}

\par\end{center}

Since we chose values of $\nu_{0}$ that correspond to (\ref{eq:3.1.39})
for the plot of (Fig. \ref{Revivals}), we observe revivals in the
relevant component of the density matrix. This means that during the
state transfer, our system loses information, and then gains it back
at the end of the transfer time. Consequently, if we could engineer
a system where the spectral density of the bath degrees of freedom
corresponded to a Lorentzian peaked at a non-zero frequency, then
we could always choose a frequency given by (\ref{eq:3.1.39}) that
would cancel the effects of a coupling to an environment. Of course
if we add the effect of damping ($\Gamma\neq0$) then this would no
longer be the case. In this case we would observe ``damped'' revivals
so instead of having the coherence be unity at the end of the state
transfer, it would be smaller by a factor that depends on $\Gamma$. 

\begin{center}
\begin{figure}[b]
\begin{centering}
\includegraphics{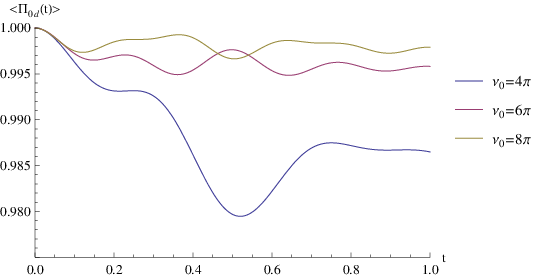}
\par\end{centering}

\caption{\ Density matrix revivals with damping}

\begin{doublespace}
\centering{}\textit{Plot of $\left\langle \tilde{\hat{\Pi}}_{0d}\left(t\right)\right\rangle $
as a function of time. We set $t_{f}=1$, $\sigma=1$ and $\Gamma=1$.
Since we chose values of $\nu_{0}$ which correspond to those of (\ref{eq:3.1.39}),
we observe partial revivals at the end of the state transfer time.
We also set the geometric phase $\gamma_{d}\left(t\right)=0$. In
the presence of damping, the coherence does not come back to unity
at the end of the state transfer protocol. Some information is inevitably
lost. }\end{doublespace}
\end{figure}

\par\end{center}

\section{Environment in thermal equilibrium}

\begin{doublespace}
Recall that for an environment in thermal equilibrium, the ``environment''
density matrix takes the form $\hat{\rho}_{env}=\frac{1}{Z_{env}}e^{-\beta\hat{H}_{env}}$.
In this section we will consider the case where the environment mode
$\delta\hat{\omega}_{1}$ can be written as 
\begin{equation}
\delta\hat{\omega}_{1}=\sum_{k}\gamma_{k}\left(\hat{b}_{k}+\hat{b}_{k}^{\dagger}\right),\label{eq:3.2.1}
\end{equation}
and that $\hat{H}_{env}\left(\delta\hat{\omega}_{1}\right)$ is quadratic
in the bosonic annihilation and creation operators
\begin{equation}
\hat{H}_{env}\left(\delta\hat{\omega}_{1}\right)=\sum_{k}\epsilon_{k}\hat{b}_{k}^{\dagger}\hat{b}_{k}.\label{eq:3.2.2}
\end{equation}
This system is often referred to as the independent boson model \cite{key-42}.
Given the above constraints it will be possible to apply Wick's theorem
to evaluate the average in Eq. (\ref{eq:3.1.18}). For simplicity,
we define 
\begin{equation}
C\left(t\right)=\left\langle T_{t}e^{i\hat{\Theta}\left(t\right)}\right\rangle ,\label{eq:3.2.3}
\end{equation}
where 
\begin{equation}
\hat{\Theta}\left(t\right)\equiv\int_{0}^{t}dt'\cos\left(2\theta\left(t'\right)\right)\delta\hat{\omega}_{1}\left(t'\right).\label{eq:3.2.4}
\end{equation}
It will prove to be convenient to expand the exponential in Eq. (\ref{eq:3.2.3})
in its Taylor series since it will allow us to use Wick's theorem
in a convenient way. Doing so, we find 
\begin{equation}
C\left(t\right)=\sum_{l=0}^{\infty}\frac{\left(i\right)^{2l}}{\left(2l\right)!}\left\langle T_{t}\hat{\Theta}^{2l}\left(t\right)\right\rangle .\label{eq:3.2.5}
\end{equation}
Here we used the fact that only even powers of $\hat{\Theta}\left(t\right)$
give non-vanishing terms when performing the time ordered average.
Using Wick's theorem, we start with 
\begin{equation}
\left\langle T_{t}\hat{\Theta}^{2k}\left(t\right)\right\rangle =\left[\mathrm{Np}\right]\left\langle T_{t}\hat{\Theta}\left(t\right)\hat{\Theta}\left(t\right)\right\rangle ^{k}.\label{eq:3.2.6}
\end{equation}
In this case $\mathrm{Np}$ represents the number of ways of finding
$k$ pairs from $2k$ identical elements. For 4 elements, there would
be three of these pairs so that $\mathrm{Np}=3$. For 6 elements,
there would be $5\times3$ pairs, for 8 there would be $7\times5\times3$
pairs and so on. In general, for $2k$ pairs we would have 
\begin{equation}
\mathrm{Np}\left|_{2k}\right.=\left(2k-1\right)!!=\frac{1}{2^{k}}\frac{\left(2k\right)!}{k!}.\label{eq:3.2.7}
\end{equation}
Using this result and (\ref{eq:3.2.6}), we find
\begin{equation}
C\left(t\right)=\sum_{k}\frac{1}{k!}\left[-\frac{1}{2}\left\langle T_{t}\hat{\Theta}\left(t\right)\hat{\Theta}\left(t\right)\right\rangle \right]^{k},\label{eq:3.2.8}
\end{equation}
so that
\begin{equation}
C\left(t\right)=e^{-\frac{1}{2}\left\langle T_{t}\hat{\Theta}\left(t\right)\hat{\Theta}\left(t\right)\right\rangle }.\label{eq:3.2.9}
\end{equation}

At this stage we still need to compute $\left\langle T_{t}\hat{\Theta}\left(t\right)\hat{\Theta}\left(t\right)\right\rangle $.
Using the definition of $\hat{\Theta}\left(t\right)$ we have that
\begin{equation}
\left\langle T_{t}\hat{\Theta}\left(t\right)\hat{\Theta}\left(t\right)\right\rangle =\int_{0}^{t}dt_{1}\int_{0}^{t}dt_{2}\cos2\theta\left(t_{1}\right)\cos2\theta\left(t_{2}\right)\left\langle T_{t}\delta\hat{\omega}_{1}\left(t_{1}\right)\delta\hat{\omega}_{1}\left(t_{2}\right)\right\rangle .\label{eq:3.2.10}
\end{equation}
For a circular path in $\left\{ \Omega_{1}\left(t\right),\Omega_{2}\left(t\right)\right\} $
space, $\theta\left(t\right)=-bt$ (where we define $b\equiv\frac{2\pi}{t_{f}}$)
so that (\ref{eq:3.2.10}) reduces to
\begin{equation}
\left\langle T_{t}\hat{\Theta}\left(t\right)\hat{\Theta}\left(t\right)\right\rangle =\int_{0}^{t}dt_{1}\int_{0}^{t}dt_{2}\cos2bt_{1}\cos2bt_{2}\left\langle T_{t}\delta\hat{\omega}_{1}\left(t_{1}\right)\delta\hat{\omega}_{1}\left(t_{2}\right)\right\rangle .\label{eq:3.2.11}
\end{equation}
We can replace the operator $\delta\hat{\omega}_{1}$ by its sum over
boson creation and annihilation operators (given in Eq. (\ref{eq:3.2.1}))
into the expression above to find
\begin{equation}
\left\langle T_{t}\hat{\Theta}\left(t\right)\hat{\Theta}\left(t\right)\right\rangle =\sum_{k}\gamma_{k}^{2}\int_{0}^{t}dt_{1}\int_{0}^{t}dt_{2}\cos2bt_{1}\cos2bt_{2}\left[\left\langle T_{t}\hat{b}_{k}\left(t_{1}\right)\hat{b}_{k}^{\dagger}\left(t_{2}\right)\right\rangle +\left\langle T_{t}\hat{b}_{k}^{\dagger}\left(t_{1}\right)\hat{b}_{k}\left(t_{2}\right)\right\rangle \right].\label{eq:3.2.12}
\end{equation}
Since the Hamiltonian is quadratic in the bosonic creation and annihilation
operators, it is straightforward to obtain their time dependence.
From a Heisenberg equation of motion, we can write 
\begin{equation}
\dot{\hat{b}}_{l}\left(t\right)=i\left[\hat{H}_{env},\hat{b}_{l}\left(t\right)\right]=ie^{i\hat{H}_{env}t}\left[\hat{H}_{env},\hat{b}_{l}\right]e^{-i\hat{H}_{env}t}.\label{eq:3.2.13}
\end{equation}
Using Eq. (\ref{eq:3.2.2}) the commutator is straightforward to compute
and we find
\begin{equation}
\dot{\hat{b}}_{l}\left(t\right)=-i\epsilon_{l}\hat{b}_{l}\left(t\right),\label{eq:3.2.14}
\end{equation}
which has the solution
\begin{equation}
\hat{b}_{l}\left(t\right)=e^{-i\epsilon_{l}t}\hat{b}_{l}.\label{eq:3.2.15}
\end{equation}
Since we are taking a thermal average of the bosonic operators $\left\langle \hat{b}_{l}^{\dagger}\hat{b}_{l}\right\rangle $
for an environment in thermal equilibrium, then we simply have a Bose-Einstein
distribution function 
\begin{equation}
\left\langle \hat{b}_{l}^{\dagger}\hat{b}_{l}\right\rangle =n_{B}\left(\epsilon_{l}\right)=\frac{1}{e^{\beta\epsilon_{l}}-1}.\label{eq:3.2.16}
\end{equation}
Using (\ref{eq:3.2.15}), (\ref{eq:3.2.16}) and taking explicit consideration
of the time ordering operator present in (\ref{eq:3.2.9}), we find
\begin{equation}
\left\langle T_{t}\hat{\Theta}\left(t\right)\hat{\Theta}\left(t\right)\right\rangle =\sum_{k}\gamma_{k}^{2}\int_{0}^{t}dt_{1}\int_{0}^{t}dt_{2}\cos2bt_{1}\cos2bt_{2}\left[e^{-i\epsilon_{k}\left|t_{1}-t_{2}\right|}+2n_{B}\left(\epsilon_{k}\right)\cos\left(\epsilon_{k}\left(t_{1}-t_{2}\right)\right)\right].\label{eq:3.2.17}
\end{equation}
We can decompose the contributions into its real and imaginary parts.
The real part will be given by 
\begin{equation}
\mathrm{Re}\left[\left\langle T_{t}\hat{\Theta}\left(t\right)\hat{\Theta}\left(t\right)\right\rangle \right]=\sum_{k}\gamma_{k}^{2}\left(1+2n_{B}\left(\epsilon_{k}\right)\right)\int_{0}^{t}dt_{1}\int_{0}^{t}dt_{2}\cos2bt_{1}\cos2bt_{2}\cos\left(\epsilon_{k}\left(t_{1}-t_{2}\right)\right).\label{eq:3.2.18}
\end{equation}
Evaluating the integral and using $1+2n_{B}\left(\epsilon_{k}\right)=\cosh\left(\frac{\beta\epsilon_{k}}{2}\right)$,
the real contribution is given by 
\begin{equation}
\mathrm{Re}\left[\left\langle T_{t}\hat{\Theta}\left(t\right)\hat{\Theta}\left(t\right)\right\rangle \right]=\sum_{k}\frac{\gamma_{k}^{2}\cosh\left(\frac{\beta\epsilon_{k}}{2}\right)}{\left(4b^{2}-\epsilon_{k}^{2}\right)^{2}}\left[\epsilon_{k}^{2}\left(1+\cos\left(2bt\right)\left\{ 1-2\cos\epsilon_{k}t\right\} \right)-4\epsilon_{k}b\sin\left(2bt\right)\sin\left(\epsilon_{k}t\right)+4b^{2}\sin^{2}\left(4bt\right)\right].\label{eq:3.2.19}
\end{equation}
Finally, we can also obtain the imaginary contribution
\begin{equation}
\mathrm{Im}\left[\left\langle T_{t}\hat{\Theta}\left(t\right)\hat{\Theta}\left(t\right)\right\rangle \right]=-\sum_{k}\gamma_{k}^{2}\int_{0}^{t}dt_{1}\int_{0}^{t}dt_{2}\cos2bt_{1}\cos2bt_{2}\sin\left(\epsilon_{k}\left|t_{1}-t_{2}\right|\right).\label{eq:3.2.20}
\end{equation}
Evaluating the integral yields
\begin{align}
\mathrm{Im}\left[\left\langle T_{t}\hat{\Theta}\left(t\right)\hat{\Theta}\left(t\right)\right\rangle \right] & =-\sum_{k}\frac{\gamma_{k}^{2}\epsilon_{k}}{4b\left(\epsilon_{k}^{2}-4b^{2}\right)^{2}}\left[16b^{2}\cos\left(\epsilon_{k}t\right)\sin\left(2bt\right)\right.\nonumber \\
 & \left.+\left(\epsilon_{k}^{2}-4b^{2}\right)\left(4bt+\sin\left(4bt\right)\right)-8b\epsilon_{k}\cos\left(2bt\right)\sin\left(\epsilon_{k}t\right)\right].\label{eq:3.2.21}
\end{align}

We could go on to evaluate these quantities by taking a continuum
limit and choosing a specific functional dependence of the coupling
strength $\gamma_{k}$. However, the objective of this section was
to show how we could apply our methods for computing the coherence
of a specific coupling to a bosonic bath. Equation (\ref{eq:3.2.19})
would give the accumulated phase for our state-transfer protocol which
would result in a correction to the closed-system geometric phase
in the case where the system would not be coupled to a bosonic bath.
In chapter 4 we will apply our methods to the case of an atom coupled
to a driven cavity. The noise will arise from fluctuations in the
number of photons inside the cavity. The goal will be to perform a
state transfer for our qubit and obtain a expression for the fidelity
of our state transfer. We will then try to optimize our path in $\left\{ \Omega_{1},\Omega_{2}\right\} $
space in order to get the best possible fidelity for this particular
quantum state transfer.
\end{doublespace}

\section{Summary}

In this section we showed how to obtain the phase of the coherence
for the case where our system was coupled to a quantum dissipative
bath. We first wrote the Hamiltonian in a superadiabatic basis by
performing a unitary transformation given by Eq. (\ref{eq:2.3.3}).
We proceeded by performing a secular approximation which amounted
to throwing away all the off-diagonal terms of the superadiabatic
Hamiltonian. This was justified in the limits where $\frac{\dot{\theta}}{G}\ll1$,
$\frac{\dot{\phi}}{G}\ll1$ and $\frac{\left|\delta\hat{\omega}_{i}\right|}{G}\ll1$.
Since the Hamiltonian was purely diagonal, it was a straightforward
matter to find the phase for $\left\langle \tilde{\hat{\Pi}}_{0d}\left(t\right)\right\rangle $
by iterating its equation of motion. We then related the fidelity
for the quantum state transfer to this phase via Eq. (\ref{eq:3.1.26}).
Using these results, we considered an example where the bath spectral
density was given by a Lorentzian peaked at a non-zero frequency.
We showed that for the values of the frequency given by Eq. (\ref{eq:3.1.39}),
these gave rise to recurrences in the coherence thus canceling the
effects of having our system coupled to a quantum dissipative bath.
We concluded this chapter by applying our methods to the independent
boson model to show how the state transfer protocol could be used
in a physical system. 

\newpage{}

\chapter{\label{chap:Atom-coupled-to}Atom coupled to a cavity}

\rule[0.5ex]{1\columnwidth}{1pt}

\section{\label{sec:Dephasing-due-to}Dephasing due to cavity shot noise}

\begin{figure}
\begin{centering}
\includegraphics[scale=0.6]{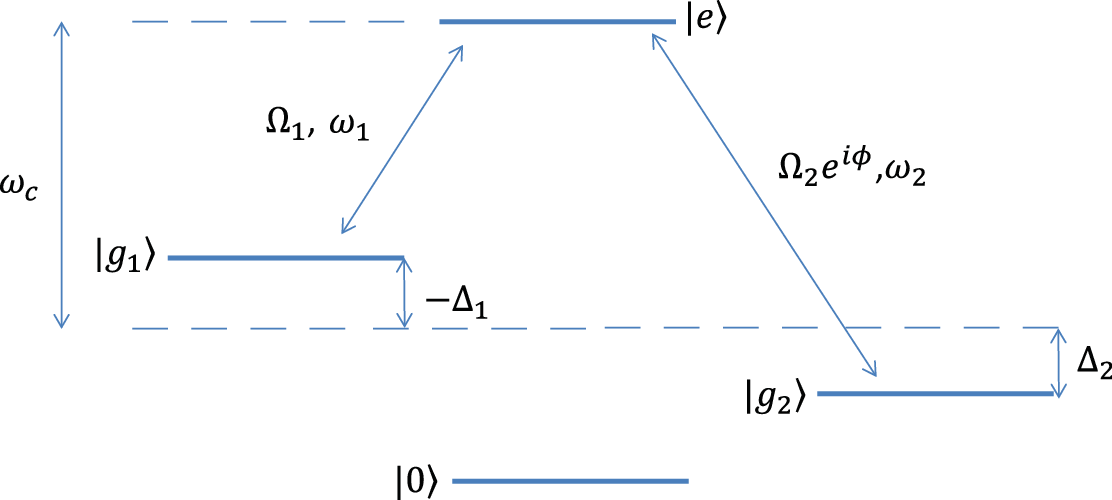}
\par\end{centering}

\centering{}\caption{\ \label{fig:Energy-level-diagram-in}Energy-level diagram in the
resonance case}
\textit{Energy-level diagram showing the structure when the laser
tones are resonant with the ($\left.|g_{1}\right\rangle $,$\left.|e\right\rangle $)
and ($\left.|g_{2}\right\rangle $,$\left.|e\right\rangle $) transitions
(see (\ref{eq:4.1.10})). }
\end{figure}

\begin{doublespace}
In this section we apply the state-transfer methods developed in chapter
3 to a four-level atom coupled to a single cavity mode. To do so,
we will drive the cavity with two classical laser fields (also containing
a quantum contribution which will add noise to the system) each detuned
from the cavity frequency (see figure (\ref{fig:Energy-level-diagram-in})).
The laser fields will provide the tunable couplings needed for the
adiabatic protocol of chapter 3. As was shown in \cite{key-40}, driving
the cavity with the two laser fields will induce fluctuations in the
number of photons inside the cavity caused by the quantum noise present
in each laser field. Consequently, decoherence and dephasing effects
will arise when performing our state transfer protocol which will
need to be accounted for. Note that this system will be particularly
well-suited to performing a phase gate which is another scenario that
we will consider at the end of this chapter.

The starting Hamiltonian will have the usual Jaynes-Cummings form
with an added contribution arising from the external bath modes and
their coupling to the cavity modes. The Hamiltonian thus takes the
form
\begin{equation}
\hat{H}\left(t\right)=\hat{H}_{0}+\hat{H}_{1}\left(t\right),\label{eq:4.1.1}
\end{equation}
where we define 
\begin{equation}
\hat{H}_{0}=\omega_{c}\hat{a}^{\dagger}\hat{a}+\omega_{g_{1}}\left.|g_{1}\right\rangle \left\langle g_{1}|\right.+\omega_{g_{2}}\left.|g_{2}\right\rangle \left\langle g_{2}|\right.+\omega_{e}\left.|e\right\rangle \left\langle e|\right.,\label{eq:4.1.2}
\end{equation}
\begin{equation}
\hat{H}_{1}\left(t\right)=g_{1}\left(\left.|e\right\rangle \left\langle g_{1}|\right.\hat{a}+\left.|g_{1}\right\rangle \left\langle e|\right.\hat{a}^{\dagger}\right)+g_{2}\left(\left.|e\right\rangle \left\langle g_{2}|\right.\hat{a}+\left.|g_{2}\right\rangle \left\langle e|\right.\hat{a}^{\dagger}\right)+\hat{H'}_{env}\left(t\right).\label{eq:4.1.3}
\end{equation}
Here, $\omega_{c}$ is the frequency associated with the cavity mode
and $\hat{H}_{env}\left(t\right)$ is given by
\begin{align}
\hat{H'}{}_{env}\left(t\right) & =\sum_{q}\hbar\omega_{q}\hat{b}_{q}^{\dagger}\hat{b}_{q}+\omega_{c}\hat{a}^{\dagger}\hat{a}+\left(-i\hbar\sqrt{\kappa}\sum_{q}e^{i\omega_{c}t}\beta\left(t\right)\hat{a}+h.c.\right)\nonumber \\
 & -i\hbar\sqrt{\frac{\kappa}{2\pi\rho}}\sum_{q}\left[\hat{a}^{\dagger}\hat{b}_{q}-\hat{b}_{q}^{\dagger}\hat{a}\right]+const.\label{eq:4.1.4}
\end{align}
 which is time-dependent due to the to the classical input field,
see (\ref{eq:A.0.16}). $\hat{H}'{}_{env}\left(t\right)$ must be
included since we are considering a driven cavity with one of its
ports being partially open (see the discussion in appendix (\ref{chap:Finding--for})).
Ergo, the cavity is being exposed to both the external drive and the
vacuum noise so that energy can leak out to the external bath modes.
To make a link between the classical drive terms present in the Hamiltonian
of (\ref{eq:3.1.1}), we consider the case where the input field is
a coherent drive with a classical and quantum part (see (\ref{eq:2.4.28}))
which for our state transfer protocol takes the form 
\begin{equation}
\hat{b}_{in}\left(t\right)=e^{-i\omega_{c}t}\left(\beta_{1}\left(t\right)+\beta_{2}\left(t\right)+\hat{\xi}\left(t\right)\right).\label{eq:4.1.5}
\end{equation}
 In this case the operator $\hat{a}$ takes the form (in the Heisenberg
picture)
\begin{equation}
\hat{a}\left(t\right)=\alpha\left(t\right)+\hat{d}\left(t\right)e^{-i\omega_{c}t}.\label{eq:4.1.6}
\end{equation}
In the above equation, $\alpha\left(t\right)$ is the classical cavity
amplitude produced by the classical drive tones and $\hat{d}$ is
the quantum part. Furthermore, we consider the case where we apply
two laser tones on the cavity enabling us to write $\alpha\left(t\right)$
as
\begin{equation}
\alpha\left(t\right)=\alpha_{1}\left(t\right)e^{-i\omega_{1}t}+\alpha_{2}\left(t\right)e^{-i\omega_{2}t},\label{eq:4.1.7}
\end{equation}
where $\alpha_{1}\left(t\right)$, $\alpha_{2}\left(t\right)$ are
proportional to the complex amplitude of the control lasers and 
\begin{equation}
\omega_{j}=\omega_{c}+\Delta_{j},\label{eq:4.1.8}
\end{equation}
 with $j\,\in\,\left\{ 1,2\right\} $. Inserting (\ref{eq:4.1.6})
along with (\ref{eq:4.1.7}) into (\ref{eq:2.4.27}), it is straightforward
to show that $\alpha_{1}\left(t\right)$ and $\alpha_{2}\left(t\right)$
are related to $\beta_{1}\left(t\right)$ and $\beta_{2}\left(t\right)$
of the input field by
\begin{equation}
\alpha_{i}\left(t\right)=-\sqrt{\kappa}e^{i\Delta_{i}t}\int_{-\infty}^{t}e^{\frac{\kappa}{2}\left(\tau-t\right)}\beta_{i}\left(\tau\right)d\tau.\label{eq:4.1.9}
\end{equation}

\end{doublespace}

The amplitudes $\alpha_{1}\left(t\right)$ and $\alpha_{2}\left(t\right)$
of the classical laser fields will be related to $\Omega_{1}\left(t\right)$
and $\Omega_{2}\left(t\right)$ in equation (\ref{eq:4.1.18}) and
(\ref{eq:4.1.19}) below. It is important to keep in mind that the
displacement transformation performed in (\ref{eq:4.1.6}) implies
that the dynamics of the $\hat{d}$ operator are described in an interaction
picture with respect to the $\omega_{c}\hat{a}^{\dagger}\hat{a}$
term in the Hamiltonian (\ref{eq:4.1.1}). 

In what follows, we pick the two laser tones to be resonant with the
two desired transitions (see figure (\ref{fig:Energy-level-diagram-in}))
so that 
\begin{equation}
\omega_{j}=\omega_{eg_{j}}=\omega_{c}+\Delta_{j},\label{eq:4.1.10}
\end{equation}
where
\begin{equation}
\omega_{eg_{j}}=\omega_{e}-\omega_{g_{j}}.\label{eq:4.1.11}
\end{equation}
 When performing the displacement transformation (\ref{eq:4.1.6})
on the Hamiltonian, there will be terms that give rise to ``unwanted''
transitions between the ground and excited states. For example, we
can have a term of the form
\begin{equation}
\hat{I}_{1}\equiv\alpha_{2}\left(t\right)g_{1}e^{-i\omega_{2}t}\left.|e\right\rangle \left\langle g_{1}|\right..\label{eq:4.1.12}
\end{equation}
that describes transitions between the levels $\left.|e\right\rangle ,\left.|g_{1}\right\rangle $
arising from the second classical laser field driving at the frequency
$\omega_{2}$. We would like to describe the situation where only
the laser field driving at frequency $\omega_{1}$ creates transitions
between the levels $\left.|e\right\rangle ,\left.|g_{1}\right\rangle $.
Similarly, we only want the field driving at frequency $\omega_{2}$
to create transitions between the levels $\left.|e\right\rangle ,\left.|g_{2}\right\rangle $.
If we go into an interaction picture with respect to the Hamiltonian
of Eq. (\ref{eq:4.1.2}), we will have 
\begin{equation}
\hat{I}_{1}=\alpha_{2}\left(t\right)g_{1}e^{-i\left(\omega_{2}-\omega_{eg_{1}}\right)t}\left.|e\right\rangle \left\langle g_{1}|\right..\label{eq:4.1.13}
\end{equation}

\begin{flushleft}
\begin{figure}
\begin{centering}
\includegraphics[scale=0.6]{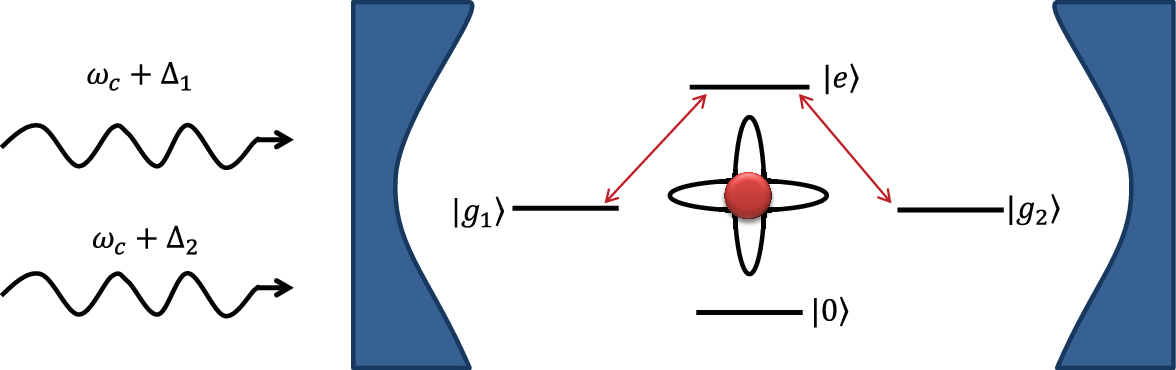}
\par\end{centering}

\caption{\ Atom-cavity coupling}

\begin{doublespace}
\begin{centering}
\textit{Figure representing a four level atom coupled to a single
cavity mode. Two laser fields are used to drive the cavity at frequencies
$\omega_{c}+\Delta_{1}$ and $\omega_{c}+\Delta_{2}$. The classical
component of the laser fields induce transitions between the levels
($\left.|e\right\rangle ,\left.|g_{1}\right\rangle $) and ($\left.|e\right\rangle ,\left.|g_{2}\right\rangle $).
This allows us to transfer the qubit state $\alpha\left.|0\right\rangle +\beta\left.|g_{1}\right\rangle $
to the qubit state $\alpha\left.|0\right\rangle +\beta e^{i\gamma_{d}\left(t_{int}\right)}\left.|g_{2}\right\rangle $
and then back to the original qubit state. The fluctuating number
of photons inside the cavity will act as the noise terms $\delta\hat{\omega}_{1}$
and $\delta\hat{\omega}_{2}$ found in the Hamiltonian (\ref{eq:3.1.1}). }
\par\end{centering}
\end{doublespace}

\end{figure}
From the resonance condition (\ref{eq:4.1.10}), we can rewrite (\ref{eq:4.1.13})
as 
\begin{equation}
\hat{I}_{1}=\alpha_{2}\left(t\right)g_{1}e^{-i\left(\omega_{g_{1}}-\omega_{g_{2}}\right)t}\left.|e\right\rangle \left\langle g_{1}|\right..\label{eq:4.1.14}
\end{equation}
When integrating over time, the terms giving rise to unwanted transitions
will scale as $\frac{\alpha_{2}\left(t\right)g_{1}}{\left|\omega_{g_{1}}-\omega_{g_{2}}\right|}$
and $\frac{\alpha_{1}\left(t\right)g_{2}}{\left|\omega_{g_{1}}-\omega_{g_{2}}\right|}$.
Consequently, if the following conditions are satisfied
\begin{equation}
\frac{\alpha_{2}\left(t\right)g_{1}}{\left|\omega_{g_{1}}-\omega_{g_{2}}\right|}\ll1,\label{eq:4.1.15}
\end{equation}
\begin{equation}
\frac{\alpha_{1}\left(t\right)g_{2}}{\left|\omega_{g_{1}}-\omega_{g_{2}}\right|}\ll1,\label{eq:4.1.16}
\end{equation}
we could suppress the ``unwanted'' transitions and so the off-resonant
terms in the Hamiltonian could safely be neglected. In what follows,
since the magnitude of the detuning frequencies (which later will
be taken to be equal magnitude, opposite sign) must be much larger
than $G$, any term of the form 
\begin{equation}
\frac{1}{\Delta_{i}\pm G}\approx\frac{1}{\Delta_{i}}.\label{eq:4.1.17}
\end{equation}
 Now, we can write $\alpha_{1}\left(t\right)=i\frac{\left|\alpha_{1}\left(t\right)\right|}{2}$,
$\alpha_{2}\left(t\right)=i\frac{\left|\alpha_{2}\left(t\right)\right|}{2}e^{-i\phi\left(t\right)}$
and define
\begin{equation}
\Omega_{1}\left(t\right)\equiv\frac{g_{1}\left|\alpha_{1}\left(t\right)\right|}{2},\label{eq:4.1.18}
\end{equation}
\begin{equation}
\Omega_{2}\left(t\right)\equiv\frac{g_{2}\left|\alpha_{2}\left(t\right)\right|}{2}.\label{eq:4.1.19}
\end{equation}
With these definitions, the full Hamiltonian of Eq. (\ref{eq:4.1.1})
in the interaction picture with respect to (\ref{eq:4.1.2}) along
with the resonance condition, (\ref{eq:4.1.10}), takes the form 
\begin{align}
\hat{H}_{I}\left(t\right) & =i\frac{\Omega_{1}\left(t\right)}{2}\left(\left.|e\right\rangle \left\langle g_{1}|\right.-\left.|g_{1}\right\rangle \left\langle e|\right.\right)+i\frac{\Omega_{2}\left(t\right)}{2}\left(e^{-i\phi\left(t\right)}\left.|e\right\rangle \left\langle g_{2}|\right.-e^{i\phi\left(t\right)}\left.|g_{2}\right\rangle \left\langle e|\right.\right)\nonumber \\
 & +g_{1}\left(\left.|e\right\rangle \left\langle g_{1}|\right.\hat{d}e^{i\Delta_{1}t}+\left.|g_{1}\right\rangle \left\langle e|\right.\hat{d}^{\dagger}e^{-i\Delta_{1}t}\right)+g_{2}\left(\left.|e\right\rangle \left\langle g_{2}|\right.\hat{d}e^{i\Delta_{2}t}+\left.|g_{2}\right\rangle \left\langle e|\right.\hat{d}^{\dagger}e^{-i\Delta_{2}t}\right)+\hat{H}'_{env}\left(t\right).\label{eq:4.1.20}
\end{align}
Note that due to the resonance condition (\ref{eq:4.1.10}), only
the detuning frequencies are present in the exponent of the interaction
picture Hamiltonian. We assume that the experimentalist can control
the relative phase $\phi\left(t\right)$ between the laser beams.
Furthermore, the terms of the form $\alpha^{*}\hat{d}$ and $\alpha\hat{d}^{\dagger}$
don't appear if we pick $\alpha\left(t\right)$ to solve the classical
equation of motion which is obtained from (\ref{eq:2.4.22}) by keeping
only the classical contribution from the displacement transformation.
We also ignore the constant $\hbar\omega_{c}\left|\alpha\right|^{2}$
since it doesn't influence the dynamics of our system and only creates
an energy shift. Without the $\hat{d}$ operator, we have the ideal
Hamiltonian studied in chapter 2. The $\hat{d}$ terms include the
effects of noise in the cavity, which could generate unwanted transitions
between the states $\left.|g_{i}\right\rangle $ and $\left.|e\right\rangle $. 
\par\end{flushleft}

\begin{doublespace}
Similar to chapter 3, we are interested in almost-perfect adiabatic
evolution where the $\Omega_{j}$'s are used to tune the wave function
of the dark state. We thus want to work in a basis of instantaneous
eigenstates of the coherent Hamiltonian of equation (\ref{eq:4.1.20}).
This can be achieved by going into a rotating frame with the unitary
operator given by 
\begin{equation}
\hat{U}=\left.|0\right\rangle \left\langle 0|\right.+\left.|d\right\rangle \left\langle d(t)|\right.+\left.|+\right\rangle \left\langle +(t)|\right.+\left.|-\right\rangle \left\langle -(t)|\right..\label{eq:4.1.21}
\end{equation}
In the first line of (\ref{eq:4.1.20}), the $\Omega_{j}$ terms are
large and $\hat{U}$ will diagonalize this part of the Hamiltonian.
The last line in (\ref{eq:4.1.20}) will describe noise-induced transitions.
We remind the reader that following the results of section (\ref{sec:Density-matrix-phase}),
it is crucial to deal with the unwanted effects of cavity noise. This
noise can cause virtual transitions from the $\left.|d\right\rangle $
to $\left.|\pm\right\rangle $ states and back. Even though these
are off resonance, at second order (via virtual transitions) they
can cause a dephasing of the dark state which is what we would like
to describe. The transitions between the dark and bright states could
be due to non-adiabatic corrections ($\dot{\theta}$, $\dot{\phi}$)
or due to noise in the cavity ($\hat{d}$ terms). For our particular
state-transfer protocol, the dephasing effects result in a reduction
of the amplitude of $\left\langle \tilde{\hat{\Pi}}_{0d}\left(t\right)\right\rangle $
(see (\ref{eq:3.1.26})). 
\end{doublespace}

For the discussion that follows, we define
\begin{equation}
\hat{H}_{I,1}\left(t\right)\equiv i\frac{\Omega_{1}\left(t\right)}{2}\left(\left.|e\right\rangle \left\langle g_{1}|\right.-\left.|g_{1}\right\rangle \left\langle e|\right.\right)+i\frac{\Omega_{2}\left(t\right)}{2}\left(e^{-i\phi\left(t\right)}\left.|e\right\rangle \left\langle g_{2}|\right.-e^{i\phi\left(t\right)}\left.|g_{2}\right\rangle \left\langle e|\right.\right),\label{eq:4.1.22}
\end{equation}
\begin{equation}
\hat{H}_{I,2}\left(t\right)\equiv g_{1}\left(\left.|e\right\rangle \left\langle g_{1}|\right.\hat{d}e^{i\Delta_{1}t}+\left.|g_{1}\right\rangle \left\langle e|\right.\hat{d}^{\dagger}e^{-i\Delta_{1}t}\right)+g_{2}\left(\left.|e\right\rangle \left\langle g_{2}|\right.\hat{d}e^{i\Delta_{2}t}+\left.|g_{2}\right\rangle \left\langle e|\right.\hat{d}^{\dagger}e^{-i\Delta_{2}t}\right)+\hat{H}'{}_{env}\left(t\right).\label{eq:4.1.23}
\end{equation}
Recall that the instantaneous eigenstates of $\hat{H}_{I,1}\left(t\right)$
were given by:
\begin{equation}
\left.|d(t)\right\rangle =\cos\theta(t)\left.|g_{1}\right\rangle +e^{i\phi\left(t\right)}\sin\theta(t)\left.|g_{2}\right\rangle ,\label{eq:4.1.24}
\end{equation}
\begin{equation}
\left.|+(t)\right\rangle =\frac{1}{\sqrt{2}}\left(\sin\theta(t)\left.|g_{1}\right\rangle +i\left.|e\right\rangle -e^{i\phi\left(t\right)}\cos\theta(t)\left.|g_{2}\right\rangle \right),\label{eq:4.1.25}
\end{equation}
\begin{equation}
\left.|-(t)\right\rangle =\frac{1}{\sqrt{2}}\left(\sin\theta(t)\left.|g_{1}\right\rangle -i\left.|e\right\rangle -e^{i\phi\left(t\right)}\cos\theta(t)\left.|g_{2}\right\rangle \right).\label{eq:4.1.26}
\end{equation}
Where
\begin{equation}
\tan\theta\left(t\right)\equiv-\frac{\Omega_{1}\left(t\right)}{\Omega_{2}\left(t\right)},\label{eq:4.1.27}
\end{equation}
\begin{equation}
G(t)\equiv\frac{1}{2}\sqrt{\Omega_{1}^{2}\left(t\right)+\Omega_{2}^{2}\left(t\right)}.\label{eq:4.1.28}
\end{equation}
We also know that the unitary transformation which diagonalizes $\hat{H}_{I,1}\left(t\right)$
in the $\left\{ \left.|\pm\right\rangle ,\left.|d\right\rangle \right\} $
basis is given in (\ref{eq:4.1.21}) with the transformation 
\begin{equation}
\hat{H}'\left(t\right)=\hat{U}\hat{H}_{I}\left(t\right)\hat{U}^{\dagger}-i\hat{U}\dot{\hat{U}}^{\dagger}.\label{eq:4.1.29}
\end{equation}
From (\ref{eq:4.1.22}) and (\ref{eq:4.1.21}) it can be shown that
\begin{equation}
\hat{U}\hat{H}_{I,1}\left(t\right)\hat{U}^{\dagger}=G(t)\left\{ \left.|+\right\rangle \left\langle +|\right.-\left.|-\right\rangle \left\langle -|\right.\right\} ,\label{eq:4.1.30}
\end{equation}
and
\begin{align}
-i\hat{U}\dot{\hat{U}}^{\dagger} & =i\frac{\dot{\theta}}{\sqrt{2}}\left(\left.|+\right\rangle \left\langle d|\right.-\left.|d\right\rangle \left\langle +|\right.+\left.|-\right\rangle \left\langle d|\right.-\left.|d\right\rangle \left\langle -|\right.\right)-\frac{\dot{\phi}}{\sqrt{2}}\left(\left.|+\right\rangle \left\langle d|\right.+\left.|d\right\rangle \left\langle +|\right.+\left.|-\right\rangle \left\langle d|\right.+\left.|d\right\rangle \left\langle -|\right.\right)\nonumber \\
 & +\frac{\dot{\phi}\cos^{2}\theta\left(t\right)}{2}\left(\left.|+\right\rangle \left\langle +|\right.+\left.|-\right\rangle \left\langle +|\right.+\left.|+\right\rangle \left\langle -|\right.+\left.|-\right\rangle \left\langle -|\right.\right)+\dot{\phi}\sin^{2}\theta\left(t\right)\left.|d\right\rangle \left\langle d|\right..\label{eq:4.1.31}
\end{align}
Now we need to calculate $\hat{U}\left.|e\right\rangle \left\langle g_{1}|\right.\hat{U}^{\dagger}$
and $\hat{U}\left.|e\right\rangle \left\langle g_{2}|\right.\hat{U}^{\dagger}$
which arise from the transformation of $\hat{H}_{I,2}\left(t\right)$.
After a bit of math, one can show that 
\begin{align}
\hat{U}\left.|e\right\rangle \left\langle g_{1}|\right.\hat{U}^{\dagger} & =-\frac{i}{\sqrt{2}}\cos\theta\left(t\right)\left(\left.|+\right\rangle \left\langle d|\right.-\left.|-\right\rangle \left\langle d|\right.\right)\nonumber \\
 & -\frac{i}{2}\sin\theta(t)\left(\left.|+\right\rangle \left\langle +|\right.+\left.|+\right\rangle \left\langle -|\right.-\left.|-\right\rangle \left\langle +|\right.-\left.|-\right\rangle \left\langle -|\right.\right),\label{eq:4.1.32}
\end{align}
and
\begin{align}
\hat{U}\left.|e\right\rangle \left\langle g_{2}|\right.\hat{U}^{\dagger} & =-\frac{i}{\sqrt{2}}e^{i\phi\left(t\right)}\sin\theta\left(t\right)\left(\left.|+\right\rangle \left\langle d|\right.-\left.|-\right\rangle \left\langle d|\right.\right)\nonumber \\
 & +\frac{i}{2}e^{i\phi\left(t\right)}\cos\theta(t)\left(\left.|+\right\rangle \left\langle +|\right.+\left.|+\right\rangle \left\langle -|\right.-\left.|-\right\rangle \left\langle +|\right.-\left.|-\right\rangle \left\langle -|\right.\right).\label{eq:4.1.33}
\end{align}
From this point on we will set both cavity coupling constants to be
identical
\begin{equation}
g_{1}=g_{2}\equiv g.\label{eq:4.1.34}
\end{equation}
The condition (\ref{eq:4.1.34}) is not strictly necessary, but simplifies
certain terms and will allow us to obtain much simpler results than
if the coupling constants $g_{1}$ and $g_{2}$ were left completely
arbitrary. In order to simplify the notation in what will follow,
we define the following operators 
\begin{equation}
\hat{F}_{1}\left(t\right)\equiv-\frac{i}{2}\sin\theta\left(t\right)g\left(e^{i\Delta_{1}t}\hat{d}-e^{-i\Delta_{1}t}\hat{d}^{\dagger}\right),\label{eq:4.1.35}
\end{equation}
\begin{equation}
\hat{F}_{2}\left(t\right)\equiv\frac{i}{2}\cos\theta\left(t\right)g\left(e^{i\phi\left(t\right)}e^{i\Delta_{2}t}\hat{d}-e^{-i\phi\left(t\right)}e^{-i\Delta_{2}t}\hat{d}^{\dagger}\right),\label{eq:4.1.36}
\end{equation}
\begin{equation}
\hat{f}\left(t\right)=-\frac{ig}{\sqrt{2}}\left(\cos\theta\left(t\right)e^{i\Delta_{1}t}+\sin\theta\left(t\right)e^{i\phi\left(t\right)}e^{i\Delta_{2}t}\right)\hat{d}.\label{eq:4.1.37}
\end{equation}
The operators in (\ref{eq:4.1.35}) to (\ref{eq:4.1.37}) describe
various ways noise in $\hat{d}$ affect the system. With these definitions,
we can write the Hamiltonian in its instantaneous eigenbasis in the
simple form given by 
\begin{equation}
\hat{H}'\left(t\right)=\hat{H}'_{0,old}\left(t\right)+\hat{H}'_{0,new}\left(t\right)+\hat{V}\left(t\right)+\hat{H'}_{env}.\label{eq:4.1.38}
\end{equation}
where the terms in the Hamiltonian are defined as
\begin{equation}
\hat{H}'_{0,old}\left(t\right)\equiv G\left(t\right)\left\{ \left.|+\right\rangle \left\langle +|\right.-\left.|-\right\rangle \left\langle -|\right.\right\} +\dot{\phi}\sin^{2}\theta\left(t\right)\left.|d\right\rangle \left\langle d|+\right.\frac{1}{2}\dot{\phi}\cos^{2}\theta\left(t\right)\left(\left.|+\right\rangle \left\langle +|\right.+\left.|-\right\rangle \left\langle -|\right.\right),\label{eq:4.1.39}
\end{equation}
\begin{equation}
\hat{H}'_{0,new}\left(t\right)\equiv\left(\hat{F}_{1}\left(t\right)+\hat{F}_{2}\left(t\right)\right)\left(\left.|+\right\rangle \left\langle +|\right.-\left.|-\right\rangle \left\langle -|\right.\right),\label{eq:4.1.40}
\end{equation}
and
\begin{equation}
\hat{V}\left(t\right)\equiv\frac{i}{\sqrt{2}}\left(\dot{\theta}+i\dot{\phi}\right)\left(\left.|+\right\rangle \left\langle d|\right.+\left.|-\right\rangle \left\langle d|\right.\right)+\hat{f}\left(t\right)\left(\left.|+\right\rangle \left\langle d|\right.-\left.|-\right\rangle \left\langle d|\right.\right)+\hat{V}_{+/-}+h.c,\label{eq:4.1.41}
\end{equation}
where $\hat{V}_{+/-}$ gives transitions between the bright states
and so will not enter in the dynamics of $\left\langle \tilde{\hat{\Pi}}_{0d}\left(t_{f}\right)\right\rangle $
at leading nontrivial order in a secular approximation. The terms
in (\ref{eq:4.1.41}) that are proportional to $\hat{f}\left(t\right)$
and $\hat{f}^{\dagger}\left(t\right)$ will (after performing the
Schrieffer-Wolff transformation) give rise to secular noise terms
acting on the dark state $\left.|d\left(t\right)\right\rangle $ accounting
for virtual transitions $\sim\mathcal{O}\left(\frac{1}{G}\right)$.
This will be understood later on when after performing the Schrieffer-Wolff
transformation, the resulting Hamiltonian will have diagonal components
that depend on these noise terms (remember that from (\ref{eq:4.1.37}),
$\hat{f}\left(t\right)$ contains operator terms proportional to $\hat{d}$
and $\hat{d}^{\dagger}$). Furthermore, it is important to state that
we will consider a situation where the cavity noise is narrow-band
enough that it cannot cause real (only virtual), energy conserving
transitions between the dark and bright states.

At this stage a great simplification can be made. When performing
a Schrieffer-Wolff transformation on the above Hamiltonian, it will
be necessary to calculate a commutator between an anti-unitary operator
$\hat{S}\left(t\right)$ with $\hat{H}_{env}$ which would not vanish.
In order to avoid having to do this, we will diagonalize $\hat{H}_{env}$
and then go into an interaction picture with respect to the resulting
Hamiltonian. To see how this can be achieved, we start by reminding
the reader that 
\begin{equation}
\hat{H}_{env}=\sum_{q}\hbar\omega_{q}\hat{b}_{q}^{\dagger}\hat{b}_{q}-i\hbar\sqrt{\frac{\kappa}{2\pi\rho}}\sum_{q}\left[\hat{d}^{\dagger}\hat{b}_{q}-\hat{b}_{q}^{\dagger}\hat{d}\right].\label{eq:4.1.42}
\end{equation}
\textcolor{black}{The next step is to diagonalize both the cavity-bath
coupling Hamiltonian and the bath Hamiltonian itself, which is the
sum of the two terms in (\ref{eq:4.1.42}). Doing so, we introduce
a new set of normal modes $\hat{f}_{j}$ (with frequency $\omega_{j}$)
such that} 
\begin{equation}
\sum_{q}\hbar\omega_{q}\hat{b}_{q}^{\dagger}\hat{b}_{q}-i\hbar\sqrt{\frac{\kappa}{2\pi\rho}}\sum_{q}\left[\hat{d}^{\dagger}\hat{b}_{q}-\hat{b}_{q}^{\dagger}\hat{d}\right]=\sum_{j}\hbar\omega_{j}\hat{f}_{j}^{\dagger}\hat{f}_{j}\label{eq:4.1.43}
\end{equation}
so that we have 
\begin{equation}
\hat{H}_{env}=\sum_{j}\hbar\omega_{j}\hat{f}_{j}^{\dagger}\hat{f}_{j}\label{eq:4.1.44}
\end{equation}
We can then write the operator $\hat{d}$ as a linear combination
of the new set of normal modes as 
\begin{equation}
\hat{d}=\sum_{j}\lambda_{j}\hat{f}_{j}\label{eq:4.1.45}
\end{equation}
The $\lambda_{j}$'s are change of basis coefficients, which for our
purposes don't need to be explicitly found. Next we go into an interaction
picture with respect to the free Hamiltonian of the $\hat{f}_{j}$
bosons (\ref{eq:4.1.44}). The $\hat{d}$ operators will now be time-dependent
\begin{equation}
e^{i\hat{H}_{env}t}\hat{d}e^{-i\hat{H}_{env}t}=\hat{d}\left(t\right).\label{eq:4.1.46}
\end{equation}
The time-dependence we get for $\hat{d}\left(t\right)$ is exactly
what we would have in the absence of any coupling to the atom and
any classical drives. We can thus get the properties of $\hat{d}\left(t\right)$
by solving the Heisenberg Langevin equations for the uncoupled cavity,
as was done in equation (\ref{eq:2.4.22}) in section (\ref{sec:Review-of-photon}).
Therefore, $\hat{d}\left(t\right)$ will have the standard noise properties
we expect given by (\ref{eq:2.4.36}). 

Now we can perform a Schrieffer-Wolff transformation on the Hamiltonian
in (\ref{eq:4.1.38}) to find a simplified effective Hamiltonian accounting
for virtual transitions due to leading-order non-secular terms. The
terms $\hat{f}\left(t\right)\left(\left.|+\right\rangle \left\langle d|\right.-\left.|-\right\rangle \left\langle d|\right.\right)+\hat{f}^{\dagger}\left(t\right)\left(\left.|d\right\rangle \left\langle +|\right.-\left.|d\right\rangle \left\langle -|\right.\right)$
in (\ref{eq:4.1.41}) are the ones which will play an important role
since after performing the Schrieffer-Wolff transformation, they will
give rise to noise terms proportional to $\left.|d\right\rangle \left\langle d|\right.$
and as we saw in (\ref{eq:3.1.17}) and (\ref{eq:3.1.18}) these terms
enter directly into the exponent of $\left\langle \tilde{\hat{\Pi}}_{0d}\left(t\right)\right\rangle $.
We start by performing the time-dependent canonical transformation
given by

\begin{align}
\tilde{\hat{H}}'\left(t\right) & =e^{\hat{S}\left(t\right)}\hat{H}'\left(t\right)e^{-\hat{S}\left(t\right)}-ie^{\hat{S}\left(t\right)}\frac{\partial}{\partial t}e^{-\hat{S}\left(t\right)}\nonumber \\
 & =\hat{H}'\left(t\right)+\left[\hat{S}\left(t\right),\hat{H}'\left(t\right)\right]+\frac{1}{2}\left[\hat{S}\left(t\right),\left[\hat{S}\left(t\right),\hat{H}'\left(t\right)\right]\right]-ie^{\hat{S}\left(t\right)}\frac{\partial}{\partial t}e^{-\hat{S}\left(t\right)}+\mathcal{O}\left(\hat{S}\left(t\right)^{3}\right)\label{eq:4.1.47}
\end{align}
 Note that in writing (\ref{eq:4.1.47}), $\left[\hat{S}\left(t\right),\dot{\hat{S}}\left(t\right)\right]\neq0$.
Consequently, one should be careful when expanding the right hand
side of (\ref{eq:4.1.47}) in powers of $\hat{S}\left(t\right)$.
We only keep leading order terms in $\hat{S}\left(t\right)$ because
as we show below in (\ref{eq:4.1.77}) and the definition of the coefficients,
$\hat{S}\left(t\right)\sim\mathcal{O}\left(\frac{\left|\left|\hat{V}\right|\right|}{G},\frac{\left|\left|\hat{V}\right|\right|}{\Delta}\right)$
and higher order terms in $\left(\frac{\left|\left|\hat{V}\right|\right|}{G},\frac{\left|\left|\hat{V}\right|\right|}{\Delta}\right)$
are neglected. Then we can approximate 
\begin{equation}
-ie^{\hat{S}\left(t\right)}\frac{\partial}{\partial t}e^{-\hat{S}\left(t\right)}=i\frac{d\hat{S}\left(t\right)}{dt}+\frac{i}{2}\left[\hat{S}\left(t\right),\frac{d\hat{S}\left(t\right)}{dt}\right]+\mathcal{O}\left(\frac{\left|\left|\dot{\hat{V}}\right|\right|\left|\left|\hat{V}\right|\right|^{2}}{G^{3}}\right)\label{eq:4.1.48}
\end{equation}
In what follows, we will approximate the last term in (\ref{eq:4.1.47})
by the first two terms in (\ref{eq:4.1.48}). To eliminate the non-secular
terms of order $\dot{\theta}$, $\dot{\phi}$ and $g$, $\hat{S}\left(t\right)$
is chosen to satisfy
\begin{equation}
\left[\hat{S}\left(t\right),\hat{H}'_{0,old}\left(t\right)\right]+\hat{V}\left(t\right)+i\frac{d\hat{S}\left(t\right)}{dt}=0.\label{eq:4.1.49}
\end{equation}
We don't include $\hat{H}'_{0,new}\left(t\right)$ into the commutator
of (\ref{eq:4.1.49}) since $\hat{H}'_{0,old}\left(t\right)\sim G\left(t\right)$
will scale much larger than $\hat{F}_{1}\left(t\right)\sim g\sqrt{N_{th}}$
and $\hat{F}_{2}\left(t\right)$. Consequently, its inclusion would
not affect the solution to the differential equation in (\ref{eq:4.1.49}).
The only role of $\hat{H}'_{0,new}\left(t\right)$ is that it can
cause the energy of the $\left.|\pm\right\rangle $ states to wiggle.
Expanding $-ie^{\hat{S}\left(t\right)}\frac{\partial}{\partial t}e^{-\hat{S}\left(t\right)}$
as in (\ref{eq:4.1.48}) and inserting the result into (\ref{eq:4.1.47}),
we find that 
\begin{align}
\tilde{\hat{H}}'\left(t\right) & =\hat{H}'_{0,old}\left(t\right)+\hat{H}'_{0,new}\left(t\right)+\hat{V}\left(t\right)+i\frac{d\hat{S}\left(t\right)}{dt}+\frac{i}{2}\left[\hat{S}\left(t\right),\frac{d\hat{S}\left(t\right)}{dt}\right]\nonumber \\
 & +\left[\hat{S}\left(t\right),\hat{H}'_{0,old}\left(t\right)+\hat{H}'_{0,new}\left(t\right)+\hat{V}\left(t\right)\right]\nonumber \\
 & +\frac{1}{2}\left[\hat{S}\left(t\right),\left[\hat{S}\left(t\right),\hat{H}'_{0,old}\left(t\right)+\hat{H}'_{0,new}\left(t\right)+\hat{V}\left(t\right)\right]\right]+\mathcal{O}\left(G\hat{S}\left(t\right)^{3},g\sqrt{N_{th}}\hat{S}\left(t\right)^{3},\hat{S}^{2}\left(t\right)\frac{d\hat{S}\left(t\right)}{dt}\right)\label{eq:4.1.50}
\end{align}
The factor of $g\sqrt{N_{th}}$ in the last term of (\ref{eq:4.1.50})
comes from the size of $\hat{H}'_{0,new}\left(t\right)$. Using (\ref{eq:4.1.49}),
and neglecting terms of $\mathcal{O}\left(G\hat{S}\left(t\right)^{3},g\sqrt{N_{th}}\hat{S}\left(t\right)^{3},\hat{S}^{2}\left(t\right)\frac{d\hat{S}\left(t\right)}{dt}\right)$,
(\ref{eq:4.1.50}) becomes 
\begin{align}
\tilde{\hat{H}}'\left(t\right) & \cong\hat{H}'_{0,old}\left(t\right)+\hat{H}'_{0,new}\left(t\right)+\left[\hat{S}\left(t\right),\hat{H}'_{0,new}\left(t\right)\right]\nonumber \\
 & +\frac{1}{2}\left[\hat{S}\left(t\right),\hat{V}\left(t\right)\right]+\mathcal{O}\left(g\sqrt{N_{th}}\hat{S}\left(t\right)^{2},\hat{S}^{2}\left(t\right)\frac{d\hat{S}\left(t\right)}{dt}\right),\label{eq:4.1.51}
\end{align}
where by choosing $\hat{S}\left(t\right)$ as per equation (\ref{eq:4.1.49})
(which depends explicitly on $\frac{d\hat{S}\left(t\right)}{dt}$),
the $\left[\hat{S}\left(t\right),\frac{d\hat{S}\left(t\right)}{dt}\right]$
commutator term arising from the transformation gets exactly canceled.
After finding the explicit form of $\hat{S}\left(t\right)$, we will
be able to determine the typical size of the sub-leading terms that
were dropped when going from (\ref{eq:4.1.50}) to (\ref{eq:4.1.51})
(i.e. terms of order $\mathcal{O}\left(g\sqrt{N_{th}}\hat{S}\left(t\right)^{2},\hat{S}^{2}\left(t\right)\frac{d\hat{S}\left(t\right)}{dt}\right)$)
self-consistently. This will be done in detail in the next subsection.

In order to find the anti-unitary operator $\hat{S}\left(t\right)$
that satisfies (\ref{eq:4.1.49}), we make the following ansatz:
\begin{equation}
\hat{S}\left(t\right)=\hat{\chi}_{1}\left(t\right)\left.|+\right\rangle \left\langle d|\right.+\hat{\chi}_{2}\left(t\right)\left.|-\right\rangle \left\langle d|\right.+\hat{\chi}_{3}\left(t\right)\left.|+\right\rangle \left\langle -|\right.-h.c.\label{eq:4.1.52}
\end{equation}
However, in what follows, we will drop the $\hat{\chi}_{3}\left(t\right)$
term since it will not contribute to $\left\langle \tilde{\hat{\Pi}}_{0d}\left(t_{f}\right)\right\rangle $
at leading nontrivial order in the secular approximation. We include
a ``hat'' on the coefficients of $\hat{S}\left(t\right)$ since
these will be operators. Enforcing that $\hat{S}\left(t\right)$ satisfies
(\ref{eq:4.1.49}), the coefficients $\hat{\chi}_{i}\left(t\right)$'s
are found to be given by the differential equations 
\begin{equation}
\hat{\chi}_{1}\left(t\right)\left(\dot{\phi}\sin^{2}\theta\left(t\right)-G\left(t\right)-\frac{1}{2}\dot{\phi}\cos^{2}\theta\left(t\right)\right)=-\frac{1}{\sqrt{2}}\left(i\dot{\theta}-\dot{\phi}\right)-\hat{f}\left(t\right)-i\frac{d}{dt}\hat{\chi}_{1}\left(t\right),\label{eq:4.1.53}
\end{equation}
\begin{equation}
\hat{\chi}_{2}\left(t\right)\left(\dot{\phi}\sin^{2}\theta\left(t\right)+G\left(t\right)-\frac{1}{2}\dot{\phi}\cos^{2}\theta\left(t\right)\right)=-\frac{1}{\sqrt{2}}\left(i\dot{\theta}-\dot{\phi}\right)+\hat{f}\left(t\right)-i\frac{d}{dt}\hat{\chi}_{2}\left(t\right),\label{eq:4.1.54}
\end{equation}

We are interested in the adiabatic regime hence the rate at which
$\theta\left(t\right)$ and $\phi\left(t\right)$ change is much,
much smaller than $G\left(t\right)$. This allows us to make the following
approximation
\begin{equation}
\dot{\phi}\sin^{2}\theta\left(t\right)\pm G\left(t\right)-\frac{1}{2}\dot{\phi}\cos^{2}\theta\left(t\right)\approx\pm G\left(t\right).\label{eq:4.1.55}
\end{equation}
The above approximation allows us to rewrite the differential equation
in (\ref{eq:4.1.53}) as 
\begin{equation}
\frac{d}{dt}\hat{\chi}_{1}\left(t\right)+iG\left(t\right)\hat{\chi}_{1}\left(t\right)=i\left[\frac{1}{\sqrt{2}}\left(i\dot{\theta}-\dot{\phi}\right)+\hat{f}\left(t\right)\right].\label{eq:4.1.56}
\end{equation}
Note that (\ref{eq:4.1.56}) is just the equation of motion of an
undamped, driven simple harmonic oscillator with a time-dependent
frequency $G\left(t\right)$. For simplification, we define the operator
\begin{equation}
\hat{P}\left(t\right)\equiv\frac{1}{\sqrt{2}}\left(i\dot{\theta}-\dot{\phi}\right)+\hat{f}\left(t\right).\label{eq:4.1.57}
\end{equation}
which is just like the driving force on the oscillator. Then, the
general solution to the differential equation in (\ref{eq:4.1.56})
is given by 
\begin{equation}
\hat{\chi}_{1}\left(t\right)=ie^{-i\int_{0}^{t}G\left(t'\right)dt'}\int_{0}^{t}e^{i\int_{0}^{t'}G\left(t''\right)dt''}\hat{P}\left(t'\right)dt'.\label{eq:4.1.58}
\end{equation}
For simplicity, we assume that $G\left(t\right)$ is time independent.
This can be achieved by using equations (\ref{eq:3.1.35}) and (\ref{eq:3.1.36})
of chapter three 
\begin{equation}
\Omega_{1}=-A\sin\theta\left(t\right),\label{eq:4.1.59}
\end{equation}
and
\begin{equation}
\Omega_{2}=A\cos\theta\left(t\right).\label{eq:4.1.60}
\end{equation}
where $A$ is a constant amplitude. Consequently, (\ref{eq:4.1.58})
can be written as 
\begin{equation}
\hat{\chi}_{1}\left(t\right)=ie^{-iGt}\int_{0}^{t}e^{iGt'}\hat{P}\left(t'\right)dt'.\label{eq:4.1.61}
\end{equation}
Inserting the definition of $\hat{P}\left(t\right)$ into (\ref{eq:4.1.61}),
we can decompose it into two parts, 
\begin{align}
\hat{\chi}_{1}\left(t\right) & =\frac{i}{\sqrt{2}}e^{-iGt}\int_{0}^{t}e^{iGt'}\left(i\dot{\theta}\left(t'\right)-\dot{\phi}\left(t'\right)\right)dt'\nonumber \\
 & +ie^{-iGt}\int_{0}^{t}e^{iGt'}\hat{f}\left(t'\right)dt'.\label{eq:4.1.62}
\end{align}
We begin by focusing on the first line in (\ref{eq:4.1.62}) and define
the function 
\begin{equation}
h\left(t\right)\equiv i\dot{\theta}\left(t\right)-\dot{\phi}\left(t\right).\label{eq:4.1.63}
\end{equation}
The natural frequency of our effective simple harmonic oscillator
is $G$. The functions $\dot{\theta}\left(t\right)$ and $\dot{\phi}\left(t\right)$
are very slow. Thus, the $h\left(t\right)$ term in (\ref{eq:4.1.63})
is like driving our simple harmonic oscillator with a frequency that
is very far from resonance so that we expect a very weak response.
We can use this fact to greatly simplify the first term in (\ref{eq:4.1.62}).
We start by writing
\begin{equation}
\hat{\chi}_{1}=\hat{\chi}_{1,1}+\hat{\chi}_{1,2},\label{eq:4.1.63-1}
\end{equation}
with
\begin{equation}
\hat{\chi}_{1,1}\left(t\right)\equiv\frac{i}{\sqrt{2}}e^{-iGt}\int_{0}^{t}e^{iGt'}h\left(t'\right)dt'\label{eq:4.1.64}
\end{equation}
and $\hat{\chi}_{1,2}\left(t\right)$ defined in (\ref{eq:4.1.70})
below. We can integrate (\ref{eq:4.1.64}) by parts to find that
\begin{equation}
\hat{\chi}_{1,1}\left(t\right)=\frac{1}{\sqrt{2}G}\left[h\left(t\right)-e^{-iGt}h\left(0\right)\right]+\mathcal{O}\left(\frac{\ddot{\theta}}{G^{2}},\frac{\ddot{\phi}}{G^{2}}\right)\label{eq:4.1.65}
\end{equation}
Next we also need to consider the second term in (\ref{eq:4.1.62}).
Recall that the function $\hat{f}\left(t\right)$ was defined as 
\begin{equation}
\hat{f}\left(t\right)=-\frac{ig}{\sqrt{2}}\left(\cos\theta\left(t\right)e^{i\Delta_{1}t}+\sin\theta\left(t\right)e^{i\phi\left(t\right)}e^{i\Delta_{2}t}\right)\hat{d}\left(t\right).\label{eq:4.1.66}
\end{equation}
With this in mind we define 
\begin{equation}
\hat{\chi}_{1,2}\left(t\right)\equiv-\frac{ig}{\sqrt{2}}e^{-iGt}\int_{0}^{t}e^{iGt'}\left(\cos\theta\left(t\right)e^{i\Delta_{1}t}+\sin\theta\left(t\right)e^{i\phi\left(t\right)}e^{i\Delta_{2}t}\right)\hat{d}\left(t'\right)dt'.\label{eq:4.1.67}
\end{equation}
We can use the same arguments that lead from (\ref{eq:4.1.64}) to
(\ref{eq:4.1.65}) in order to calculate (\ref{eq:4.1.67}). Considering
the case where the noise stemming from $\hat{d}\left(t\right)$ is
slow, meaning that the cavity-damping rate is much smaller than the
detuning frequency, we define the operators
\begin{equation}
\hat{v}_{1}\left(t\right)=\cos\theta\left(t\right)\hat{d}\left(t\right),\label{eq:4.1.68}
\end{equation}
\begin{equation}
\hat{v}_{2}\left(t\right)=\sin\theta\left(t\right)e^{i\phi\left(t\right)}\hat{d}\left(t\right),\label{eq:4.1.69}
\end{equation}
and integrate by parts to find that 
\begin{equation}
\hat{\chi}_{1,2}\left(t\right)=-\frac{g}{\sqrt{2}}\left[\frac{e^{i\Delta_{1}t}\hat{v}_{1}\left(t\right)-e^{-iGt}\hat{d}}{\Delta_{1}}+\frac{e^{i\Delta_{2}t}\hat{v}_{2}\left(t\right)}{\Delta_{2}}\right]+\mathcal{O}\left(\frac{g\dot{\hat{v}}_{1}}{\Delta_{1}^{2}},\frac{g\dot{\hat{v}}_{2}}{\Delta_{2}^{2}}\right),\label{eq:4.1.70}
\end{equation}
where we used the fact that $\left|\Delta_{i}\right|\gg G$ to simplify
the denominator of the above term. To neglect the higher order terms
in both (\ref{eq:4.1.65}) and (\ref{eq:4.1.70}), we require that
\begin{equation}
\frac{\mathrm{Max}\left(\ddot{\theta},\ddot{\phi}\right)}{G^{2}}\ll1,\label{eq:4.1.71}
\end{equation}
\begin{equation}
\frac{\mathrm{Max}\left(\dot{\theta},\dot{\phi}\right)}{\Delta_{i}}\ll1,\label{eq:4.1.72}
\end{equation}
which are consistent with the adiabatic criteria. To obtain an estimate
for the size of the $\dot{\hat{d}}\left(t\right)$ term, we remind
the reader that from section (\ref{sec:Review-of-photon}), we have
\begin{equation}
\hat{d}\left(t\right)=-\sqrt{\kappa}\int_{-\infty}^{t}d\tau e^{-\kappa/2\left(t-\tau\right)}\hat{\xi}\left(\tau\right).\label{eq:4.1.72-1}
\end{equation}
Notice that the detuning frequency does not appear in our chosen frame
(see (\ref{eq:4.1.6})). We also remind the reader that the field
$\hat{\xi}\left(t\right)$ satisfies
\begin{equation}
\left\langle \hat{\xi}^{\dagger}\left(t\right)\hat{\xi}\left(t'\right)\right\rangle =N_{th}\delta\left(t-t'\right).\label{eq:4.1.72-2}
\end{equation}
Using (\ref{eq:4.1.72-1}) and (\ref{eq:4.1.72-2}), it is straightforward
to show that 
\begin{equation}
\left\langle \hat{d}^{\dagger}\left(t\right)\hat{d}\left(t\right)\right\rangle =N_{th},\label{eq:4.1.72-3}
\end{equation}
and
\begin{equation}
\left\langle \dot{\hat{d}}^{\dagger}\left(t\right)\dot{\hat{d}}\left(t\right)\right\rangle =-\frac{3}{4}N_{th}\kappa^{2}.\label{eq:4.1.72-4}
\end{equation}
Thus we conclude that 
\begin{equation}
\sqrt{\left|\frac{\left\langle \dot{\hat{d}}^{\dagger}\left(t\right)\dot{\hat{d}}\left(t\right)\right\rangle }{\left\langle \hat{d}^{\dagger}\left(t\right)\hat{d}\left(t\right)\right\rangle }\right|}\sim\kappa.\label{eq:4.1.72-5}
\end{equation}
Going back to (\ref{eq:4.1.70}) and using (\ref{eq:4.1.72-5}), we
also require the condition 
\begin{equation}
\frac{\kappa}{\Delta_{i}}\ll1,\label{eq:4.1.74}
\end{equation}
to be satisfied in order to neglect the higher order terms in (\ref{eq:4.1.70}).
Using (\ref{eq:4.1.65}), (\ref{eq:4.1.70}) and the initial condition
$\theta\left(0\right)=0$, we find that 
\begin{equation}
\hat{\chi}_{1}\left(t\right)=\frac{1}{\sqrt{2}G}\left[h\left(t\right)-e^{-iGt}h\left(0\right)\right]-i\frac{g}{\sqrt{2}}\left[\frac{e^{i\Delta_{1}t}\cos\theta\left(t\right)\hat{d}\left(t\right)-e^{-iGt}\hat{d}}{\Delta_{1}}+\frac{\sin\theta\left(t\right)e^{i\phi\left(t\right)}e^{i\Delta_{2}t}\hat{d}\left(t\right)}{\Delta_{2}}\right].\label{eq:4.1.75}
\end{equation}

From (\ref{eq:4.1.54}), we can perform the same set of approximations
that we used to obtain $\hat{\chi}_{1}\left(t\right)$, and we find
that $\hat{\chi}_{2}\left(t\right)$ is obtained from $\hat{\chi}_{1}\left(t\right)$
by replacing $G\rightarrow-G$ and $\hat{f}\left(t\right)\rightarrow-\hat{f}\left(t\right).$
The physical origin of these differences is as follows: $\hat{\chi}_{1/2}\left(t\right)$
correspond to virtual transitions (respectively) to the $\left.|\pm\right\rangle $
state. The energy of the virtual state is different, as is the relevant
matrix element. The origin of the sign difference can be traced back
in the matrix element of the form of $\hat{V}\left(t\right)$ in the
instantaneous eigenstate basis (see (\ref{eq:4.1.41})). Thus we have
\begin{equation}
\hat{\chi}_{2}\left(t\right)=-\frac{1}{\sqrt{2}G}\left[h\left(t\right)-e^{iGt}h\left(0\right)\right]+i\frac{g}{\sqrt{2}}\left[\frac{e^{i\Delta_{1}t}\cos\theta\left(t\right)\hat{d}\left(t\right)-e^{iGt}\hat{d}}{\Delta_{1}}+\frac{\sin\theta\left(t\right)e^{i\phi\left(t\right)}e^{i\Delta_{2}t}\hat{d}\left(t\right)}{\Delta_{2}}\right].\label{eq:4.1.76}
\end{equation}
The anti-unitary operator $\hat{S}\left(t\right)$ (omitting the $\hat{\chi}_{3}$
term) can then be written as 
\begin{equation}
\hat{S}\left(t\right)=\hat{\chi}_{1}\left(t\right)\left.|+\right\rangle \left\langle d|\right.+\hat{\chi}_{2}\left(t\right)\left.|-\right\rangle \left\langle d|\right.-h.c.\label{eq:4.1.77}
\end{equation}
with $\hat{\chi}_{1}\left(t\right)$ given by (\ref{eq:4.1.75}) and
$\hat{\chi}_{2}\left(t\right)$ by (\ref{eq:4.1.76}). 

Next, we need to calculate the commutator that will give rise to the
term proportional to $\left.|d\right\rangle \left\langle d|\right.$
in (\ref{eq:4.1.51}) ($\frac{1}{2}\left[\hat{S}\left(t\right),\hat{V}\left(t\right)\right]$).
Writing only the $\left.|d\right\rangle \left\langle d|\right.$ term,
it is found to be 
\begin{align}
\frac{1}{2}\left[\hat{S}\left(t\right),\hat{V}\left(t\right)\right] & =\frac{1}{2}\left[\frac{\left(\dot{\phi}+i\dot{\theta}\right)}{\sqrt{2}}\left(\hat{\chi}_{1}\left(t\right)+\hat{\chi}_{2}\left(t\right)\right)+\frac{\left(\dot{\phi}-i\dot{\theta}\right)}{\sqrt{2}}\left(\hat{\chi}_{1}^{\dagger}\left(t\right)+\hat{\chi}_{2}^{\dagger}\left(t\right)\right)\right.\nonumber \\
 & \left.+\hat{f}^{\dagger}\left(t\right)\left(\hat{\chi}_{2}\left(t\right)-\hat{\chi}_{1}\left(t\right)\right)+\left(\hat{\chi}_{2}^{\dagger}\left(t\right)-\hat{\chi}_{1}^{\dagger}\left(t\right)\right)\hat{f}\left(t\right)\right]\left.|d\right\rangle \left\langle d|\right.\label{eq:4.1.78}
\end{align}
Note that since the function $h\left(t\right)$ has the opposite sign
in $\hat{\chi}_{1}\left(t\right)$ than in $\hat{\chi}_{2}\left(t\right)$
it disappears from the first line of (\ref{eq:4.1.78}). This means
that there will be no new terms of order $\left(\dot{\theta}+\dot{\phi}\right)^{2}$
which modify the energy of the $\left.|d\right\rangle $ state.

Before proceeding further, we can perform a simplification that will
greatly reduce the complexity of the resulting Hamiltonian. When calculating
$\left\langle \tilde{\hat{\Pi}}_{0d}\left(t_{f}\right)\right\rangle $,
we will have to integrate the term proportional to $\left.|d\right\rangle \left\langle d|\right.$
over time. Terms that oscillate with the phase $e^{i\left(G\pm\Delta_{i}\right)t}$
or $e^{iGt}$ will scale as (for a sufficiently slow-varying function
$f\left(t\right)$)
\begin{equation}
\int_{0}^{t_{f}}f\left(t\right)e^{iGt}dt=\frac{1}{iG}\left[f\left(t_{f}\right)e^{iGt_{f}}-f\left(0\right)-\mathcal{O}\left(\frac{\dot{f}}{G}\right)\right],\label{eq:4.1.79}
\end{equation}
whereas
\begin{equation}
\int_{0}^{t_{f}}f\left(t\right)dt=f\left(t_{f}\right)t_{f}-\int_{0}^{t_{f}}\dot{f}\left(t\right)tdt.\label{eq:4.1.80}
\end{equation}
Given that the adiabatic criteria require that $Gt_{f}\gg1$ and $\left|G\pm\Delta_{i}\right|t_{f}\gg1$,
this guarantees that we can safely drop terms with oscillating phases
$e^{i\left(G\pm\Delta_{i}\right)t}$ and $e^{iGt}$ in (\ref{eq:4.1.78}).
An immediate consequence is that the term in the first line of (\ref{eq:4.1.78})
can be dropped. The reason is that the sum of $\hat{\chi}_{1}\left(t\right)+\hat{\chi}_{2}\left(t\right)$
has a resulting contribution proportional to the phases $e^{\pm i\Delta_{j}t}$.
Physically, this means that we can just add the effective energy shifts
associated with each perturbation that gives us transitions to the
$\left.|\pm\right\rangle $ states. For the second line in (\ref{eq:4.1.78}),
we can use (\ref{eq:4.1.75}) and (\ref{eq:4.1.76}) for the operators
$\hat{\chi}_{1}\left(t\right)$, $\hat{\chi}_{2}\left(t\right)$ as
well as (\ref{eq:4.1.66}) for the function $\hat{f}\left(t\right)$
to show that (keeping terms with no oscillating phase)
\begin{equation}
\int_{0}^{t}\left[\hat{f}^{\dagger}\left(t'\right)\hat{\chi}_{1}\left(t'\right)+\hat{\chi}_{1}^{\dagger}\left(t'\right)\hat{f}\left(t'\right)\right]dt'\cong g^{2}\int_{0}^{t}\left[\frac{\cos^{2}\theta\left(t'\right)}{\Delta_{1}}+\frac{\sin^{2}\theta\left(t'\right)}{\Delta_{2}}\right]\hat{d}^{\dagger}\left(t'\right)\hat{d}\left(t'\right)dt'\label{eq:4.1.81}
\end{equation}
and
\begin{equation}
\int_{0}^{t}\left[\hat{f}^{\dagger}\left(t'\right)\hat{\chi}_{2}\left(t'\right)+\hat{\chi}_{2}^{\dagger}\left(t'\right)\hat{f}\left(t'\right)\right]dt'\cong-g^{2}\int_{0}^{t}\left[\frac{\cos^{2}\theta\left(t'\right)}{\Delta_{1}}+\frac{\sin^{2}\theta\left(t'\right)}{\Delta_{2}}\right]\hat{d}^{\dagger}\left(t'\right)\hat{d}\left(t'\right)dt'\label{eq:4.1.82}
\end{equation}
We thus conclude that 
\begin{equation}
\frac{1}{2}\left[\hat{S}\left(t\right),\hat{V}\left(t\right)\right]\cong-g^{2}\left[\frac{\cos^{2}\theta\left(t\right)}{\Delta_{1}}+\frac{\sin^{2}\theta\left(t\right)}{\Delta_{2}}\right]\hat{d}^{\dagger}\left(t\right)\hat{d}\left(t\right)\label{eq:4.1.83}
\end{equation}
Performing a secular approximation so that we keep only the diagonal
terms in the transformed Hamiltonian and writing only the dephasing
term proportional to $\left.|d\right\rangle \left\langle d|\right.$,
we find that 
\begin{align}
\tilde{\hat{H}}'\left(t\right) & \approx G\left(t\right)\left\{ \left.|+\right\rangle \left\langle +|\right.-\left.|-\right\rangle \left\langle -|\right.\right\} +\dot{\phi}\sin^{2}\theta\left(t\right)\left.|d\right\rangle \left\langle d|+\right.\nonumber \\
 & +\frac{1}{2}\dot{\phi}\cos^{2}\theta\left(t\right)\left(\left.|+\right\rangle \left\langle +|\right.+\left.|-\right\rangle \left\langle -|\right.\right)-g^{2}\left\{ \frac{\cos^{2}\theta\left(t\right)}{\Delta_{1}}+\frac{\sin^{2}\theta\left(t\right)}{\Delta_{2}}\right\} \hat{d}^{\dagger}\left(t\right)\hat{d}\left(t\right)\left.|d\right\rangle \left\langle d|\right.\nonumber \\
 & +other\label{eq:4.1.84}
\end{align}
 To make further progress, we consider the case where the two detuning
frequencies are equal and opposite in magnitude 
\begin{equation}
\Delta=\Delta_{1}=-\Delta_{2}.\label{eq:4.1.88}
\end{equation}
In this case the Hamiltonian of (\ref{eq:4.1.84}) then becomes 
\begin{align}
\tilde{\hat{H}}'\left(t\right) & \approx G\left(t\right)\left\{ \left.|+\right\rangle \left\langle +|\right.-\left.|-\right\rangle \left\langle -|\right.\right\} +\dot{\phi}\sin^{2}\theta\left(t\right)\left.|d\right\rangle \left\langle d|+\right.\nonumber \\
 & +\frac{1}{2}\dot{\phi}\cos^{2}\theta\left(t\right)\left(\left.|+\right\rangle \left\langle +|\right.+\left.|-\right\rangle \left\langle -|\right.\right)-\frac{g^{2}}{\Delta}\cos2\theta\left(t\right)\hat{d}^{\dagger}\left(t\right)\hat{d}\left(t\right)\left.|d\right\rangle \left\langle d|\right.\nonumber \\
 & +other,\label{eq:4.1.89}
\end{align}
where the other terms in (\ref{eq:4.1.89}) are proportional to $\left.|+\right\rangle \left\langle +|\right.$
and $\left.|-\right\rangle \left\langle -|\right.$ which don't influence
the dynamics of the coherence. Using (\ref{eq:4.1.89}), we finally
have 
\begin{equation}
\left\langle \tilde{\hat{\Pi}}_{0d}\left(t\right)\right\rangle =\alpha\beta^{*}e^{-i\gamma_{d}\left(t\right)}\left\langle T_{t}e^{i\frac{g^{2}}{\Delta}\int_{0}^{t}\cos2\theta\left(t'\right)\hat{d}^{\dagger}\left(t'\right)\hat{d}\left(t'\right)dt'}\right\rangle .\label{eq:4.1.90}
\end{equation}
It is important to keep in mind that in order for the above expression
to hold, we need to assume that the laser detunings are much larger
than the amplitude $G$ and that the changes to the initial state
$\hat{\rho}\left(0\right)$ are negligible (so that only the $\left.|d\right\rangle \left\langle d|\right.$
in the Hamiltonian of (\ref{eq:4.1.89}) is relevant for the phase
of $\left\langle \tilde{\hat{\Pi}}_{0d}\left(t\right)\right\rangle $).
In the next section we will show that indeed the changes to the initial
state under the limits that we considered will not affect the dynamics
of our system so that we can safely use (\ref{eq:4.1.90}) to describe
the fidelity of our state transfer protocol.

\subsection{A note on $g_{1}\neq g_{2}$}

Recall that the Hamiltonian of equation (\ref{eq:4.1.89}) was derived
in the regime where $g_{1}=g_{2}=g$ and $\Delta_{1}=-\Delta_{2}$.
As it turns out, it is also possible to obtain an expression analogous
to (\ref{eq:4.1.89}) without requiring that both cavity-coupling
constants to be identical. To see this, we go back to the definition
of the operator $\hat{f}\left(t\right)$ first written in (\ref{eq:4.1.37}).
If we let both cavity-coupling constants to be independent of each
other, it is easy to verify that in this case $\hat{f}\left(t\right)$
becomes 
\begin{equation}
\hat{f}\left(t\right)=-\frac{i}{\sqrt{2}}\left(g_{1}\cos\theta\left(t\right)e^{i\Delta_{1}t}+g_{2}\sin\theta\left(t\right)e^{i\phi\left(t\right)}e^{i\Delta_{2}t}\right)\hat{d}.\label{eq:4.1.90-1}
\end{equation}
Following the same steps that led to equations (\ref{eq:4.1.67})
and (\ref{eq:4.1.70}), the operators $\hat{\chi}_{1}\left(t\right)$
and $\hat{\chi}_{2}\left(t\right)$ now become 
\begin{equation}
\hat{\chi}_{1}\left(t\right)=\frac{1}{\sqrt{2}G}\left[h\left(t\right)-e^{-iGt}h\left(0\right)\right]-i\frac{1}{\sqrt{2}}\left[\frac{g_{1}\left(e^{i\Delta_{1}t}\cos\theta\left(t\right)\hat{d}\left(t\right)-e^{-iGt}\hat{d}\right)}{\Delta_{1}}+\frac{g_{2}\sin\theta\left(t\right)e^{i\phi\left(t\right)}e^{i\Delta_{2}t}\hat{d}\left(t\right)}{\Delta_{2}}\right],\label{eq:4.1.90-2}
\end{equation}

\begin{equation}
\hat{\chi}_{2}\left(t\right)=-\frac{1}{\sqrt{2}G}\left[h\left(t\right)-e^{iGt}h\left(0\right)\right]+i\frac{g}{\sqrt{2}}\left[\frac{g_{1}\left(e^{i\Delta_{1}t}\cos\theta\left(t\right)\hat{d}\left(t\right)-e^{iGt}\hat{d}\right)}{\Delta_{1}}+\frac{g_{2}\sin\theta\left(t\right)e^{i\phi\left(t\right)}e^{i\Delta_{2}t}\hat{d}\left(t\right)}{\Delta_{2}}\right].\label{eq:4.1.90-3}
\end{equation}
Then following the same steps of (\ref{eq:4.1.81}), (\ref{eq:4.1.82})
and (\ref{eq:4.1.83}), the Hamiltonian becomes 
\begin{align}
\tilde{\hat{H}}'\left(t\right) & \approx G\left(t\right)\left\{ \left.|+\right\rangle \left\langle +|\right.-\left.|-\right\rangle \left\langle -|\right.\right\} +\dot{\phi}\sin^{2}\theta\left(t\right)\left.|d\right\rangle \left\langle d|+\right.\nonumber \\
 & +\frac{1}{2}\dot{\phi}\cos^{2}\theta\left(t\right)\left(\left.|+\right\rangle \left\langle +|\right.+\left.|-\right\rangle \left\langle -|\right.\right)-\left\{ \frac{g_{1}^{2}\cos^{2}\theta\left(t\right)}{\Delta_{1}}+\frac{g_{2}^{2}\sin^{2}\theta\left(t\right)}{\Delta_{2}}\right\} \hat{d}^{\dagger}\left(t\right)\hat{d}\left(t\right)\left.|d\right\rangle \left\langle d|\right.\nonumber \\
 & +other\label{eq:4.1.90-4}
\end{align}
Consequently, we can obtain the same $\cos2\theta\left(t\right)$
dependence as in (\ref{eq:4.1.89}) by setting 
\begin{equation}
\frac{g_{1}^{2}}{\Delta_{1}}=-\frac{g_{2}^{2}}{\Delta_{2}}\label{eq:4.1.90-5}
\end{equation}
The condition (\ref{eq:4.1.90-5}) is much less restrictive than requiring
that the cavity-coupling constants be identical and the detuning frequencies
be equal in magnitude with opposite sign.

\subsection{Estimates for the size of $\hat{S}\left(t\right)$}

As was mentioned in the paragraph below (\ref{eq:4.1.51}), we now
determine the size for the terms that were neglected by going from
(\ref{eq:4.1.50}) to (\ref{eq:4.1.51}). Recall that we found (omitting
the $\hat{\chi}_{3}$ term) 
\begin{equation}
\hat{S}\left(t\right)=\hat{\chi}_{1}\left(t\right)\left.|+\right\rangle \left\langle d|\right.+\hat{\chi}_{2}\left(t\right)\left.|-\right\rangle \left\langle d|\right.-h.c.\label{eq:4.1.91}
\end{equation}
with
\begin{equation}
\hat{\chi}_{1}\left(t\right)=\frac{1}{\sqrt{2}G}\left(i\dot{\theta}\left(t\right)-\dot{\phi}\left(t\right)\right)-i\frac{g}{\sqrt{2}}\left[\frac{e^{i\Delta_{1}t}\cos\theta\left(t\right)\hat{d}\left(t\right)}{\Delta_{1}}+\frac{\sin\theta\left(t\right)e^{i\phi\left(t\right)}e^{i\Delta_{2}t}\hat{d}\left(t\right)}{\Delta_{2}}\right],\label{eq:4.1.92}
\end{equation}
We don't bother writing down $\hat{\chi}_{2}\left(t\right)$ since
it scales the same way as $\hat{\chi}_{1}\left(t\right)$. When we
performed the Schrieffer-Wolff transformation, the terms we threw
away in (\ref{eq:4.1.51}) scaled as 
\begin{equation}
\mathcal{O}\left(G\hat{S}^{3}\left(t\right),g\sqrt{N_{th}}\hat{S}\left(t\right)^{2},\hat{S}^{2}\left(t\right)\frac{d\hat{S}\left(t\right)}{dt}\right),\label{eq:4.1.93}
\end{equation}
where the factor of $g\sqrt{N_{th}}$, being the size of $\hat{H}'_{0,new}\left(t\right)$
(see (\ref{eq:4.1.40}) along with (\ref{eq:4.1.35}) and (\ref{eq:4.1.36}))
arises from the term $\left[\hat{S}\left(t\right),\left[\hat{S}\left(t\right),\hat{H}'_{0,new}\left(t\right)\right]\right]$.
The terms we keep in (\ref{eq:4.1.51}) are of the order $\mathcal{O}\left(g\sqrt{N_{th}}\hat{S}\left(t\right),\frac{d\hat{S}\left(t\right)}{dt}\right)$.
Now, the operator $\frac{d\hat{S}\left(t\right)}{dt}$ will scale
as 
\begin{equation}
\frac{d\hat{S}\left(t\right)}{dt}\sim\mathrm{Max}\left\{ \frac{\ddot{\theta}}{G},\frac{\ddot{\phi}}{G},g,\frac{g\dot{\hat{d}}}{\Delta_{1,2}},\frac{g\dot{\theta}\sqrt{N_{th}}}{\Delta_{1,2}},\frac{g\dot{\phi}\sqrt{N_{th}}}{\Delta_{2}},\right\} \label{eq:4.1.94}
\end{equation}
Using the same reasoning that led to (\ref{eq:4.1.72-5}), we summarize
in a table the conditions that have to be satisfied so that the terms
in (\ref{eq:4.1.93}) can be safely neglected and that give rise to
(\ref{eq:4.1.90}):

\begin{center}
\begin{tabular}{|c|c|}
\hline 
Adiabatic criteria & $\frac{\dot{\phi}}{G}\ll1$;$\frac{\dot{\theta}}{G}\ll1$;$\frac{1}{Gt_{f}}\ll1$\tabularnewline
\hline 
Adiabatic criteria (2) & $\frac{\mathrm{Max}\left(\ddot{\theta},\ddot{\phi}\right)}{G^{2}}\ll1$\tabularnewline
\hline 
\hline 
Small coupling & $\frac{g\sqrt{N_{th}}}{G}\ll1$\tabularnewline
\hline 
Large detuning & $\frac{G}{\Delta}\ll1$\tabularnewline
\hline 
Small cavity damping rate & $\frac{\kappa}{\Delta}\ll1$\tabularnewline
\hline 
Adiabatic/large detuning criteria & $\frac{g\mathrm{Max}\left(\dot{\theta},\dot{\phi}\right)}{\Delta^{2}}\ll1$\tabularnewline
\hline 
Detuning condition & $\Delta=\Delta_{1}=-\Delta_{2}$\tabularnewline
\hline 
\end{tabular}
\par\end{center}

\begin{center}
\textbf{\label{Table-1:-List}Table 1}: List of conditions for the
range of validity of our state transfer protocol.
\par\end{center}

We add a last note on the upper bound of $t_{f}$. Recall that in
section (\ref{sec:Adiabatic-approximation-and}), we showed that the
adiabatic approximation was only valid for time scales such that 
\begin{equation}
\int_{0}^{t}\sum_{n}c_{n}(t')e^{i\left[\theta_{n}(t')-\theta_{m}(t')\right]}\frac{\left\langle \phi_{m}\left(t'\right)|\right.\dot{\hat{H}}\left(t'\right)\left.|\phi_{n}\left(t'\right)\right\rangle }{\left[E_{n}(t')-E_{m}(t')\right]}dt'\lesssim1,\label{eq:4.1.99}
\end{equation}
where $\theta_{n}(t)$ is the dynamical phase of the instantaneous
eigenstate $\left.|\phi_{n}\left(t\right)\right\rangle $ of the Hamiltonian.
If we use only the first line in (\ref{eq:4.1.20}), then we can show
that 
\begin{equation}
\left\langle -\left(t\right)|\right.\dot{\hat{H}}\left(t\right)\left.|+\left(t\right)\right\rangle \sim G\mathrm{Max}\left\{ \dot{\theta},\dot{\phi}\right\} \label{eq:4.1.100}
\end{equation}
Using the fact that the eigenvalues of the bright states are $E_{\pm}=\pm G$
and that $\left\{ \dot{\theta},\dot{\phi}\right\} \sim\frac{1}{t_{f}}$,
then it is straightforward to show that 
\begin{equation}
\int_{0}^{t}\sum_{n}c_{n}(t')e^{i\left[\theta_{n}(t')-\theta_{m}(t')\right]}\frac{\left\langle \phi_{m}\left(t'\right)|\right.\dot{\hat{H}}\left(t'\right)\left.|\phi_{n}\left(t'\right)\right\rangle }{\left[E_{n}(t')-E_{m}(t')\right]}dt'\sim\frac{1}{Gt_{f}}\ll1\label{eq:4.1.101}
\end{equation}
by the adiabatic criteria. Therefore, as long as the adiabatic criteria
are satisfied, we never need to worry about time scales where non-adiabatic
corrections start kicking in. This will be true for the leading order
corrections. Subleading corrections still grow, but much more slowly
as $\left|\frac{\dot{\phi}}{G}\right|,\left|\frac{\dot{\theta}}{G}\right|\ll1$.

\section{\label{sec:Changes-to-the}Changes to the initial state}

In deriving (\ref{eq:4.1.90}), we assumed that the changes from the
Schrieffer-Wolff transformation to the initial state are negligible.
More specifically, applying the Schrieffer-Wolff transformation changes
$\hat{\rho}\left(0\right)\rightarrow e^{\hat{S}\left(t\right)}\hat{\rho}\left(0\right)e^{-\hat{S}\left(t\right)}$
which will affect $\left\langle \tilde{\hat{\Pi}}_{0d}\left(t\right)\right\rangle $
in a way that will be described in (\ref{eq:4.3.19}) below. In this
section we give conditions for when we can ignore these changes. Recall
that at initial times $t_{i}=0$, the density matrix is given by 
\begin{equation}
\hat{\rho}\left(0\right)=\left.|\psi_{s}\left(0\right)\right\rangle \left\langle \psi_{s}\left(0\right)|\right.\otimes\hat{\rho}_{env}\left(0\right),\label{eq:4.3.1}
\end{equation}
where $\hat{\rho}_{env}\left(0\right)$ describes the cavity degrees
of freedom being in a thermal state. The initial state of the system
is given by 
\begin{equation}
\left.|\psi_{s}\left(0\right)\right\rangle =\alpha\left.|0\right\rangle +\beta\left.|g_{1}\right\rangle \label{eq:4.3.4}
\end{equation}
Using (\ref{eq:4.3.4}) it is straightforward to see that
\begin{equation}
\hat{\rho}\left(0\right)=\left(\left|\alpha\right|^{2}\left.|0\right\rangle \left\langle 0|\right.+\alpha\beta^{*}\left.|0\right\rangle \left\langle g_{1}|\right.+\alpha^{*}\beta\left.|g_{1}\right\rangle \left\langle 0|\right.+\left|\beta\right|^{2}\left.|g_{1}\right\rangle \left\langle g_{1}|\right.\right)\otimes\hat{\rho}_{env}\left(0\right)\label{eq:4.3.5}
\end{equation}
Now, we want to see how the transformation $\hat{U}\left(t\right)=e^{\hat{S}\left(t\right)}$
affects the density matrix at initial times. This is important because
if the changes to the density matrix at initial times arising from
the Schrieffer-Wolff transformation cannot be neglected relative to
$\hat{\rho}\left(0\right)$, then there would be terms in the Hamiltonian
(other than $\left.|d\right\rangle \left\langle d|\right.$ in (\ref{eq:4.1.89}))
that would contribute to the phase of $\left\langle \tilde{\hat{\Pi}}_{0d}\left(t\right)\right\rangle $. 

Recall that we found that
\begin{equation}
\hat{S}\left(t\right)=\hat{\chi}_{1}\left(t\right)\left.|+\right\rangle \left\langle d|\right.+\hat{\chi}_{2}\left(t\right)\left.|-\right\rangle \left\langle d|\right.+\hat{\chi}_{3}\left(t\right)\left.|+\right\rangle \left\langle -|\right.-h.c.\label{eq:4.3.6}
\end{equation}
with
\begin{equation}
\hat{\chi}_{1}\left(t\right)=\frac{1}{\sqrt{2}G}\left[h\left(t\right)-e^{-iGt}h\left(0\right)\right]-i\frac{g}{\sqrt{2}}\left[\frac{e^{i\Delta_{1}t}\cos\theta\left(t\right)\hat{d}\left(t\right)-e^{-iGt}\hat{d}}{\Delta+G}-\frac{\sin\theta\left(t\right)e^{i\phi\left(t\right)}e^{i\Delta_{2}t}\hat{d}\left(t\right)}{\Delta-G}\right],\label{eq:4.3.7}
\end{equation}
and
\begin{equation}
\hat{\chi}_{2}\left(t\right)=-\frac{1}{\sqrt{2}G}\left[h\left(t\right)-e^{-iGt}h\left(0\right)\right]+i\frac{g}{\sqrt{2}}\left[\frac{e^{i\Delta_{1}t}\cos\theta\left(t\right)\hat{d}\left(t\right)-e^{iGt}\hat{d}}{\Delta-G}-\frac{\sin\theta\left(t\right)e^{i\phi\left(t\right)}e^{i\Delta_{2}t}\hat{d}\left(t\right)}{\Delta+G}\right].\label{eq:4.3.8}
\end{equation}
We don't need to specify $\hat{\chi}_{3}\left(t\right)$ since as
we will see below it commutes with $\hat{\rho}\left(0\right)$. At
this stage we need to evaluate 
\begin{equation}
e^{\hat{S}\left(t\right)}\hat{\rho}\left(0\right)e^{-\hat{S}\left(t\right)}=\rho\left(0\right)+\left[\hat{S}\left(t\right),\hat{\rho}\left(0\right)\right]+\frac{1}{2}\left[\hat{S}\left(t\right),\left[\hat{S}\left(t\right),\hat{\rho}\left(0\right)\right]\right]+...\label{eq:4.3.9}
\end{equation}
At time $t_{i}=0$, the instantaneous eigenstates of the Hamiltonian
are given by 
\begin{equation}
\left.|d\right\rangle =\left.|g_{1}\right\rangle \label{eq:4.3.10}
\end{equation}
\begin{equation}
\left.|+\right\rangle =\frac{1}{\sqrt{2}}\left(+i\left.|e\right\rangle -\left.|g_{2}\right\rangle \right),\label{eq:4.3.11}
\end{equation}
\begin{equation}
\left.|-\right\rangle =-\frac{1}{\sqrt{2}}\left(i\left.|e\right\rangle +\left.|g_{2}\right\rangle \right).\label{eq:4.3.12}
\end{equation}
Using (\ref{eq:4.3.6}) and (\ref{eq:4.3.10}) to (\ref{eq:4.3.12}),
it is straightforward to show that 
\begin{align}
\left[\hat{S}\left(t\right),\hat{\rho}\left(0\right)\right] & =\left[\alpha^{*}\beta\left(\hat{\chi}_{1}\left(t\right)\left.|+\right\rangle \left\langle 0|\right.+\hat{\chi}_{2}\left(t\right)\left.|-\right\rangle \left\langle 0|\right.\right)\right.\nonumber \\
 & \left.+\left|\beta\right|^{2}\left(\hat{\chi}_{1}\left(t\right)\left.|+\right\rangle \left\langle d|\right.+\hat{\chi}_{2}\left(t\right)\left.|-\right\rangle \left\langle d|\right.\right)\right]\otimes\hat{\rho}_{env}\left(0\right)+h.c.\label{eq:4.3.13}
\end{align}
Now the prefactor in (\ref{eq:4.3.13}) will scale as (using that
$\hat{\chi}_{1}\left(t\right)$ scales the same way as $\hat{\chi}_{2}\left(t\right)$)
\begin{equation}
\hat{\chi}_{1}\left(t\right)\sim\mathrm{max}\left\{ \frac{\dot{\theta}}{G},\frac{\dot{\phi}}{G},\frac{g\sqrt{N_{th}}}{\Delta}\right\} \label{eq:4.3.15}
\end{equation}
However, the observable that we are interested in calculating is $\left\langle \tilde{\hat{\Pi}}_{0d}\left(t\right)\right\rangle $.
We can use the results of (\ref{eq:3.1.6}) to (\ref{eq:3.1.12})
with the exception that the state $\hat{\rho}\left(0\right)$ has
to be replaced with $e^{\hat{S}\left(t\right)}\hat{\rho}\left(0\right)e^{-\hat{S}\left(t\right)}$
so that 
\begin{equation}
\left\langle \tilde{\hat{\Pi}}_{0d}\left(t\right)\right\rangle =\mathrm{Tr}\left\{ e^{\hat{S}\left(t\right)}\hat{\rho}\left(0\right)e^{-\hat{S}\left(t\right)}e^{-i\int_{0}^{t}dt'L_{v}\left(t'\right)}\left.|0\right\rangle \left\langle d|\right.\right\} ,\label{eq:4.3.16}
\end{equation}
where 
\begin{equation}
L_{v}\left(t'\right)\left.|0\right\rangle \left\langle d|\right.=\left[\tilde{\hat{H}}'\left(t\right),\left.|0\right\rangle \left\langle d|\right.\right].\label{eq:4.3.17}
\end{equation}
 From (\ref{eq:4.3.5}), we will focus on the $\alpha^{*}\beta\left.|d\right\rangle \left\langle 0|\right.$
contribution to $\hat{\rho}\left(0\right)$ in (\ref{eq:4.3.16})
since we are only interested in calculating how the corrections to
$\left\langle \tilde{\hat{\Pi}}_{0d}\left(t\right)\right\rangle $
scale. Using the cyclic permutation property of the trace and the
fact that the operation
\begin{equation}
e^{-i\int_{0}^{t}dt'L_{v}\left(t'\right)}\left.|0\right\rangle \left\langle d|\right.=e^{-i\hat{\phi}_{0d}\left(t\right)}\left.|0\right\rangle \left\langle d|\right.,\label{eq:4.3.18}
\end{equation}
with the operator $\hat{\phi}_{0d}\left(t\right)$ satisfying the
property
\begin{equation}
\mathrm{Tr}\left\{ e^{-i\hat{\phi}_{0d}\left(t\right)}\hat{\rho}_{env}\left(0\right)\right\} =\alpha\beta^{*}e^{-i\gamma_{d}\left(t\right)}\left\langle T_{t}e^{i\frac{g^{2}}{\Delta}\int_{0}^{t}\cos2\theta\left(t'\right)\hat{d}^{\dagger}\left(t'\right)\hat{d}\left(t'\right)dt'}\right\rangle ,
\end{equation}
which is what we found in (\ref{eq:4.1.90}). Now we can write (\ref{eq:4.3.16})
as 
\begin{equation}
\left\langle \tilde{\hat{\Pi}}_{0d}\left(t\right)\right\rangle =\mathrm{Tr}\left\{ \left\langle d|\right.e^{\hat{S}\left(t\right)}\left.|d\right\rangle e^{-i\hat{\phi}_{0d}\left(t\right)}\hat{\rho}_{env}\left(0\right)\right\} \label{eq:4.3.19}
\end{equation}
where we used the fact that $e^{\hat{S}\left(t\right)}\left.|0\right\rangle =\left.|0\right\rangle $.
The prefactor $\left\langle d|\right.e^{\hat{S}\left(t\right)}\left.|d\right\rangle $
in (\ref{eq:4.3.19}) will not cause decay to zero in the chosen parameter
regime. To understand this, we note that the exponential decay arising
from $\hat{\phi}_{0d}\left(t\right)$ is due to the fact that we integrate
over terms proportional to $\hat{d}^{\dagger}\left(t\right)\hat{d}\left(t\right)$.
However, $\hat{S}\left(t\right)$ only contains terms proportional
to $\hat{d}\left(t\right)$ (from $\hat{\chi}_{1,2}\left(t\right)$)
which are not being integrated over in$\left\langle d|\right.e^{\hat{S}\left(t\right)}\left.|d\right\rangle $.
Thus the term $\left\langle d|\right.e^{\hat{S}\left(t\right)}\left.|d\right\rangle $
will not give rise to exponential decay for the chosen parameter regime.

From (\ref{eq:4.3.6}), a quick calculation shows that 
\begin{align}
\left\langle d|\right.e^{\hat{S}\left(t\right)}\left.|d\right\rangle  & =\sum_{k=0}^{\infty}\frac{1}{\left(2k\right)!}\left\langle d|\right.\hat{S}^{2k}\left.|d\right\rangle \nonumber \\
 & =\sum_{k=0}^{\infty}\frac{(-1)^{k}}{\left(2k\right)!}\left(\hat{\chi}_{1}^{\dagger}\left(t\right)\hat{\chi}_{1}\left(t\right)+\hat{\chi}_{2}^{\dagger}\left(t\right)\hat{\chi}_{2}\left(t\right)\right)^{k}\nonumber \\
 & =\cos\sqrt{\hat{\chi}_{1}^{\dagger}\left(t\right)\hat{\chi}_{1}\left(t\right)+\hat{\chi}_{2}^{\dagger}\left(t\right)\hat{\chi}_{2}\left(t\right)}\nonumber \\
 & =1-\frac{1}{2}\left(\hat{\chi}_{1}^{\dagger}\left(t\right)\hat{\chi}_{1}\left(t\right)+\hat{\chi}_{2}^{\dagger}\left(t\right)\hat{\chi}_{2}\left(t\right)\right)+\mathcal{O}\left(\hat{S}\left(t\right)^{4}\right)\label{eq:4.3.20}
\end{align}
Going back to the definitions of $\hat{\chi}_{1}\left(t\right)$ and
$\hat{\chi}_{2}\left(t\right)$ given in (\ref{eq:4.3.7}) and (\ref{eq:4.3.8}),
we can show that 
\begin{equation}
\left|\left|\hat{\chi}_{1}^{\dagger}\left(t\right)\hat{\chi}_{1}\left(t\right)+\hat{\chi}_{2}^{\dagger}\left(t\right)\hat{\chi}_{2}\left(t\right)\right|\right|\sim\mathrm{Max}\left\{ \frac{1}{t_{f}^{2}G^{2}},\,\frac{g\sqrt{N_{th}}}{Gt_{f}\Delta},\,\frac{g^{2}N_{th}}{\Delta^{2}}\right\} \label{eq:4.3.21}
\end{equation}
In the next section, we will see that on short time scales the fidelity
that we obtain (which will not include the corrections in (\ref{eq:4.3.21}))
will be very close to unity. If we included the leading order correction
terms in (\ref{eq:4.3.20}), these terms would create small oscillations
about the reported fidelity. Ergo, when we report an error rate $\epsilon$
at a time $t_{f}$ due to dephasing effects, we need to ensure that
the correction terms in (\ref{eq:4.3.21}) are much smaller than $\epsilon$.

\section{\label{sec:Calculation-of-the}Calculation of the fidelity for the
case where $g_{1}=g_{2}$}

The goal of this section will be to obtain the fidelity for our state
transfer protocol when the two cavity coupling constants are identical.
It is important to remember that (\ref{eq:4.1.90}) is only valid
when $g_{1}=g_{2}$. To keep the analysis simple, the state will be
transferred using a path such that $\theta\left(t\right)$ is a linear
function of time. Consequently, the path in $\left\{ \Omega_{1},\Omega_{2}\right\} $
space will be described by (\ref{eq:3.1.35}) and (\ref{eq:3.1.36})
with $\theta\left(t\right)=-bt$ where
\begin{equation}
b\equiv\frac{2n\pi}{t_{f}},\label{eq:4.4.1}
\end{equation}
and $n$ is a parameter that controls how many loops will be done
during the evolution. To calculate the fidelity, we must first calculate
the correlation function found in (\ref{eq:4.1.90}) at the final
time of the state transfer protocol. We start by defining
\begin{equation}
\hat{X}\left(t\right)\equiv\frac{g^{2}}{\Delta}\int_{0}^{t}dt'\cos\left(\frac{4\pi nt'}{t_{f}}\right)\hat{d}^{\dagger}\left(t'\right)\hat{d}\left(t'\right).\label{eq:4.4.2}
\end{equation}
From this definition we can rewrite (\ref{eq:4.1.90}) when $t=t_{f}$
as 
\begin{equation}
\left\langle \tilde{\hat{\Pi}}_{0d}\left(t_{f}\right)\right\rangle =\alpha\beta^{*}e^{-i\gamma_{d}\left(t\right)}\left\langle T_{t}e^{i\hat{X}\left(t_{f}\right)}\right\rangle .\label{eq:4.4.3}
\end{equation}
The quantum noise contribution to the cavity lowering operator for
a cavity driven by a single classical laser field was given in Eq.
(\ref{eq:2.4.31}). When we performed the displacement transformation
in (\ref{eq:4.1.6}), only the cavity frequency appears in the time
dependent exponential. This is because we are working in an interaction
picture at the cavity resonance frequency. With this particular type
of transformation we have that 
\begin{equation}
\hat{d}\left(t\right)=-\sqrt{\kappa}\int_{-\infty}^{t}e^{-\frac{\kappa}{2}\left(t-\tau\right)}\hat{\xi}\left(\tau\right)d\tau.\label{eq:4.4.4}
\end{equation}
We remind the reader that $\kappa$ represents the cavity damping
rate and $\hat{\xi}\left(t\right)$ describes both thermal and vacuum
noise incident on the cavity through the drive port. Also, it has
the auto-correlation function given by 
\begin{equation}
\left\langle \hat{\xi}^{\dagger}\left(t\right)\hat{\xi}\left(t'\right)\right\rangle =N_{th}\delta\left(t-t'\right),\label{eq:4.4.5}
\end{equation}
where $N_{th}$ is the bosonic thermal equilibrium occupation number.
Given these results, a straightforward calculation shows that
\begin{equation}
\left\langle \hat{d}^{\dagger}\left(t_{1}\right)\hat{d}\left(t_{2}\right)\right\rangle =N_{th}e^{-\frac{\kappa}{2}\left|t_{1}-t_{2}\right|}.\label{eq:4.4.6}
\end{equation}

\begin{doublespace}
The rest of this section will be devoted to calculating $\left\langle T_{t}e^{-i\hat{X}\left(t_{f}\right)}\right\rangle $
and finding the fidelity as a function of $t_{f}$ for the state transfer
protocol. We will expand the exponential in its Taylor series and
perform a moment expansion allowing us to keep only the second order
term ($\left\langle T_{t}\hat{X}\left(t_{f}\right)\hat{X}\left(t_{f}\right)\right\rangle $).
This can be justified by choosing the right conditions on our parameters
such that higher order terms will give rise to much smaller contributions.
These conditions will be established below. 
\end{doublespace}

The leading-order term in the moment expansion will involve the correlation
function 
\begin{equation}
\left\langle \hat{X}\left(t_{f}\right)\right\rangle =-\frac{g^{2}}{\Delta}\int_{0}^{t_{f}}dt\cos\left(\frac{4n\pi t}{t_{f}}\right)\left\langle \hat{d}^{\dagger}\left(t\right)\hat{d}\left(t\right)\right\rangle .\label{eq:4.4.7}
\end{equation}
From (\ref{eq:4.4.6}), we see that 
\begin{equation}
\left\langle \hat{X}\left(t_{f}\right)\right\rangle =-\frac{N_{th}g^{2}}{\Delta}\int_{0}^{t_{f}}dt\cos\left(\frac{4n\pi t}{t_{f}}\right)=0.\label{eq:4.4.8}
\end{equation}
Consequently, the leading order term vanishes. The next order term
(and also the most important one) is given by 
\begin{equation}
\left\langle T_{t}\hat{X}\left(t_{f}\right)\hat{X}\left(t_{f}\right)\right\rangle =\left(-\frac{g^{2}}{\Delta}\right)^{2}\int_{0}^{t_{f}}dt_{1}dt_{2}\cos\left(\frac{4\pi nt_{1}}{t_{f}}\right)\cos\left(\frac{4\pi nt_{2}}{t_{f}}\right)\left\langle T_{t}\hat{d}^{\dagger}\left(t_{1}\right)\hat{d}\left(t_{1}\right)\hat{d}^{\dagger}\left(t_{2}\right)\hat{d}\left(t_{2}\right)\right\rangle .\label{eq:4.4.9}
\end{equation}
We can use Wick's theorem to calculate the correlation function appearing
in the above expression. However, since there are integral factors
appearing in Eq. (\ref{eq:4.4.9}), we can simplify much of the notation
by defining the operator 
\begin{equation}
L^{\left(n\right)}\left(t_{f}\right)\mathcal{A}\equiv\int_{0}^{t_{f}}\prod_{i=1}^{n}dt_{i}\cos\left(\frac{4\pi t_{i}}{t_{f}}\right)\mathcal{A}.\label{eq:4.4.10}
\end{equation}
Then using this definition and Wick's theorem we find
\begin{align}
L^{\left(2\right)}\left(t_{f}\right)\left\langle T_{t}\hat{d}^{\dagger}\left(t_{1}\right)\hat{d}\left(t_{1}\right)\hat{d}^{\dagger}\left(t_{2}\right)\hat{d}\left(t_{2}\right)\right\rangle  & =L^{\left(2\right)}\left(t_{f}\right)\left\{ \left\langle \hat{d}^{\dagger}\left(t_{1}\right)\hat{d}\left(t_{2}\right)\right\rangle \left\langle \hat{d}\left(t_{1}\right)\hat{d}^{\dagger}\left(t_{2}\right)\right\rangle \Theta\left(t_{1}-t_{2}\right)\right.\nonumber \\
 & \left.+\left\langle \hat{d}^{\dagger}\left(t_{2}\right)\hat{d}\left(t_{1}\right)\right\rangle \left\langle \hat{d}\left(t_{2}\right)\hat{d}^{\dagger}\left(t_{1}\right)\right\rangle \Theta\left(t_{2}-t_{1}\right)\right\} .\label{eq:4.4.11}
\end{align}
Note that correlation functions of the $\hat{d}$ operator evaluated
at equal times will always vanish when evaluated under the integral
(just like what we showed in (\ref{eq:4.4.8})). Using (\ref{eq:4.4.6})
along with (\ref{eq:2.4.35}), the product of correlation functions
appearing in the above equation is found to be 
\begin{equation}
\left\langle \hat{d}^{\dagger}\left(t_{1}\right)\hat{d}\left(t_{2}\right)\right\rangle \left\langle \hat{d}\left(t_{1}\right)\hat{d}^{\dagger}\left(t_{2}\right)\right\rangle =N_{th}\left(N_{th}+1\right)e^{-\kappa\left|t_{1}-t_{2}\right|}.\label{eq:4.4.12}
\end{equation}
Putting everything together we find that
\begin{equation}
\left\langle T_{t}\hat{X}\left(t_{f}\right)\hat{X}\left(t_{f}\right)\right\rangle =\left(\frac{g^{2}}{\Delta}\right)^{2}N_{th}\left(N_{th}+1\right)\int_{0}^{t_{f}}dt_{1}dt_{2}\cos\left(\frac{4\pi nt_{1}}{t_{f}}\right)\cos\left(\frac{4\pi nt_{2}}{t_{f}}\right)e^{-\kappa\left|t_{1}-t_{2}\right|}.\label{eq:4.4.13}
\end{equation}
This integral is straightforward to compute and yields one of the
central results of this thesis
\begin{equation}
\left\langle T_{t}\hat{X}\left(t_{f}\right)\hat{X}\left(t_{f}\right)\right\rangle =N_{th}\left(N_{th}+1\right)\left(\frac{g^{2}}{\Delta}\right)^{2}\frac{\kappa t_{f}^{3}\left[4n^{2}\pi^{2}+2\kappa t_{f}\left(e^{-\kappa t_{f}}-1\right)+\kappa^{2}t_{f}^{2}\right]}{\left(4n^{2}\pi^{2}+\kappa^{2}t_{f}^{2}\right)^{2}}.\label{eq:4.4.14}
\end{equation}
We will come back to the implications of this expression shortly.
For now, we will focus on determining how the third-order term scales
in terms of the relevant parameters. This can be determined from 
\begin{equation}
\left\langle T_{t}\hat{X}\left(t_{f}\right)\hat{X}\left(t_{f}\right)\hat{X}\left(t_{f}\right)\right\rangle =\left(-\frac{g^{2}}{\Delta}\right)^{3}L^{\left(3\right)}\left(t_{f}\right)\left\langle T_{t}\hat{d}^{\dagger}\left(t_{1}\right)\hat{d}\left(t_{1}\right)\hat{d}^{\dagger}\left(t_{2}\right)\hat{d}\left(t_{2}\right)\hat{d}^{\dagger}\left(t_{3}\right)\hat{d}\left(t_{3}\right)\right\rangle .\label{eq:4.4.15}
\end{equation}
When expanding the term on the right hand side of the above equation,
there will be $3!=6$ terms coming from the time ordering symbol.
We can evaluate one of them to determine how they scale with respect
to the relevant parameters. As an example, we pick 
\begin{equation}
J_{1}\equiv L^{\left(3\right)}\left(t_{f}\right)\left\langle \hat{d}^{\dagger}\left(t_{1}\right)\hat{d}\left(t_{1}\right)\hat{d}^{\dagger}\left(t_{2}\right)\hat{d}\left(t_{2}\right)\hat{d}^{\dagger}\left(t_{3}\right)\hat{d}\left(t_{3}\right)\right\rangle \Theta\left(t_{1}-t_{2}\right)\Theta\left(t_{2}-t_{3}\right).\label{eq:4.4.16}
\end{equation}
Expanding using Wick's theorem we get
\begin{align}
J_{1} & =L^{\left(3\right)}\left(t_{f}\right)\left\langle \hat{d}^{\dagger}\left(t_{1}\right)\hat{d}\left(t_{2}\right)\right\rangle \left\langle \hat{d}\left(t_{1}\right)\hat{d}^{\dagger}\left(t_{3}\right)\right\rangle \left\langle \hat{d}^{\dagger}\left(t_{2}\right)\hat{d}\left(t_{3}\right)\right\rangle \Theta\left(t_{1}-t_{2}\right)\Theta\left(t_{2}-t_{3}\right)\nonumber \\
 & +L^{\left(3\right)}\left(t_{f}\right)\left\langle \hat{d}^{\dagger}\left(t_{1}\right)\hat{d}\left(t_{3}\right)\right\rangle \left\langle \hat{d}\left(t_{1}\right)\hat{d}^{\dagger}\left(t_{2}\right)\right\rangle \left\langle \hat{d}\left(t_{2}\right)\hat{d}^{\dagger}\left(t_{3}\right)\right\rangle \Theta\left(t_{1}-t_{2}\right)\Theta\left(t_{2}-t_{3}\right).\label{eq:4.4.17}
\end{align}
All the other terms will vanish since at least one correlation function
would be evaluated at equal times thus vanishing under the action
of the integral. We can evaluate the products using (\ref{eq:4.4.6})
and find that
\begin{equation}
\left\langle \hat{d}^{\dagger}\left(t_{1}\right)\hat{d}\left(t_{2}\right)\right\rangle \left\langle \hat{d}\left(t_{1}\right)\hat{d}^{\dagger}\left(t_{3}\right)\right\rangle \left\langle \hat{d}^{\dagger}\left(t_{2}\right)\hat{d}\left(t_{3}\right)\right\rangle =N_{th}^{2}\left(N_{th}+1\right)e^{-\frac{\kappa}{2}\left(\left|t_{1}-t_{2}\right|+\left|t_{1}-t_{3}\right|+\left|t_{2}-t_{3}\right|\right)}.\label{eq:4.4.18}
\end{equation}
Evaluating this term under the integral, we find that (for $n=1$)
\begin{equation}
N_{th}^{2}\left(N_{th}+1\right)L^{\left(3\right)}\left(t_{f}\right)e^{-\frac{\kappa}{2}\left(\left|t_{1}-t_{2}\right|+\left|t_{1}-t_{3}\right|+\left|t_{2}-t_{3}\right|\right)}\Theta\left(t_{1}-t_{2}\right)\Theta\left(t_{2}-t_{3}\right)=-\frac{2N_{th}^{2}\left(N_{th}+1\right)\left(1-e^{-\kappa t_{f}}\right)\kappa t_{f}^{4}}{1024\pi^{4}+80\pi^{2}\kappa^{2}t_{f}^{2}+\kappa^{4}t_{f}^{4}}.\label{eq:4.4.19}
\end{equation}
We can now determine how the third-order contribution scales relative
to equation (\ref{eq:4.4.14}). We first begin by considering the
long time limit, where $\kappa t_{f}\gg1$. Then the second-order
term will scale as 
\begin{equation}
\left\langle T_{t}\hat{X}\left(t_{f}\right)\hat{X}\left(t_{f}\right)\right\rangle \sim\frac{N_{th}\left(N_{th}+1\right)g^{4}}{\Delta^{2}}\frac{t_{f}}{\kappa}.\label{eq:4.4.20}
\end{equation}
The third-order term is seen to scale as 
\begin{equation}
\left\langle T_{t}\hat{X}\left(t_{f}\right)\hat{X}\left(t_{f}\right)\hat{X}\left(t_{f}\right)\right\rangle \sim\frac{N_{th}^{2}\left(N_{th}+1\right)g^{6}}{\Delta^{3}}\frac{1}{\kappa^{3}}.\label{eq:4.4.21}
\end{equation}
Consequently, to keep only leading order terms in the large time limit,
we require 
\begin{equation}
\frac{N_{th}g^{2}}{\Delta\kappa^{2}t_{f}}\ll1.\label{eq:4.4.22}
\end{equation}
In the opposite (short-time) limit, $\kappa t_{f}\ll1$, it can be
shown that the leading-order term will scale as
\begin{equation}
\left\langle T_{t}\hat{X}\left(t_{f}\right)\hat{X}\left(t_{f}\right)\right\rangle \sim\frac{N_{th}\left(N_{th}+1\right)g^{4}\kappa n^{2}t_{f}^{3}}{\Delta^{2}}.\label{eq:4.4.23}
\end{equation}
whereas the third-order term scales as 
\begin{equation}
\left\langle T_{t}\hat{X}\left(t_{f}\right)\hat{X}\left(t_{f}\right)\hat{X}\left(t_{f}\right)\right\rangle \sim\frac{N_{th}^{2}\left(N_{th}+1\right)g^{6}\kappa^{2}n^{3}t_{f}^{5}}{\Delta^{3}}.\label{eq:4.4.24}
\end{equation}
To keep only leading-order terms in this limit, we require that 
\begin{equation}
\frac{N_{th}g^{2}\kappa t_{f}^{2}n}{\Delta}\ll1.\label{eq:4.4.25}
\end{equation}
Assuming that conditions (\ref{eq:4.4.22}) and (\ref{eq:4.4.25})
are satisfied, we can perform a moment expansion in evaluating $\left\langle T_{t}e^{i\hat{X}}\right\rangle $.
To do so we use the following identity
\begin{equation}
\left\langle T_{t}e^{-i\hat{X}}\right\rangle =e^{\ln\left\langle T_{t}e^{-i\hat{X}}\right\rangle }.\label{eq:4.4.26}
\end{equation}
Expanding we have that 
\begin{equation}
\ln\left\langle T_{t}e^{-i\hat{X}}\right\rangle =\ln\left[1-\left(\frac{1}{2}\left\langle T_{t}\hat{X}^{2}\right\rangle -\frac{i}{6}\left\langle T_{t}\hat{X}^{3}\right\rangle +\mathcal{O}\left(\left\langle \hat{X}^{4}\right\rangle \right)\right)\right].\label{eq:4.4.27}
\end{equation}
Expanding the logarithm, it can be approximated by 
\begin{equation}
\ln\left\langle T_{t}e^{i\hat{X}}\right\rangle \approx-\frac{1}{2}\left\langle T_{t}\hat{X}^{2}\right\rangle -\frac{1}{8}\left\langle T_{t}\hat{X}^{2}\right\rangle ^{2}+\frac{i}{6}\left\langle T_{t}\hat{X}^{3}\right\rangle .\label{eq:4.4.28}
\end{equation}
Thus, if (\ref{eq:4.4.22}) is satisfied we can write
\begin{equation}
\left\langle T_{t}e^{i\hat{X}}\right\rangle \approx e^{-\frac{1}{2}\left\langle T_{t}\hat{X}\left(t_{f}\right)\hat{X}\left(t_{f}\right)\right\rangle }.\label{eq:4.4.29}
\end{equation}
This means that under this approximation, we only need to compute
the correlation function $\left\langle T_{t}\hat{X}\left(t_{f}\right)\hat{X}\left(t_{f}\right)\right\rangle $
to obtain the correction to the geometric phase arising from the fluctuating
number of photons inside the cavity. Using (\ref{eq:4.4.14}) and
(\ref{eq:3.1.18}), the average of the relevant component of the density
matrix for our state transfer protocol is found to be 
\begin{equation}
\left\langle \tilde{\hat{\Pi}}_{0d}\left(t_{f}\right)\right\rangle =\alpha\beta^{*}e^{i\gamma_{d}\left(t\right)}\exp\left[-\frac{N_{th}\left(N_{th}+1\right)}{2}\left(\frac{g^{2}}{\Delta}\right)^{2}\frac{\kappa t_{f}^{3}\left[4n^{2}\pi^{2}+2\kappa t_{f}\left(e^{-\kappa t_{f}}-1\right)+\kappa^{2}t_{f}^{2}\right]}{\left(4n^{2}\pi^{2}+\kappa^{2}t_{f}^{2}\right)^{2}}\right].\label{eq:4.4.30}
\end{equation}
From this result, the fidelity is found to be given by
\begin{equation}
F=\left|\alpha\right|^{4}+\left|\beta\right|^{4}+2\left|\alpha\right|^{2}\left|\beta\right|^{2}\cos\left(\gamma_{d}\left(t_{f}\right)\right)\exp\left(-\frac{N_{th}\left(N_{th}+1\right)}{2}\left(\frac{g^{2}}{\Delta}\right)^{2}\frac{\kappa t_{f}^{3}\left[4n^{2}\pi^{2}+2\kappa t_{f}\left(e^{-\kappa t_{f}}-1\right)+\kappa^{2}t_{f}^{2}\right]}{\left(4n^{2}\pi^{2}+\kappa^{2}t_{f}^{2}\right)^{2}}\right),\label{eq:4.4.31}
\end{equation}
where we reimd the reader that $\gamma_{d}\left(t\right)$ was found
to be 
\begin{equation}
\gamma_{d}\left(t\right)=\int_{0}^{t}\dot{\phi}\left(t'\right)\sin^{2}\theta\left(t'\right)dt'.\label{eq:4.3.31-1}
\end{equation}
To get more intuition on the behavior of the fidelity, we will consider
certain limits. If we first consider the limit where $\kappa t_{f}\gg1$
and $n=1$, the fidelity will reduce to
\begin{equation}
F\approx\left|\alpha\right|^{4}+\left|\beta\right|^{4}+2\left|\alpha\right|^{2}\left|\beta\right|^{2}\cos\left(\gamma_{d}\left(t_{f}\right)\right)\exp\left(-\frac{N_{th}\left(N_{th}+1\right)}{2}\left(\frac{g^{2}}{\Delta}\right)^{2}\frac{t_{f}}{\kappa}\right),\label{eq:4.4.32}
\end{equation}
and so the exponent depends linearly on the state transfer time. Recall
that in the $\kappa t_{f}\gg1$ limit, the condition to neglect the
third moment was $\frac{N_{th}g^{2}}{\Delta\kappa^{2}t_{f}}\ll1.$
Since the exponential in (\ref{eq:4.4.32}) scales as $\frac{N_{th}^{2}g^{4}t_{f}}{\Delta^{2}\kappa}$,
it can still be made large as long as $\Delta^{3}\kappa^{3}\sim N_{th}^{3}g^{6}$. 

In the opposite limit where $\kappa t_{f}\ll1$, the fidelity is approximately
given by
\begin{equation}
F\approx\left|\alpha\right|^{4}+\left|\beta\right|^{4}+2\left|\alpha\right|^{2}\left|\beta\right|^{2}\cos\left(\gamma_{d}\left(t_{f}\right)\right)\exp\left(-\frac{N_{th}\left(N_{th}+1\right)}{2}\left(\frac{g^{2}}{\Delta}\right)^{2}\frac{\kappa t_{f}^{3}}{16\pi^{2}}\right).\label{eq:4.4.33}
\end{equation}
 Coming back to the condition $\frac{N_{th}g^{2}\kappa t_{f}^{2}}{\Delta}\ll1$
that allowed us to neglect the third moment, the exponent in (\ref{eq:4.4.33})
scales as $\frac{N_{th}^{2}g^{4}\kappa t_{f}^{3}}{\Delta^{2}}$. Since
neglecting the third moment requires having $\frac{N_{th}g^{2}t_{f}}{\Delta}\lesssim1$,
we conclude that the terms in the exponent will always be much smaller
than 1. In the short-time regime, (\ref{eq:4.4.33}) shows that the
decay is super-exponential ($\sim e^{-\left(t_{f}/\tau\right)^{3}}$).
This is analogous to the case of a Hahn-echo decay with a Lorentzian
spectral density since in the short-time limit the behavior is similar
(apart for constant prefactors). The reader is referred to \cite{key-45}
for more details.

In the next section we will propose a scheme that allows us to improve
the fidelity by performing the state transfer sequence over many cycles,
directly analogous to an n-pulse Carr-Purcell-Meiboom-Gill (CPMG)
sequence (see section (\ref{sec:Many-cycle-evolution}) for more details).
Consequently, it would be useful to compare the fidelity arising from
a many cycle evolution to the case where we left the system in its
initial superposition state $\left.|\psi\left(0\right)\right\rangle =\alpha\left.|0\right\rangle +\beta\left.|g_{1}\right\rangle $
(this would be equivalent to setting the function $\cos2\theta\left(t\right)=1$).
In other words, if we prepared our qubit state in the linear combination
$\alpha\left.|0\right\rangle +\beta\left.|g_{1}\right\rangle $, the
system would not pick up a geometric phase (since there would be no
lasers creating transition between the different levels and so there
would be no closed loop evolution) but it would still decohere due
to the noise inside the cavity. We assume that even though the cavity
isn't being driven (so as to leave the qubit state in the superposition
$\alpha\left.|0\right\rangle +\beta\left.|g_{1}\right\rangle $),
there is still noise inside the cavity arising from fluctuations in
the number of photons. Consequently, our scheme would be extremely
useful since it would offer a way of prolonging the life of the qubit.
Of course, one must keep in mind that the ultimate goal is to perform
a phase gate and to do this the state $\alpha\left.|0\right\rangle +\beta\left.|g_{1}\right\rangle $
needs to pick a geometric phase. So the whole idea behind our state
transfer protocol is to minimize the effects of dephasing with the
intention of performing a phase gate. 

\begin{doublespace}
To determine the dephasing effects from keeping the qubit state in
its initial state, we only have to consider a Hamiltonian (in an interaction
picture) with the noise terms 
\begin{equation}
\hat{H}\left(t\right)=\delta\hat{\omega}_{1}\left.|g_{1}\right\rangle \left\langle g_{1}|\right.+\delta\hat{\omega}_{2}\left.|g_{2}\right\rangle \left\langle g_{2}|\right.+\hat{H}_{env}\left(\delta\hat{\omega}_{1},\delta\hat{\omega}_{2}\right).\label{eq:4.4.34}
\end{equation}
Since the above Hamiltonian is purely diagonal, we don't need to write
it in terms of a superadiabatic basis. Instead we can work in the
original basis and compute the average of the operator 
\begin{equation}
\tilde{\hat{\Pi}}_{0g_{1}}\equiv\left.|0\right\rangle \left\langle g_{1}|\right..\label{eq:4.4.35}
\end{equation}
Note however that $\left.|g_{1}\right\rangle =\left.|d\left(t_{i}\right)\right\rangle =\left.|d\right\rangle $
so that $\tilde{\hat{\Pi}}_{0g_{1}}=\tilde{\hat{\Pi}}_{0d}$. We can
apply the same procedure as we did in equations (\ref{eq:3.1.5})
to (\ref{eq:3.1.17}). Doing so we find 
\begin{equation}
\left\langle \tilde{\hat{\Pi}}_{0g_{1}}\left(t\right)\right\rangle =\left\langle \tilde{\hat{\Pi}}_{0g_{1}}\left(0\right)\right\rangle \left\langle T_{t}e^{-i\int_{0}^{t}dt'\delta\hat{\omega}_{1}\left(t'\right)}\right\rangle .\label{eq:4.4.36}
\end{equation}
where we set $g_{1}=g_{2}$ so that $\delta\hat{\omega}_{1}=-\delta\hat{\omega}_{2}$
and $\delta\hat{\omega}_{1}=\frac{g^{2}}{\Delta}\hat{d}^{\dagger}\hat{d}$.
If we perform a moment expansion on the term $\left\langle T_{t}e^{-i\int_{0}^{t}dt'\delta\hat{\omega}_{1}\left(t'\right)}\right\rangle $
and keep the second order term as we did previously, and using (\ref{eq:4.3.5}),
the fidelity is found to be 
\begin{equation}
F=\left|\alpha\right|^{4}+\left|\beta\right|^{4}+2\left|\alpha\right|^{2}\left|\beta\right|^{2}\exp\left(-\frac{N_{th}\left(N_{th}+1\right)}{\kappa^{2}}\left(\frac{g^{2}}{\Delta}\right)^{2}\left(e^{-\kappa t_{f}}-1+\kappa t_{f}\right)\right).\label{eq:4.4.37}
\end{equation}

Before comparing the fidelity obtained for our state-transfer protocol
to the one obtained by keeping the qubit in its original state, we
will consider certain limits of the above expression. If we first
consider the limit where $\kappa t_{f}\gg1$, the fidelity will reduce
to
\begin{equation}
F\approx\left|\alpha\right|^{4}+\left|\beta\right|^{4}+2\left|\alpha\right|^{2}\left|\beta\right|^{2}\exp\left(-\frac{N_{th}\left(N_{th}+1\right)}{\kappa}\left(\frac{g^{2}}{\Delta}\right)^{2}t_{f}\right).\label{eq:4.4.38}
\end{equation}
Just as in the state-transfer case, the long-time limit corresponds
to a linear dependence on time (total state transfer time) for the
exponent. Notice the factor of $\frac{1}{2}$ appearing in (\ref{eq:4.4.32})
not present in (\ref{eq:4.4.38}). This can be understood from the
fact that if the state remains in $\left.|g_{1}\right\rangle $ for
the entire state transfer protocol, it will only see the noise from
$\delta\hat{\omega}_{1}$. However, for our state transfer protocol,
half of the time the state will see as much of $\left.|g_{1}\right\rangle $
and $\left.|g_{2}\right\rangle $. As we mentioned in section (\ref{sec:Antisymmetric-noise}),
the average of the noise Hamiltonian in the state $\left.|g_{1}\right\rangle +\left.|g_{2}\right\rangle $
is $\delta\hat{\omega}_{1}+\delta\hat{\omega}_{2}=0$. Consequently,
we expect the fidelity to improve by a factor of 2 relative to the
case where we stay in $\left.|g_{1}\right\rangle $ for the entire
state transfer.
\end{doublespace}

In the opposite limit where $\kappa t_{f}\ll1$, the fidelity is approximately
given by
\begin{equation}
F\approx\left|\alpha\right|^{4}+\left|\beta\right|^{4}+2\left|\alpha\right|^{2}\left|\beta\right|^{2}\exp\left(-\frac{N_{th}\left(N_{th}+1\right)}{2}\left(\frac{g^{2}}{\Delta}\right)^{2}t_{f}^{2}\right).\label{eq:4.4.39}
\end{equation}
Contrary to the state transfer case, in the short-time limit the exponent
depends on $t_{f}^{2}$ instead of $t_{f}^{3}$ and also has no dependence
on the damping rate.

\section{\label{sec:Many-cycle-evolution}Many-cycle evolution }

Careful observation of Eq. (\ref{eq:4.4.31}) shows that in the short
time limit the exponent scales as $1/n^{2}$ where the integer $n$
corresponds to the number of state-transfer sequences performed in
parameter space. It is thus seen that as $n$ gets larger the exponential
will become closer to unity giving rise to a better fidelity. This
suggests that it is very advantageous to perform the state-transfer
protocol over many cycles, especially for state preservation. The
reason being that having a longer coherence time (arising from transferring
information back and fourth) would allow one to store information
for longer time scales. The closed path in parameter space effectively
averages the noise over progressively shorter time scales as $n$
is increased, similar to an n-pulse CPMG sequence, for which $s\left(t\right)=\cos2\theta\left(t\right)$
alternates between $\pm1$.

To trust the results derived in this chapter, one must be careful
to obey the adiabatic criteria (which could fail if $n$ is chosen
to be large enough). We know that one of the relevant conditions that
needs to be satisfied is $\frac{\dot{\theta}}{G}\ll1$. For a circular
state transfer in parameter space, this condition reduces to 
\begin{equation}
\frac{n\pi}{Gt_{f}}\ll1.\label{eq:4.5.1}
\end{equation}
Thus, as long as $G$ is chosen to satisfy the condition (\ref{eq:4.5.1})
and the short-time condition is satisfied, we can safely apply our
adiabatic theory developed above for the quantum state transfer protocol
over many evolution cycles. Using (\ref{eq:4.4.31}) and (\ref{eq:4.4.37})
we can plot the fidelity as a function of the total state transfer
time allowing us to compare the improvement due to a many-cycle evolution
over leaving the state in its initial superposition.

\begin{flushleft}
\begin{figure}
\begin{centering}
\includegraphics[scale=0.85]{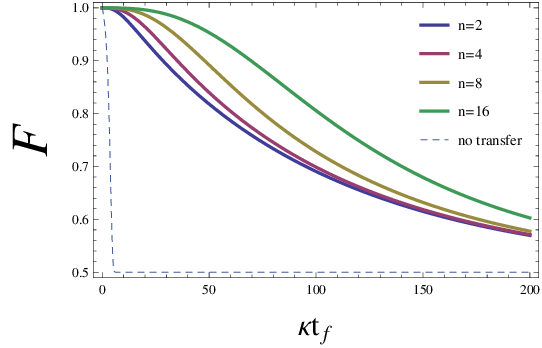}
\par\end{centering}

\caption{\ Many-cycle evolution\label{fig:n-cycle}}

\begin{doublespace}
\centering{}\textit{Plot of the fidelity as a function of $t_{f}$
(the total state transfer time). We have chosen the values $\alpha=\beta=\frac{1}{2}$,
$N_{th}=1$, $\frac{g^{2}}{\Delta}=0.1$ and $\kappa=1$. However,
each curve corresponds to a particular value of the state transfer
cycles ($n$) illustrated in the plot legend. The dashed curve corresponds
to the fidelity that one would get if no state transfer was performed.
Note that we set the phase $\phi\left(t\right)=0$ so that the closed
system Berry's phase ($\gamma_{d}\left(t_{f}\right)$) vanishes. Note
that the chosen values are consistent with the conditions (\ref{eq:4.4.22})
and (\ref{eq:4.4.25}) allowing us to neglect the third moment in
section (\ref{sec:Calculation-of-the}). }\end{doublespace}
\end{figure}
Figure (\ref{fig:n-cycle}) clearly shows a significant advantage
in performing the state-transfer sequence compared to leaving the
qubit in its initial state. Furthermore, performing the state-transfer
protocol over many sequences allows one to fight decoherence even
further. Consequently, to preserve a quantum state for longer time
periods before all the information is lost due to decoherence effects,
one can perform a state transfer sequence by driving the cavity with
two laser fields which will induce transitions between the states
of the two atoms inside the cavity. To preserve the qubit state of
atom 1 on a longer time scale, one can keep driving the cavity over
many cycles which will improve its coherence time. Of course, we remind
the reader that we are limited by the adiabatic regime in the number
of times one can perform the state transfer sequence.
\par\end{flushleft}

We end this section by considering a many-cycle evolution of the four-level
system using experimental values for the relevant physical parameters.
Following \cite{key-49}, we take the value for the cavity coupling
constant to be 
\begin{equation}
g=50\times10^{3}Hz,\label{eq:4.5.2}
\end{equation}
the cavity damping rate is chosen to be
\begin{equation}
\kappa=10^{5}Hz.\label{eq:4.5.3}
\end{equation}
To ensure that the conditions in table (\ref{Table-1:-List}) are
satisfied, we choose the detuning frequency to be given by 
\begin{equation}
\Delta=10^{6}Hz,\label{eq:4.5.4}
\end{equation}
the thermal occupation number
\begin{equation}
N_{th}=1,\label{eq:4.5.5}
\end{equation}
and $G$ to be 
\begin{equation}
G=10^{5}Hz.\label{eq:4.4.5-1}
\end{equation}
In plotting the fidelity, we use the expression obtained in (\ref{eq:4.4.31})
rewritten using the dimensionless parameter $\kappa t_{f}$:
\begin{align}
F & =\left|\alpha\right|^{4}+\left|\beta\right|^{4}+2\left|\alpha\right|^{2}\left|\beta\right|^{2}\left\{ \cos\left(\gamma_{d}\left(t_{f}\right)\right)\right.\nonumber \\
 & \left.\exp\left(-\frac{N_{th}\left(N_{th}+1\right)}{2}\left(\frac{g^{2}}{\Delta}\right)^{2}\frac{\left(\kappa t_{f}\right)^{3}\left[4n^{2}\pi^{2}+2\kappa t_{f}\left(e^{-\kappa t_{f}}-1\right)+\left(\kappa t_{f}\right)^{2}\right]}{\kappa^{2}\left(4n^{2}\pi^{2}+\left(\kappa t_{f}\right)^{2}\right)^{2}}\right)\right\} \label{eq:4.4.5-2}
\end{align}
and for the case where we don't perform the state transfer, the expression
for the fidelity was found in (\ref{eq:4.4.37}) to be given by 
\begin{equation}
F=\left|\alpha\right|^{4}+\left|\beta\right|^{4}+2\left|\alpha\right|^{2}\left|\beta\right|^{2}\exp\left(-\frac{N_{th}\left(N_{th}+1\right)}{\kappa^{2}}\left(\frac{g^{2}}{\Delta}\right)^{2}\left(e^{-\kappa t_{f}}-1+\kappa t_{f}\right)\right).\label{eq:4.5.7}
\end{equation}

\begin{figure}
\begin{centering}
\includegraphics[scale=0.85]{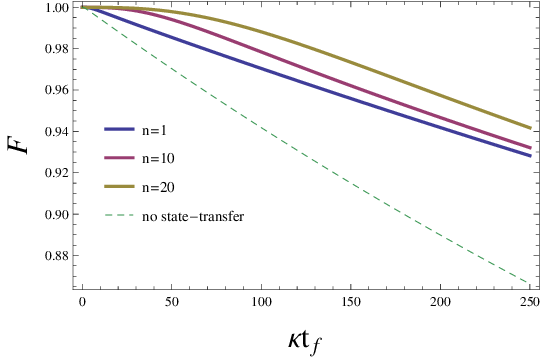}
\par\end{centering}

\begin{centering}
\caption{\label{fig:Fidelity-using-experimental-1}\ Fidelity using experimental
values for the system parameters}
\textit{Plot of the fidelity as a function of $t_{f}$ (the total
state transfer time) using the experimental values of (\ref{eq:4.5.2})
to (\ref{eq:4.5.5}). We set the parameters $\alpha=\beta=\frac{1}{2}$.
However, each curve corresponds to a particular value of the state
transfer cycles ($n$) illustrated in the plot legend. The dashed
curve corresponds to the fidelity that one would get if no state transfer
was performed. Note that we set the phase $\phi\left(t\right)=0$
so that the closed system Berry's phase ($\gamma_{d}\left(t_{f}\right)$)
vanishes.}
\par\end{centering}

\end{figure}

As expected, figure (\ref{fig:Fidelity-using-experimental-1}) shows
that as we increase the number of state-transfer cycles ($n$), the
fidelity remains closer to unity for longer time scales. Also, the
dashed curve representing dephasing effects for the case where no
state-transfer is performed decays much more quickly than the state-transfer
protocol curves (thick curves). Consequently, we see that our state-transfer
protocol offers a significant advantage for fighting dephasing effects
when using experimental values for the parameters of interest.  

\begin{figure}
\begin{centering}
\includegraphics[scale=0.85]{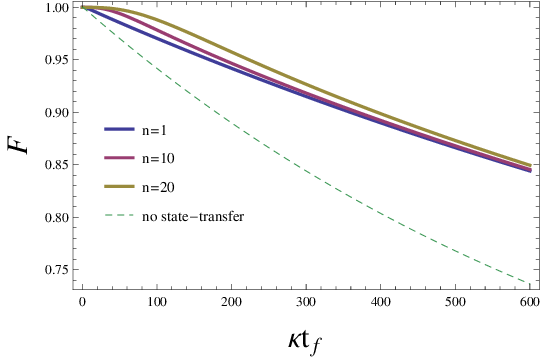}
\par\end{centering}

\caption{\ Fidelity using experimental values for the system parameters with
extended state-transfer time}

\begin{centering}
\textit{Plot of the fidelity as a function of $t_{f}$ (the total
state transfer time) using the experimental values of (\ref{eq:4.5.2})
to (\ref{eq:4.5.5}). We set the parameters $\alpha=\beta=\frac{1}{2}$.
We set the phase $\phi\left(t\right)=0$ so that the closed system
Berry's phase ($\gamma_{d}\left(t_{f}\right)$) vanishes. We see that
for longer times, the improvement of the fidelity by increasing the
number of state transfer cycles decreases. }
\par\end{centering}

\end{figure}

\begin{figure}
\begin{centering}
\includegraphics[scale=0.85]{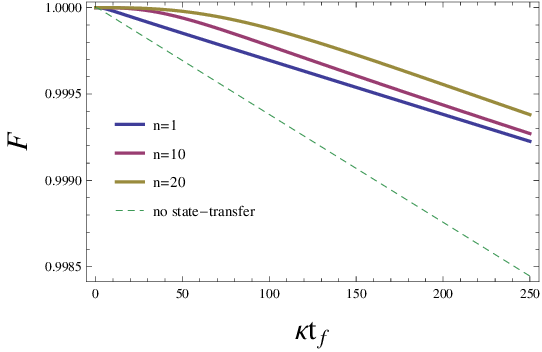}
\par\end{centering}

\caption{\label{fig:Fidelity-using-experimental}\ Fidelity using experimental
values for the system parameters with increased detuning frequency}

\centering{}\textit{Plot of the fidelity as a function of $t_{f}$
(the total state transfer time) using the experimental values of (\ref{eq:4.5.2})
to (\ref{eq:4.5.5}). We set the parameters $\alpha=\beta=\frac{1}{2}$.
We set the phase $\phi\left(t\right)=0$ so that the closed system
Berry's phase ($\gamma_{d}\left(t_{f}\right)$) vanishes. We increased
the value of the detuning frequency to $\Delta=10^{7}Hz$. It is thus
observed that for larger detuning frequencies, the fidelity remains
closer to unity for longer time scales. }
\end{figure}

\section{\label{sec:Short-time-Expansion}Short-time Expansion}

\begin{flushleft}
Upon careful consideration of the short time limit of the fidelity
for our state transfer protocol (Eq. (\ref{eq:4.4.33})), the $t_{f}^{3}$
dependence of the exponent is analogous to the case of a Carr-Purcell
sequence for coherence control \cite{key-45}. This scheme uses a
sequence of Pi-pulses to fight decoherence (extend the coherence time
of the qubit). In the short-time limit, the average of the spin 1/2
matrix $\left\langle \sigma_{+}\right\rangle $ (also known as spin-coherence)
also exhibits a super-exponential $\sim e^{-\left(t_{f}/\tau\right)^{3}}$
decay. However, the CPMG cycle uses a sequence of evenly spaced Pi-pulses
in order to extend the life of the qubit \cite{key-46}. To further
extend the life of the qubit, an optimal sequence of Pi-pulses was
proposed by G.S. Uhrig \cite{key-47}. Based on these findings, we
propose a scheme for finding an optimal path to perform our state
transfer protocol via a short time expansion.
\par\end{flushleft}

\begin{flushleft}
The first step is to perform a short time expansion on $C\left(t_{f}\right)\equiv\left\langle T_{t}e^{i\hat{X}\left(t_{f}\right)}\right\rangle $.
We remind the reader that 
\begin{equation}
\hat{X}\left(t\right)=\int_{0}^{t}dt'\cos\left(2\theta\left(t',t_{f}\right)\right)\delta\hat{\omega}_{1}\left(t'\right),\label{eq:4.6.1}
\end{equation}
where 
\begin{equation}
\delta\hat{\omega}_{1}=\frac{g^{2}}{\Delta}\hat{d}^{\dagger}\hat{d}.\label{eq:4.6.2}
\end{equation}
Notice that $\theta$ is also a function of $t_{f}$ which is reflected
in (\ref{eq:4.6.1}). For example, if we had a circular evolution,
we know that $\cos\left(2\theta\left(t,t_{f}\right)\right)=\cos\left(\frac{4n\pi t}{t_{f}}\right)$.
For simplicity let us also define the function 
\begin{equation}
s\left(t,t_{f}\right)\equiv\cos\left(2\theta\left(t,t_{f}\right)\right).\label{eq:4.6.3}
\end{equation}
Also, for a purely symmetric evolution the mean $\left\langle \hat{X}\left(t_{f}\right)\right\rangle =0$.
Performing the same set of approximations as we did in previous sections,
we can write $C\left(t_{f}\right)\approx e^{-\frac{1}{2}\left\langle T_{t}\hat{X}^{2}\left(t_{f}\right)\right\rangle }$where
we have that
\begin{equation}
\left\langle T_{t}\hat{X}^{2}\left(t_{f}\right)\right\rangle =\left(\frac{g^{2}}{\Delta}\right)^{2}\int_{0}^{t_{f}}dt_{1}\int_{0}^{t_{f}}dt_{2}s\left(t_{1},t_{f}\right)s\left(t_{2},t_{f}\right)\left\langle \hat{d}^{\dagger}\left(t_{1}\right)\hat{d}\left(t_{1}\right)\hat{d}^{\dagger}\left(t_{2}\right)\hat{d}\left(t_{2}\right)\right\rangle .\label{eq:4.6.4}
\end{equation}
Notice that we are keeping things completely general since we have
not specified a function for $s\left(t,t_{f}\right)$. Let us define
\begin{equation}
S_{\delta\omega}\left(t_{1}-t_{2}\right)\equiv\left(\frac{g^{2}}{\Delta}\right)^{2}\left\langle \hat{d}^{\dagger}\left(t_{1}\right)\hat{d}\left(t_{1}\right)\hat{d}^{\dagger}\left(t_{2}\right)\hat{d}\left(t_{2}\right)\right\rangle .\label{eq:4.6.5}
\end{equation}
We can introduce its Fourier transform $S_{\delta\omega}\left(t\right)=\int_{-\infty}^{\infty}\frac{d\omega}{2\pi}e^{i\omega t}S_{\delta\omega}\left(\omega\right)$
into equation (\ref{eq:4.6.4}) and rewrite it as 
\begin{equation}
\left\langle T_{t}\hat{X}^{2}\left(t_{f}\right)\right\rangle =\int_{-\infty}^{\infty}\frac{d\omega}{2\pi}S_{\delta\omega}\left(\omega\right)\int_{0}^{t_{f}}dt_{1}e^{i\omega t_{1}}s\left(t_{1},t_{f}\right)\int_{0}^{t_{f}}dt_{2}e^{-i\omega t_{2}}s\left(t_{2},t_{f}\right).\label{eq:4.6.6}
\end{equation}
Now let us define 
\begin{equation}
f\left(\omega,t_{f}\right)=\int_{0}^{t_{f}}dte^{i\omega t}s\left(t,t_{f}\right).\label{eq:4.6.7}
\end{equation}
With this definition and defining the filter function $\mathcal{F}\left(\omega,t_{f}\right)\equiv\left|f\left(\omega,t_{f}\right)\right|^{2}$,
we can rewrite equation (\ref{eq:4.6.6}) as
\begin{equation}
\left\langle T_{t}\hat{X}^{2}\left(t_{f}\right)\right\rangle =\int_{-\infty}^{\infty}\frac{d\omega}{2\pi}S_{\delta\omega}\left(\omega\right)\mathcal{F}\left(\omega,t_{f}\right).\label{eq:4.6.8}
\end{equation}

\par\end{flushleft}

We now use a similar procedure to what was done in \cite{key-47}
in order to find an optimized sequence of Pi-pulses. Instead of choosing
the function $\cos\left(2\theta\left(t,t_{f}\right)\right)=\cos\left(\frac{2\pi nt}{t_{f}}\right)$
which corresponds to an evolution over n cycles, we will represent
$\cos\left(2\theta\left(t,t_{f}\right)\right)$ by a Fourier series
with a finite frequency cutoff given by 
\begin{equation}
\cos\left[2\theta\left(t,t_{f}\right)\right]=\sum_{m=1}^{n}\left(A_{m}\cos\left(\frac{2\pi mt}{t_{f}}\right)+B_{m}\sin\left(\frac{2\pi mt}{t_{f}}\right)\right).\label{eq:4.6.9}
\end{equation}
Using (\ref{eq:4.6.7}) and (\ref{eq:4.6.9}), we find that 
\begin{equation}
\bar{f}\left(z\right)=z\sum_{m=1}^{n}\frac{\left(1-e^{iz}\right)\left(izA_{m}-2\pi mB_{m}\right)}{z^{2}-4\pi^{2}m^{2}}.\label{eq:4.6.10}
\end{equation}
where $z=\omega t_{f}$. In terms of the function $\bar{f}\left(z\right)$,
we can rewrite (\ref{eq:4.6.8}) as 
\begin{equation}
\left\langle T_{\tau}\hat{X}^{2}\left(t_{f}\right)\right\rangle =\int_{-\infty}^{\infty}\frac{d\omega}{2\pi}\frac{S_{\delta_{\omega}}\left(\omega\right)}{\omega^{2}}\left|\bar{f}\left(\omega t_{f}\right)\right|^{2}.\label{eq:4.6.11}
\end{equation}
Note that the coefficients $A_{m}$ and $B_{m}$ must be chosen such
that they ensure that (\ref{eq:4.6.9}) is bounded between $\pm1$.
For now, we will ignore this and focus on determining them to make
the first $n$ derivatives of (\ref{eq:4.6.10}) vanish (see the procedure
below). In order to have the fidelity as close to unity as possible
for short time scales, we perform a short-time expansion by setting
as many derivatives of (\ref{eq:4.6.10}) to zero (when $z=0$). We
give here the first few results:
\begin{equation}
\bar{f}^{\left(1\right)}\left(z\right)\left|_{z=0}\right.=0,\label{eq:4.6.12}
\end{equation}
\begin{equation}
\bar{f}^{\left(2\right)}\left(z\right)\left|_{z=0}\right.=-\frac{1}{\pi}\sum_{m=1}^{n}\frac{B_{m}}{m},\label{eq:4.6.13}
\end{equation}
\begin{equation}
\bar{f}^{\left(3\right)}\left(z\right)\left|_{z=0}\right.=-\frac{3}{2\pi^{2}}\sum_{m=1}^{n}\left(\frac{A_{m}}{m^{2}}-\frac{B_{m}}{m}\right),\label{eq:4.6.14}
\end{equation}
\begin{equation}
\bar{f}^{\left(4\right)}\left(z\right)\left|_{z=0}\right.=\frac{i}{\pi^{3}}\sum_{m=1}^{n}\left(-3\pi\frac{A_{m}}{m^{2}}-3\frac{B_{m}}{m^{3}}+2\pi^{2}\frac{B_{m}}{m}\right),\label{eq:4.6.15}
\end{equation}
\begin{equation}
\bar{f}^{\left(5\right)}\left(z\right)\left|_{z=0}\right.=-\frac{5}{2\pi^{4}}\sum_{m=1}^{n}\left(\frac{A_{m}}{m^{4}}-2\pi^{2}\frac{A_{m}}{m^{2}}-3\pi\frac{B_{m}}{m^{3}}+\pi^{3}\frac{B_{m}}{m}\right),\label{eq:4.6.16}
\end{equation}
\begin{equation}
\bar{f}^{\left(6\right)}\left(z\right)\left|_{z=0}\right.=\frac{3i}{2\pi^{5}}\sum_{m=1}^{n}\left(-15\pi\frac{A_{m}}{m^{4}}+5\pi^{3}\frac{A_{m}}{m^{2}}-15\frac{B_{m}}{m^{5}}+10\pi^{2}\frac{B_{m}}{m^{3}}-2\pi^{4}\frac{B_{m}}{m}\right).\label{eq:4.6.17}
\end{equation}
We want all the derivatives up to $n^{th}$ order to vanish. From
(\ref{eq:4.6.12}) to (\ref{eq:4.6.17}), we see that if $n$ is even,
then the following sufficient, but not necessary, conditions must
be satisfied:
\begin{equation}
\sum_{m=1}^{n}\frac{A_{m}}{m^{j}}=0,\,\forall\, j\,\in\left\{ 2,4,...,n-2\right\} ,\label{eq:4.6.18}
\end{equation}
and
\begin{equation}
\sum_{m=1}^{n}\frac{B_{m}}{m^{l}}=0,\,\forall\, l\,\in\left\{ 1,3,...,n-1\right\} .\label{eq:4.6.19}
\end{equation}
On the other hand, if $n$ is odd, then the following conditions must
be satisfied:
\begin{equation}
\sum_{m=1}^{n}\frac{A_{m}}{m^{j}}=0,\,\forall\, j\,\in\left\{ 2,4,...,n-1\right\} ,\label{eq:4.6.20}
\end{equation}
and
\begin{equation}
\sum_{m=1}^{n}\frac{B_{m}}{m^{l}}=0,\,\forall\, l\,\in\left\{ 1,3,...,n-2\right\} .\label{eq:4.6.21}
\end{equation}

Since we mentioned that the function $\cos\left[2\theta\left(t,t_{f}\right)\right]$
must be bounded between $\pm1$, this imposes constraints on the coefficients
$A_{m}$ and $B_{m}$ for all times $t\,\in\left[0,t_{f}\right]$.
For $t=0$, we use $\cos\left[2\theta\left(0,t_{f}\right)\right]=1$
to conclude that 
\begin{equation}
\sum_{m=1}^{n}A_{m}=1.\label{eq:4.6.22}
\end{equation}
Note that $\cos\left[2\theta\left(t_{f},t_{f}\right)\right]=1$ will
also be satisfied in this case. We could include constraint equations
in equations (\ref{eq:4.6.18}) to (\ref{eq:4.6.21}) ensuring that
$\cos\left[2\theta\left(t,t_{f}\right)\right]$ always has the correct
bound, but for the examples we consider below these will not be necessary. 

At this stage it is important to realize that there are many types
of functions that can satisfy the conditions (\ref{eq:4.6.18}) to
(\ref{eq:4.6.22}). The procedure will always be the same: We must
solve linear equations for the coefficients $A_{j}$ and $B_{j}$
that come from (\ref{eq:4.6.18}) to (\ref{eq:4.6.22}). However,
these will not be sufficient to fix all the coefficients. So we could
use the remaining freedom in the coefficients $A_{j}$ and $B_{j}$
by inserting the expanded function $\cos\left[2\theta\left(t,t_{f}\right)\right]$
into (\ref{eq:4.6.11}) and minimizing the result.

To see how this works, we will consider an example. To begin, we set
$n=2$. So from (\ref{eq:4.6.19}) and (\ref{eq:4.6.22}) we require
that 
\begin{equation}
B_{1}+\frac{1}{2}B_{2}=0,\label{eq:4.6.23}
\end{equation}
and
\begin{equation}
A_{1}+A_{2}=1.\label{eq:4.6.24}
\end{equation}
Using the above results and (\ref{eq:4.6.9}), we can write 
\begin{equation}
\cos\left[2\theta\left(t,t_{f}\right)\right]=A_{1}\left[\cos\left(\frac{2\pi t}{t_{f}}\right)-\cos\left(\frac{4\pi t}{t_{f}}\right)\right]+B_{2}\left[\sin\left(\frac{4\pi t}{t_{f}}\right)-\frac{1}{2}\sin\left(\frac{2\pi t}{t_{f}}\right)\right]+\cos\left(\frac{4\pi t}{t_{f}}\right).\label{eq:4.6.25}
\end{equation}
One possible approach that we can take is to set $B_{2}=0$. Also,
the reader is reminded that for cavity-photon shot noise, (\ref{eq:4.6.8})
can be written in the time domain as 
\begin{equation}
\left\langle T_{t}\hat{X}\left(t_{f}\right)\hat{X}\left(t_{f}\right)\right\rangle =\left(\frac{g^{2}}{\Delta}\right)^{2}N_{th}\left(N_{th}+1\right)\int_{0}^{t_{f}}dt_{1}dt_{2}\cos\left[2\theta\left(t_{1},t_{f}\right)\right]\cos\left[2\theta\left(t_{2},t_{f}\right)\right]e^{-\kappa\left|t_{1}-t_{2}\right|}.\label{eq:4.6.26}
\end{equation}
Using (\ref{eq:4.6.25}), (\ref{eq:4.6.26}) can be calculated and
the result will depend on the parameters $A_{1}$ and $B_{2}$ (the
expression is a bit laborious and will not be written here). However,
In the short-time limit, we find 
\begin{equation}
\left\langle T_{t}\hat{X}\left(t_{f}\right)\hat{X}\left(t_{f}\right)\right\rangle =\left(\frac{g^{2}}{\Delta}\right)^{2}N_{th}\left(N_{th}+1\right)\frac{\kappa\left(1-2A_{1}+5A_{1}^{2}+2B_{2}^{2}\right)t_{f}^{3}}{16\pi^{2}}+\mathcal{O}\left(t_{f}^{4}\right).\label{eq:4.6.27}
\end{equation}
Since the $B_{2}$ coefficient only adds a positive contribution,
we clearly see that choosing $B_{2}=0$ was the best choice. Note
that the above term cannot be made to vanish since the equation $1-2A_{1}+5A_{1}^{2}=0$
has imaginary solutions. However, we can find the coefficient such
that $f\left(A_{1}\right)=1-2A_{1}+5A_{1}^{2}$ is as small as possible.
The answer is 
\begin{equation}
A_{1}=\frac{1}{5}.\label{eq:4.6.28}
\end{equation}
Consequently, for the class of functions where $B_{1}=B_{2}=0$, the
optimal path to get the best fidelity for short times is given by
\begin{equation}
\cos\left[2\theta\left(t,t_{f}\right)\right]=\frac{1}{5}\left[\cos\left(\frac{2\pi t}{t_{f}}\right)+4\cos\left(\frac{4\pi t}{t_{f}}\right)\right].\label{eq:4.6.29}
\end{equation}
It is also worth mentioning that the function given by (\ref{eq:4.6.29})
is bounded between $\pm1$ and so can accurately describe a state
transfer protocol. 

\begin{figure}
\begin{centering}
\includegraphics[scale=0.65]{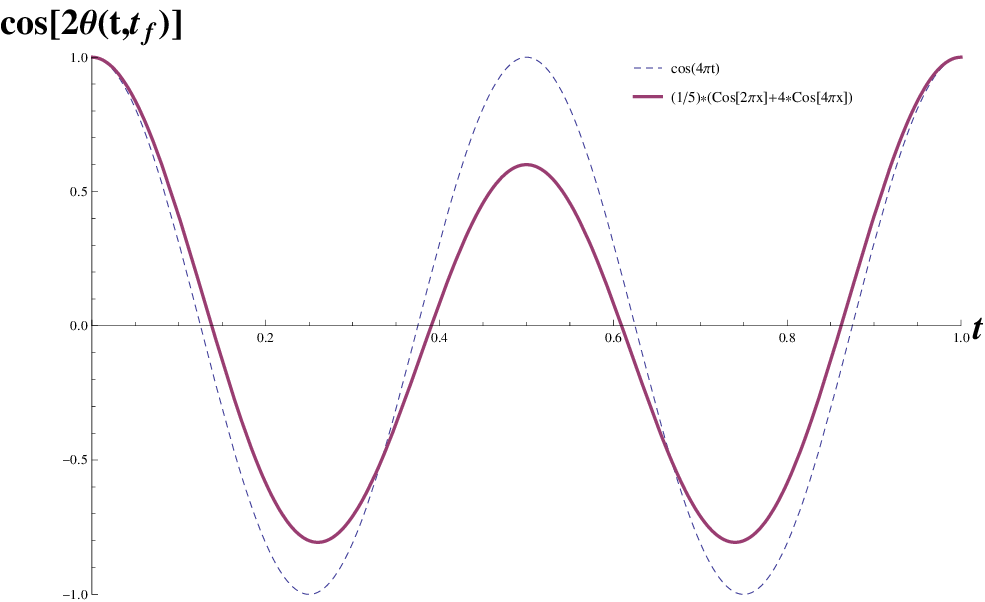}
\par\end{centering}

\caption{\label{fig:Non-trivial-path-for}\ Non-trivial path for $n=2$}

\begin{centering}
\textit{Plot of the path described by (\ref{eq:4.6.29}) keeping only
the second harmonic in the Fourier expansion. We chose $t_{f}=1$. }
\par\end{centering}

\end{figure}

It is also worthwhile drawing a plot that compares the previous solution
we took ($\cos\left[2\theta\left(t,t_{f}\right)\right]=\cos\left(\frac{4\pi t}{t_{f}}\right)$)
to the one found in (\ref{eq:4.6.29}). 

\begin{center}
\begin{figure}
\begin{centering}
\includegraphics[scale=0.8]{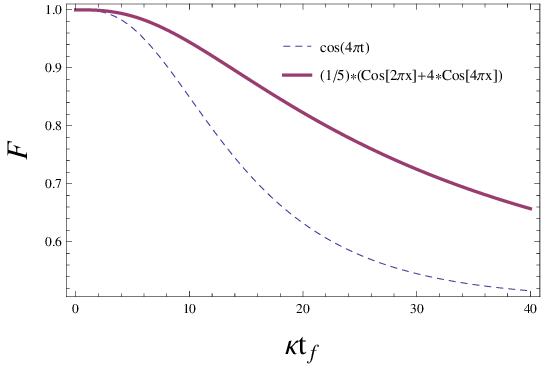}
\par\end{centering}

\caption{\label{fig:Optimal-path-comparison}\ Optimal path comparison}

\centering{}\textit{Plot of the fidelity as a function of $t_{f}$
comparing a circular evolution to the path found in (\ref{eq:4.6.29}).
We have chosen the values $\alpha=\beta=\frac{1}{2}$, $N=1$, $\frac{g^{2}}{\Delta}=0.1$
and $\kappa=1$. Note that we also turned off the phase $\phi\left(t\right)$
so that the closed system Berry's phase ($\gamma_{d}\left(t_{f}\right)$)
vanishes.}
\end{figure}

\par\end{center}

Our optimization procedure applied to the case where $n=2$ gives
rise to a non-trivial path, in the sense that the state is never fully
transferred to $\left.|g_{2}\right\rangle $ (see figure (\ref{fig:Non-trivial-path-for})).
To understand this, we follow the noise correlation argument that
we gave in the paragraph below (\ref{eq:3.1.18}). There we argued
that it is most advantageous to do the adiabatic evolution in such
a way to spend as much time as possible with the state (\ref{eq:2.2.26})
being in the equal superposition $\left.|g_{1}\right\rangle +\left.|g_{2}\right\rangle $.
This means that it would be favorable to avoid having the state (\ref{eq:2.2.26})
being all $\left.|g_{2}\right\rangle $ so as to maximise the presence
of simultaneous $\delta\hat{\omega}_{1}$ and $\delta\hat{\omega}_{2}$
noise. Increasing $n$, it is expected that similar behavior will
be exhibited where optimal solutions would differ from the usual linear
paths ($\theta\left(t\right)=\frac{n\pi t}{t_{f}}$) that we considered
throughout most of this thesis.

As a last note, since for both the paths $\cos\left[2\theta\left(t,t_{f}\right)\right]=\cos\left(\frac{4\pi t}{t_{f}}\right)$
and $\cos\left[2\theta\left(t,t_{f}\right)\right]=\frac{1}{5}\left[\cos\left(\frac{2\pi t}{t_{f}}\right)+4\cos\left(\frac{4\pi t}{t_{f}}\right)\right]$,
$\left\langle \tilde{\hat{\Pi}}_{0d}\left(t\right)\right\rangle \sim e^{-\left(t/\tau\right)^{3}}$
in the short time limit $\kappa t_{f}\ll1$, we can compute the ratio
of the decay times for both paths. Defining $\tau_{n=2}$ as the decay
time for the path $\cos\left[2\theta\left(t,t_{f}\right)\right]=\cos\left(\frac{4\pi t}{t_{f}}\right)$
and $\tau_{opt}$ as the decay time for the path in (\ref{eq:4.6.29}),
we can use (\ref{eq:4.4.33}) and (\ref{eq:4.6.27}) to show that
\begin{equation}
\frac{\tau_{opt}}{\tau_{n}}=\left(\frac{5}{4}\right)^{1/3}\label{eq:4.5.30}
\end{equation}

\section{\label{sec:Phase-gate}Phase gate}

Throughout this thesis, the whole idea behind considering our state-transfer
protocol was to perform a phase gate. We could have avoided dephasing
effects altogether if the four-level atom was kept in a superposition
state given by a linear combination of $\left.|g_{1}\right\rangle +\left.|g_{2}\right\rangle $,
since this state is insensitive to the noise. The goal here is as
follows: for the closed-system case, we know that when performing
our state transfer protocol, the state of interest at time $t_{i}=0$
will have the form 
\begin{equation}
\left.|\psi\left(t\right)\right\rangle =\alpha\left.|0\right\rangle +\beta\left.|g_{1}\right\rangle ,\label{eq:4.7.1}
\end{equation}
and that at the end of the state transfer 
\begin{equation}
\left.|\psi\left(t_{f}\right)\right\rangle =\alpha\left.|0\right\rangle +\beta e^{i\gamma_{d}\left(t_{f}\right)}\left.|g_{1}\right\rangle .\label{eq:4.7.2}
\end{equation}
We remind the reader that for a general path $\theta\left(t\right)$,
the geometric phase is given by 
\begin{equation}
\gamma_{d}\left(t\right)=\int_{0}^{t}\dot{\phi}\left(t'\right)\sin^{2}\theta\left(t'\right)dt',\label{eq:4.7.3}
\end{equation}
Now suppose that at the end of the state transfer protocol we want
$\gamma_{d}\left(t_{f}\right)$ to be given by a fixed phase $\alpha$.
We can rewrite (\ref{eq:4.7.3}) as 
\begin{equation}
\alpha=\gamma_{d}\left(t_{f}\right)=\int\sin^{2}\theta d\phi.\label{eq:4.7.4}
\end{equation}
We can think of (\ref{eq:4.7.4}) as being proportional to the area
projected onto the x-y plane by a spin-half vector $\vec{S}$ executing
a closed-path evolution in 3-d space. 

\begin{center}
\begin{figure}
\begin{centering}
\includegraphics[scale=0.7]{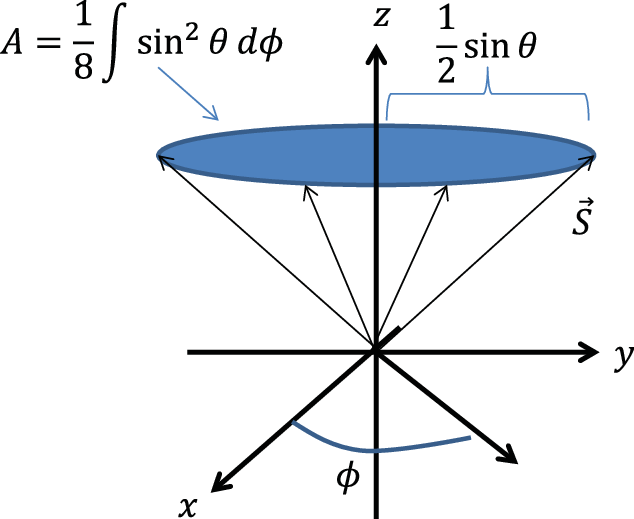}
\par\end{centering}

\caption{\ Area enclosed by geometric phase}

\begin{doublespace}
\begin{centering}
\textit{Figure showing the relation between the area projected onto
the x-y plane of a spin-half vector to the geometric phase.}
\par\end{centering}
\end{doublespace}

\end{figure}

\par\end{center}

In order to see this, we start by writing $\left\langle \overrightarrow{S}\right\rangle $
in its three dimensional components 
\begin{align}
\left\langle \overrightarrow{S}\right\rangle  & =\left(\left\langle S_{x}\right\rangle ,\left\langle S_{y}\right\rangle ,\left\langle S_{z}\right\rangle \right)\nonumber \\
 & =\frac{1}{2}\left(\sin\theta\cos\phi,\sin\theta\sin\phi,\cos\theta\right).\label{eq:4.7.5}
\end{align}
The area projected onto the x-y plane is given by 
\begin{equation}
dA_{\perp}=\frac{1}{2}\vec{S}_{perp}dt,\label{eq:4.7.6}
\end{equation}
where 
\begin{equation}
\vec{S}_{perp}=\left|\vec{S}_{\perp}\times\dot{\vec{S}}_{\perp}\right|\label{eq:4.7.6-1}
\end{equation}
Then a straightforward calculation shows that 
\begin{equation}
\left|\vec{S}_{\perp}\times\dot{\vec{S}}_{\perp}\right|=\frac{1}{4}\dot{\phi}\sin^{2}\theta.\label{eq:4.7.7}
\end{equation}
Thus we conclude that 
\begin{equation}
A_{\perp}=\frac{1}{8}\int\sin^{2}\theta d\phi.\label{eq:4.7.8}
\end{equation}
If we identify $\alpha$ with $A_{\perp}$, (\ref{eq:4.7.8}) shows
that we can choose many paths $\theta\left(t\right)$ that give the
same area projected onto the x-y plane. When noise is present in our
system, there will be dephasing effects which will reduce the fidelity
of our state transfer protocol (as was shown in (\ref{eq:4.4.31}))
but it will not modify the geometric phase. Consequently, the idea
will be to find an optimal path for a given cutoff frequency (as was
done using the short-time expansion protocol of the previous section)
in order to minimize dephasing effects. The optimal solution will
then fix the function $\theta\left(t\right)$. Thus to obtain the
desired phase $\alpha=\gamma_{d}\left(t_{f}\right)$ at the end of
the state transfer protocol, the function $\phi\left(t\right)$ is
chosen such that the integral in (\ref{eq:4.7.3}) reduces to $\alpha$.
To make things clearer, we give two examples that illustrate the general
procedure. 

For $n=1$ in (\ref{eq:4.6.9}), then it is easy to check that the
optimal solution which minimizes $\left\langle T_{\tau}\hat{X}^{2}\left(t_{f}\right)\right\rangle $
for short times is given by 
\begin{equation}
\cos\left[2\theta\left(t,t_{f}\right)\right]=\cos\left(\frac{2\pi t}{t_{f}}\right).\label{eq:4.7.9}
\end{equation}
If we choose the function $\phi\left(t\right)$ to satisfy
\begin{equation}
\dot{\phi}\left(t\right)=ct,\label{eq:4.7.10}
\end{equation}
which is simply a linear function of time with slope $c$, the the
integral in (\ref{eq:4.7.3}) is easy to compute allowing us to fix
the parameter $c$ to be 
\begin{equation}
c=\frac{4\alpha}{t_{f}^{2}}.\label{eq:4.7.11}
\end{equation}
Consequently, we are left with
\begin{equation}
\phi\left(t\right)=\frac{2\alpha}{t_{f}^{2}}t^{2},\label{eq:4.7.12}
\end{equation}
where we choose the constant such that at $t=0$, $\phi\left(0\right)=0$.
Thus, the function $\phi\left(t\right)$ chosen in (\ref{eq:4.7.12})
will ensure that we get the desired phase at the end of the state
transfer protocol when $\theta\left(t\right)$ is given by (\ref{eq:4.7.9}).

For $n=2$ in (\ref{eq:4.6.9}), we showed that the optimal solution
was given by (\ref{eq:4.6.29}). With the same quadratic function
for the phase $\phi\left(t\right)$ as in (\ref{eq:4.7.10}), the
integral in (\ref{eq:4.7.3}) is now found to be 
\begin{align}
\alpha=\gamma_{d}\left(t\right) & =c\int_{0}^{t}t\sin^{2}\left\{ \frac{1}{2}\arccos\left[\frac{1}{5}\left[\cos\left(\frac{2\pi t}{t_{f}}\right)+4\cos\left(\frac{4\pi t}{t_{f}}\right)\right]\right]\right\} dt'\nonumber \\
 & =\frac{ct_{f}^{2}}{4}.\label{eq:4.7.13}
\end{align}
Just as in (\ref{eq:4.7.11}), we conclude that 
\begin{equation}
c=\frac{4\alpha}{t_{f}^{2}}.\label{eq:4.7.14}
\end{equation}
Thus although the path is different, the area of the integral (\ref{eq:4.7.3})
is the same for both paths where $n=1$ and $n=2$ which means that
we can use the same function $\phi\left(t\right)$ in both cases to
obtain the desired phase during the state transfer protocol. Note
however that we could choose a different functional form for the phase
$\phi\left(t\right)$ which could also be well suited for performing
a phase gate. For example, if we had chosen the simple constant 
\begin{equation}
\dot{\phi}\left(t\right)=c,\label{eq:4.7.15}
\end{equation}
then the integral of (\ref{eq:4.7.3}) would evaluate to be 
\begin{align}
\alpha=\gamma_{d}\left(t\right) & =c\int_{0}^{t}\sin^{2}\left\{ \frac{1}{2}\arccos\left[\frac{1}{5}\left[\cos\left(\frac{2\pi t}{t_{f}}\right)+4\cos\left(\frac{4\pi t}{t_{f}}\right)\right]\right]\right\} dt'\nonumber \\
 & =\frac{ct_{f}}{2}.\label{eq:4.7.16}
\end{align}
In this case we would have 
\begin{equation}
\phi\left(t\right)=\frac{2\alpha}{t_{f}}t.\label{eq:4.7.17}
\end{equation}
In general, after finding the optimal path for a particular value
of the frequency cutoff, we are free to choose a convenient functional
dependence for the phase $\phi\left(t\right)$ such that the integral
in (\ref{eq:4.7.3}) is as simple as possible. As was shown above,
for the case where $n=1$ or $n=2$, linear or quadratic functions
of time are convenient choices.

\section{Summary}

We considered a system where a four-level atom was coupled to a cavity
being driven by two coherent drives (each with its own detuning frequency).
These created transitions between the atomic states allowing us to
perform our state transfer protocol. In order to avoid off-resonant
terms, we required that $\left|\Delta_{i}\right|\gg G$. To get the
desired Hamiltonian that would allow us to calculate dephasing effects,
we first wrote our Hamiltonian in a superadiabatic basis and then
performed a Schrieffer-Wolff transformation. In doing so, table (\ref{Table-1:-List})
gave a list of conditions that needs to be satisfied in order for
our theory to be valid. The Schrieffer-Wolff transformation enabled
us to get the appropriate noise terms that allowed us to apply the
methods of chapter 3 to calculate the fidelity given by Eq. (\ref{eq:4.4.31})
which depended on the number of cycles for the state-transfer protocol.
The plot of figure (\ref{fig:n-cycle}) clearly showed that performing
the state-transfer protocol over many cycles extended the coherence
time of the atomic qubit state. It also demonstrated that the lifetime
of the qubit was greatly enhanced by performing the state-transfer
protocol versus leaving the system in its qubit state. It was also
noted that one had to be careful not to break the adiabatic criteria
by choosing an arbitrarily large value for the number of state-transfer
cycles. Using known experimental values for the cavity and drive parameters,
the plot of figure (\ref{fig:Fidelity-using-experimental-1}) showed
the same behavior as (\ref{fig:n-cycle}) thus enforcing the idea
that our state-transfer protocol is a viable way of fighting decoherence
effects. By increasing the value of the detuning frequency, figure
(\ref{fig:Fidelity-using-experimental}) showed that the fidelity
remains closer to unity for longer times.

We addressed the issue of finding an optimal path for our state-transfer
protocol. Using a short-time expansion method along with a class of
functions described by Eq. (\ref{eq:4.6.9}), we saw that for $n\geq2$,
a linear combination of cosine and sine functions with coefficients
determined by the equations of (\ref{eq:4.6.18}) to (\ref{eq:4.6.21}),
and by ensuring that $\cos\left[2\theta\left(t,t_{f}\right)\right]$
is appropriately normalized, gave the optimal path. Finally, we showed
how one could choose a phase $\phi\left(t\right)$ to perform a phase
gate. Choosing (\ref{eq:4.7.12}) would allow us to get the desired
geometric phase (using the path obtained in (\ref{eq:4.6.29})) but
that other convenient choices were also possible, such as in (\ref{eq:4.7.17}).
One very important remark is that we performed a phase gate by going
from the state (\ref{eq:4.7.1}) to (\ref{eq:4.7.2}) and so for this
particular state-transfer protocol, dephasing effects are inevitable. 

\newpage{}

\chapter{Conclusion}

In conclusion, we developed theoretical methods for calculating dephasing
effects when a four-level system (following our state-transfer protocol)
is coupled to a quantum dissipative environment. The first step is
to write the Hamiltonian in a superadiabatic basis by going into a
rotating frame with the appropriate unitary transformation given by
Eq. (\ref{eq:2.3.3}). Then, we performed a secular approximation
which amounts to dropping all the off-diagonal terms of the superadiabatic
Hamiltonian. This is valid as long as the conditions $\frac{\dot{\theta}}{G}\ll1$,
$\frac{\dot{\phi}}{G}\ll1$ and $\frac{\left|\delta\hat{\omega}_{i}\right|}{G}\ll1$
are satisfied. With a purely diagonal Hamiltonian, the phase of the
``0d'' component of the density matrix was obtained by iterating
its equation of motion. This particular component of the density matrix
contains all the phase information that is relevant to the state-transfer
scheme that was proposed and thus contains the information relevant
for dephasing effects. However, our methods can be used to obtain
the phase information for any component of the density matrix. In
section (\ref{sec:Density-matrix-phase}), we showed that when the
spectral density of the bath degrees of freedom obey a Lorentzian
peaked at a non-zero frequency (Eq.(\ref{eq:3.1.31})), the phase
of the relevant component of the density matrix would come back to
unity at the end of the state-transfer protocol. This is only true
if the frequency is given by $\nu_{0}=\frac{2n\pi}{t_{f}}\,\left\{ n\neq2\right\} $.
Consequently, if it is possible to find a mechanism such that this
frequency can be controlled to take on this particular value, it would
always be possible to reduce the dephasing effects of having the system
coupled to a quantum dissipative bath (if damping is involved, then
choosing $\nu_{0}=\frac{2n\pi}{t_{f}}\,\left\{ n\neq2\right\} $ only
gives rise to partial recurrences). 

After developing these general theoretical methods, we applied them
to the case of a four-level atom coupled to a driven cavity. Two laser
tones were used to drive the cavity, each detuned from the cavity
frequency. The classical component of the laser fields were used to
create transitions allowing us to perform the state-transfer protocol
described in chapter three. In order to ensure that each laser field
only induced transitions between the appropriate atomic levels (eliminating
the off-resonant terms), the conditions $\frac{\alpha_{2}\left(t\right)g_{1}}{\left|\Delta_{2}-\Delta_{1}\right|}\ll1$
and $\frac{\alpha_{1}\left(t\right)g_{2}}{\left|\Delta_{2}-\Delta_{1}\right|}\ll1$
needed to be satisfied. Since the cavity interacted with its environment
leading to fluctuations in the number of photons inside the cavity,
this induced dephasing/dissipation effects when performing our state-transfer
protocol. We rewrote the initial Hamiltonian of Eq.(\ref{eq:4.1.1})
by first writing it in a superadiabatic basis and then performing
a Schrieffer-Wolff transformation. Doing so allowed us to apply our
theoretical methods to the state-transfer protocol arising from driving
the cavity. For our theory to apply, we needed that the conditions
in table (\ref{Table-1:-List}) to be satisfied. We calculated the
fidelity of the state-transfer protocol using a circular path in $\left\{ \Omega_{1},\Omega_{2}\right\} $
space (with $\theta\left(t\right)$ being a linear function of time)
allowing the freedom to repeat the state-transfer cycle an arbitrary
number of times. We showed that doing so would extend the coherence
time of the qubit state being transferred. 

To ensure that the adiabatic criteria remained satisfied for this
state-transfer cycle, the condition $\frac{n\pi}{Gt_{f}}\ll1$ must
be satisfied. Finally, for the class of functions considered in Eq.(\ref{eq:4.6.9}),
we gave an algorithm for finding an optimal path that minimized dephasing
effects for our particular state-transfer protocol. We considered
an example where we chose the frequency cutoff at $n=2$ and in this
case the optimal path was given by (\ref{eq:4.6.29}). The algorithm
was found using a short-time expansion after writing the function
$\cos2\theta\left(t,t_{f}\right)$ in a Fourier series expansion with
a finite-frequency cutoff. As the figure in (\ref{fig:Optimal-path-comparison})
shows, there is a significant improvement of the fidelity for the
path given by (\ref{eq:4.6.29}) over the function $\cos2\theta\left(t,t_{f}\right)=\cos\left(\frac{4\pi nt}{t_{f}}\right)$. 

Finally, for the optimal paths found in section (\ref{sec:Many-cycle-evolution}),
we were able to find a simple functional form for the phase $\phi\left(t\right)$
given by (\ref{eq:4.7.12}) and (\ref{eq:4.7.17}) which was particularly
well suited for performing a phase gate. We showed that many convenient
choices were possible for the phase $\phi\left(t\right)$ once the
optimal path of $\cos2\theta\left(t,t_{f}\right)$ had been found.

Our work goes beyond the usual Bloch-Redfield master equation approach
for studying state-transfer protocols and performing phase gates as
in \cite{key-29,key-48} by performing a secular approximation in
the superadiabatic basis. This ensures that our results are not constrained
to environments with short correlation times compared to the coherence
time of the ``$0d$'' component of the density matrix and we don't
need to assume weak dissipation. We also considered a particular noise
model arising from photon cavity shot noise instead of adding them
by hand which is crucial when performing state-transfer protocols/phase
gates for an atom coupled to a cavity. 

\appendix

\chapter{\label{chap:Finding--for}Finding $H_{env}\left(t\right)$ for a
driven cavity}

In section (\ref{sec:Review-of-photon}), we considered a cavity driven
by an external field. Making a distinction between the internal cavity
modes and the external bath modes, we wrote the Hamiltonian as 
\begin{equation}
\hat{H}=\hat{H}_{sys}+\hat{H}_{bath}+\hat{H}_{int}.\label{eq:A.0.1}
\end{equation}
The bath Hamiltonian was described by a collection of harmonic modes
\begin{equation}
\hat{H}_{bath}=\sum_{q}\hbar\omega_{q}\hat{b}_{q}^{\dagger}\hat{b}_{q},\label{eq:A.0.2}
\end{equation}
which obeyed the commutation relations 
\begin{equation}
\left[\hat{b}_{q},\hat{b}_{q'}^{\dagger}\right]=\delta_{q,q'}.\label{eq:A.0.3}
\end{equation}
Within a rotating wave approximation and using (\ref{eq:2.4.15}),
the coupling Hamiltonian was described as 
\begin{equation}
\hat{H}_{int}=-i\hbar\sqrt{\frac{\kappa}{2\pi\rho}}\sum_{q}\left[\hat{a}^{\dagger}\hat{b}_{q}-\hat{b}_{q}^{\dagger}\hat{a}\right].\label{eq:A.0.4}
\end{equation}
Now, we consider the situation of section (\ref{sec:Dephasing-due-to})
where an atom is coupled to a driven cavity. Starting at a time $t=t_{0}<0$,
we turn on the laser drive tones. In this situation, the density matrix
for the atom and environment (cavity + bath) can be written as a product
state 
\begin{equation}
\hat{\rho}\left(t_{0}\right)=\hat{\rho}_{at}\left(t_{0}\right)\otimes\hat{\overline{\rho}}_{env},\label{eq:A.0.5}
\end{equation}
with the environmental part to the density matrix being in thermal
equilibrium
\begin{equation}
\hat{\overline{\rho}}_{env}=\frac{1}{Z}e^{-\beta\hat{H}_{env}}.\label{eq:A.0.6}
\end{equation}
The Hamiltonian $\hat{H}_{env}$ will incorporate the dynamics of
the cavity and bath modes as well as the cavity-bath coupling contributions.
Consequently, we can write 
\begin{equation}
\hat{H}_{env}=\sum_{q}\hbar\omega_{q}\hat{b}_{q}^{\dagger}\hat{b}_{q}+\omega_{c}\hat{a}^{\dagger}\hat{a}-i\hbar\sqrt{\frac{\kappa}{2\pi\rho}}\sum_{q}\left[\hat{a}^{\dagger}\hat{b}_{q}-\hat{b}_{q}^{\dagger}\hat{a}\right].\label{eq:A.0.7}
\end{equation}
After turning on the laser drive tones, the laser will populate the
environment modes so that $\hat{\rho}_{env}\left(t\right)$ will no
longer be in thermal equilibrium described by (\ref{eq:A.0.6}) for
$t>t_{0}$. Defining 
\begin{equation}
\beta\left(t\right)\equiv\beta_{1}\left(t\right)+\beta_{2}\left(t\right),\label{eq:A.0.8}
\end{equation}
with the input field being a coherent drive with a classical and quantum
part (see (\ref{eq:4.1.7})), the environment can approximately be
described by a coherent state with a non-zero expectation value for
the input field given by 
\begin{equation}
\left\langle \hat{b}_{in}\left(t\right)\right\rangle =\mathrm{Tr}\left\{ \hat{b}_{in}\hat{\rho}_{env}\left(t\right)\right\} =e^{i\omega_{c}t}\beta\left(t\right),
\end{equation}
where, following (\ref{eq:2.4.16}), we have 
\begin{equation}
\hat{b}_{in}=\frac{1}{\sqrt{2\pi\rho}}\sum_{q}\hat{b}_{q}.\label{eq:A.0.10}
\end{equation}
\begin{figure}
\begin{centering}
\includegraphics[scale=0.6]{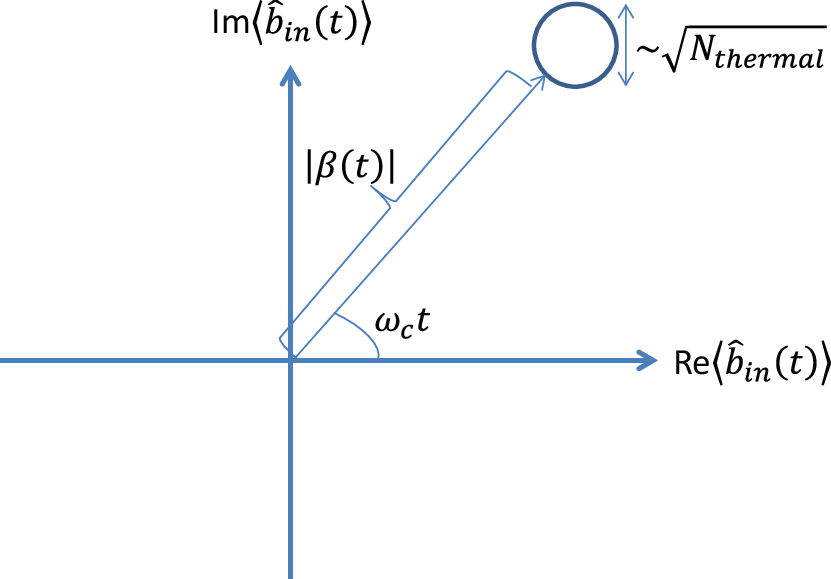}
\par\end{centering}

\centering{}\caption{\ Phase space plot of the coherent state $\left\langle \hat{b}_{in}\left(t\right)\right\rangle $ }
\textit{The above figure corresponds to the phase space plot of the
coherent state }$\left\langle \hat{b}_{in}\left(t\right)\right\rangle $.
\textit{The blob represents vacuum noise arising from thermal fluctuations
in $\hat{b}_{in}\left(t\right)$ with the amplitude being determined
by $\sqrt{N_{th}}$.}
\end{figure}

We can get a time-independent $\hat{\rho}_{env}\left(t\right)$ by
performing a displacement transformation using the unitary operator
\begin{equation}
\hat{U}_{\beta}\left(t\right)=e^{-ie^{i\omega_{c}t}\beta\left(t\right)\hat{p}},\label{eq:A.0.11}
\end{equation}
where
\begin{equation}
\hat{p}=\alpha\left(\hat{b}_{in}-i\hat{b}_{in}^{\dagger}\right),\label{eq:A.0.12}
\end{equation}
and $\alpha$ is a proportionality constant. Indeed, one can verify
that the unitary operator written in (\ref{eq:A.0.11}) displaces
$\hat{b}_{in}$ in the following way
\begin{equation}
\hat{U}_{\beta}\left(t\right)\hat{b}_{in}\hat{U}_{\beta}^{\dagger}\left(t\right)=e^{i\omega_{c}t}\beta\left(t\right)+\hat{b}_{in}.\label{eq:A.0.13}
\end{equation}
With this displacement transformation, $\hat{\rho}_{env}$ transforms
as 
\begin{equation}
\hat{U}_{\beta}\left(t\right)\hat{\rho}_{env}\left(t\right)\hat{U}_{\beta}^{\dagger}\left(t\right)=\hat{\bar{\rho}}_{\beta}\label{eq:A.0.14}
\end{equation}
which is now time-independent. The Hamiltonian (\ref{eq:A.0.7}) will
also transform under the unitary operator (\ref{eq:A.0.11}) as 
\begin{equation}
\hat{H}'_{env}\left(t\right)=\hat{U}_{\beta}\left(t\right)\hat{b}_{in}\hat{U}_{\beta}^{\dagger}\left(t\right)-i\hat{U}_{\beta}\left(t\right)\dot{\hat{U}}_{\beta}^{\dagger}\left(t\right).\label{eq:A.0.15}
\end{equation}
Choosing the complex amplitude $\beta\left(t\right)$ to solve the
classical equation of motion and taking $t_{0}\rightarrow-\infty$,
the transformed Hamiltonian (\ref{eq:A.0.15}) becomes
\begin{equation}
\hat{H}'_{env}\left(t\right)=\sum_{q}\hbar\omega_{q}\hat{b}_{q}^{\dagger}\hat{b}_{q}+\omega_{c}\hat{a}^{\dagger}\hat{a}+\left(-i\hbar\sqrt{\kappa}\sum_{q}e^{i\omega_{c}t}\beta\left(t\right)\hat{a}+h.c.\right)-i\hbar\sqrt{\frac{\kappa}{2\pi\rho}}\sum_{q}\left[\hat{a}^{\dagger}\hat{b}_{q}-\hat{b}_{q}^{\dagger}\hat{a}\right]+const\label{eq:A.0.16}
\end{equation}
where $-i\hbar\sqrt{\kappa}\sum_{q}e^{i\omega_{c}t}\beta\left(t\right)\hat{a}+h.c.$
corresponds to the classical drive term. 

\newpage{}

\end{document}